\documentclass[11pt]{elsarticle}

    \long\def\comment#1{ }

  \usepackage{latexsym,bm,amsmath,amssymb,amsfonts}
  \usepackage{epsfig,graphics,graphicx}
  \usepackage{slashed}
  \usepackage{cite}
  \usepackage[latin1]{inputenc}
  \usepackage{mathrsfs}
\usepackage{mathcomp}

  \long\def\comment#1{ }

  \newcommand{\eqnum}[1]{Eq.~\eqref{#1}}

  \newcommand{\abar}{\bar{\alpha}_s}
  
  \newcommand{\del}{\partial}

  \newcommand{\mcal}{\mathcal}
  \newcommand{\rme}{{\rm e}}
  \newcommand{\rmd}{{\rm d}}   

  \newcommand{\Lam}{\Lambda_{{\rm QCD}}}
  
  \newcommand{\nn}{\nonumber\\}
   
  \newcommand{\order}[1]{\mcal{O}{(#1)}}

  \newcommand{\beq}{\begin{eqnarray}}
  \newcommand{\eeq}{\end{eqnarray}}
  
 \def\simge{\mathrel{%
   \rlap{\raise 0.511ex \hbox{$>$}}{\lower 0.511ex \hbox{$\sim$}}}}
\def\simle{\mathrel{
   \rlap{\raise 0.511ex \hbox{$<$}}{\lower 0.511ex \hbox{$\sim$}}}}

\vspace{1.5cm}

\begin{document}
\begin{frontmatter}

\title{CCFM Evolution with Unitarity Corrections}

\author{Emil Avsar, Edmond Iancu}\ead{Emil.Avsar@cea.fr, Edmond.Iancu@cea.fr}

\address{Institut de Physique Th\'eorique de Saclay,
 F-91191 Gif-sur-Yvette, France}

\date{\today}
\vspace{1.2cm}
\begin{abstract}
We considerably extend our previous analysis of the implementation of an
absorptive  boundary condition, which mimics saturation effects, on the
linear CCFM evolution.  We present detailed results for the evolution of
the unintegrated gluon density in the presence of saturation and extract
the energy dependence of the emerging saturation momentum. We show that
CCFM and BFKL evolution lead to almost identical predictions after
including the effects of gluon saturation and of the running of the
coupling. We moreover elucidate several important and subtle aspects of
the CCFM formalism, such as its relation to BFKL, the structure of the
angular ordered cascade, and the derivation of more inclusive versions of
CCFM. We also propose non--leading modifications of the standard CCFM
evolution which may play an important role for phenomenological studies.
\end{abstract}

\end{frontmatter}

\section{Introduction}
\setcounter{equation}{0}

In a previous work, Ref.~\citep{Avsar:2009pv}, we have proposed a method
for effectively implementing saturation and unitarity within a generic
linear evolution equation for the unintegrated gluon distribution, so
like the BFKL \citep{BFKL} and the CCFM \citep{Ciafaloni:1987ur,
Catani:1989sg, Catani:1989yc} equations. The method is based on enforcing
an absorptive boundary condition at low transverse momenta which prevents
the gluon phase--space occupation numbers to grow beyond their physical
values at saturation. Our method is the extension of a strategy
originally introduced in relation with analytic studies of the BFKL
evolution in the presence of saturation
\citep{Iancu:2002tr,Mueller:2002zm}, whose deeper justification
\citep{Munier:2003vc} lies in the correspondence between high--energy
evolution in QCD and the reaction--diffusion process in statistical
physics \citep{Iancu:2004es}. This correspondence is however limited to
asymptotically high energies and to a fixed coupling
\citep{Dumitru:2007ew}, whereas our analysis in Ref.~\citep{Avsar:2009pv}
shows that the absorptive boundary method is in fact more general and
also very powerful. After reformulating this method in such a way to be
suitable for numerical simulations, we have demonstrated its efficiency
by comparing the numerical solutions to the BFKL equation with absorptive
boundary condition against those of the proper non--linear generalization
of BFKL which includes saturation --- the Balitsky--Kovchegov (BK)
equation \citep{Balitsky:1995ub,Kovchegov:1999yj}. We have thus shown
that the absorptive boundary method successfully reproduces the results
of the BK equation for both fixed and running coupling, and for {\em all}
the energies, and not only the asymptotic ones. This success, together
with its relative simplicity, makes this method a very compelling tool
for phenomenological studies at LHC and, in particular, for implementing
saturation within Monte Carlo based event generators, such as CASCADE
\citep{Jung:2000hk} and LDCMC \citep{Kharraziha:1997dn}. Most
importantly, our effective method can be also implemented within
formalisms whose non--linear generalizations are not known, so like the
CCFM formalism  \citep{Ciafaloni:1987ur, Catani:1989sg, Catani:1989yc}
which lies at the basis of the above mentioned generators. It is our main
purpose in this paper to provide an extensive numerical study of the CCFM
evolution supplemented with the absorptive boundary condition, and thus
demonstrate the role of the saturation effects in that context. Some
preliminary results in that sense were already presented in
Ref.~\citep{Avsar:2009pv}, but out present analysis will considerably
enlarge that previous analysis, in particular by exploring the CCFM
formalism in much more detail.

Based on our current theoretical understanding and on extrapolations from
the phenomenology at HERA and RHIC, we expect at LHC a considerably
larger phase space where saturation effects should be important. The
characteristic transverse momentum scale for the onset of unitarity
corrections is the saturation momentum $Q_s$, and is expected to grow
quite fast with increasing energy. Next--to--leading order BFKL
calculations \citep{Triantafyllopoulos:2002nz} suggest a power law
$Q_s^2\sim s^\lambda$ with $\lambda\simeq  0.2\div 0.3$, which appears to
be supported by the HERA data at small $x\le 0.01$
\citep{Stasto:2000er,Iancu:2003ge,Soyez:2007kg}. For forward jet
production in proton--proton collisions at LHC, one thus expects $Q_s$ in
the ballpark of 2 to 3 GeV. Even higher values could be reached in
nucleus--nucleus collisions, or in some rare events,  like
Mueller--Navelet jets \citep{Iancu:2008kb}.

\emph{A priori}, it seems natural to look for saturation effects in the
underlying event, that is, in the bulk of the particle production at
relatively low momenta, where the saturation effects (also viewed as
multiple scattering \citep{Acosta:2004wqa, Field:2002vt}) are clearly
important. However, the underlying event at LHC will be extremely complex
and difficult to study. Besides, for such low momenta, it may be
difficult to separate saturation physics from the genuinely
non--perturbative physics in the soft sector of QCD. (A similar ambiguity
occurs in the interpretation of the small--$x$ data at HERA.) It is
therefore useful to recall at this point that saturation effects can make
themselves felt even at relatively large momenta $Q$, well above $Q_s$,
via phenomena like the ``geometric scaling'' observed at HERA
\citep{Stasto:2000er,Marquet:2006jb, Gelis:2006bs}. Such phenomena, which
reflect the change in the unintegrated gluon distribution at high
$k_\perp\gg Q_s$ due to saturation at low $k_\perp\lesssim Q_s$
\citep{Iancu:2002tr,Mueller:2002zm,Triantafyllopoulos:2002nz,Munier:2003vc},
are particularly interesting in that they represent signatures of
saturation in the high--$Q^2$ domain which was traditionally believed to
fully lie within the realm of the ``standard'' pQCD formalism
--- the DGLAP evolution \citep{DGLAP} of the parton distributions
together with the collinear factorization of the hadronic
cross--sections.

Schemes based on the NLO DGLAP evolution have been quite successful at
HERA (at least for not too low $Q^2$) \citep{Klein:2008di}, but at LHC
one will probe much smaller values of Bjorken $x$, and this even for
relatively hard $Q^2$ (e.g., in the forward kinematics). The jets to be
measured at LHC will carry relatively large transverse momenta $Q\ge
10$~GeV, but because of the high--energy kinematics, their description
may require $k_\perp$--factorization \citep{KTFACT} together with the BFKL, or CCFM,
evolution of the unintegrated gluon distribution. Moreover, saturation
effects, like geometric scaling, could manifest themselves in hard
observables at LHC, so like the cross--section for forward jet production
\citep{Iancu:2008kb}. Thus LHC will for the first time allow us to study
saturation physics in the kinematical regime where this physics lies in
the realm of perturbative QCD.

The BFKL formalism, properly generalized to include the non--linear
effects responsible for gluon saturation
\citep{Balitsky:1995ub,Kovchegov:1999yj,JKLW,CGC,CGCreviews}, is
specially tailored to describe the evolution of the unintegrated gluon
distribution with increasing energy and its approach towards saturation.
As such, this is well--suited to study the high--energy evolution of
inclusive cross--sections, and it is able to accommodate important
phenomena, like the geometric scaling at HERA
\citep{Stasto:2000er,Marquet:2006jb, Gelis:2006bs}, or the turnover in
the DIS structure function $F_2(x,Q^2)$ at semi--hard $Q^2$
\citep{Iancu:2003ge,Soyez:2007kg}. However, this is not the right
formalism to also describe exclusive final states, because it misses the
quantum coherence between successive gluon emissions in the process of
high--energy evolution. Besides, there are additional problems, to be
discussed at the end of section \ref{sec:angord}, which make the BFKL
formalism unsuitable for studying exclusive final states. All such
problems are corrected in the CCFM formalism \citep{Ciafaloni:1987ur,
Catani:1989sg, Catani:1989yc}, which has also the advantage to apply
within a wider kinematical region, interpolating between the high energy
evolution (the realm of BFKL) and the evolution with increasing
virtuality (the realm of DGLAP). A similar formalism derived out of CCFM
but using a different physical picture for the evolution is the Linked
Dipole Chain (LDC) model \citep{Andersson:1995ju} which covers the same
kinematical region as CCFM, and our method of implementing saturation can
be equally well applied also to this formalism. Infact one of the main
equations to which we apply our method is equivalent to the master
equation in LDC. This will be discussed more below.

Like BFKL, the CCFM formalism is based on the $k_\perp$--factorization.
This makes it possible to take into account some of the NLO corrections
in the collinear approach by simply treating the kinematics of the
scattering more accurately \citep{Andersson:2002cf, Andersen:2003xj, Andersen:2006pg}. 
Still like BFKL, the CCFM evolution resums all powers
of $\alpha_s$ which are accompanied by large energy logarithms $\ln s$,
with $s$ the center of mass energy. In fact, the CCFM and BFKL evolutions
yield identical predictions for the dominant behaviour in the formal
high--energy limit $s\to\infty$. But this formal limit is conceptually
unrealistic and phenomenologically irrelevant, since it violates
unitarity. What is relevant, is the approach towards the unitarity limit
and gluon saturation with increasing energy, which is \emph{a priori}
different in the two formalisms. In the recent years, this approach has
been extensively studied within the BFKL evolution, by using its
non--linear generalizations: the BK
\citep{Balitsky:1995ub,Kovchegov:1999yj} and the JIMWLK equations
\citep{JKLW,CGC}. However, no such a study was performed within the
context of the CCFM evolution prior to our recent work
\citep{Avsar:2009pv}. As already stated, such a study is the main
objective of the present work.

The present study should be viewed as a step towards a systematic
inclusion of the effects of saturation in the description of exclusive
final states. Most studies have so far concentrated on more inclusive
observables, for which a good description can generally be obtained (for
sufficiently high $Q^2$) also with linear evolution equations alone. A
number of papers emphasized the need and importance of studying
saturation in exclusive final states \citep{Avsar:2005iz, Avsar:2006jy,
Avsar:2007xg, Flensburg:2008ag}, but so far the explicit studies were
mostly confined to inclusive observables. The possibility of looking at
more exclusive observables will undoubtedly make it easier to distinguish
the predictions of linear and non--linear evolutions.

But before we consider the effects of saturation, we shall dedicate a
large part of this paper to a detailed presentation of the respective
linear evolution, in order to clarify several non--trivial aspects of the
CCFM formalism which are important for our purposes. In doing so, we will
try to carefully motivate the steps leading to the final evolution
equations (simplified versions of the CCFM formalism) on which we shall
apply our saturation boundary. Among the different aspects of CCFM to be
discussed here, there are parts which have been already presented in
previous papers that we shall refer to, but there are also parts which to
our knowledge have never been presented before. In this paper, we shall
try to give a unified presentation of all the relevant aspects, using a
intuitive geometrical representation for the phase--space of the CCFM
evolution. To facilitate the reading of the paper, we have moved some of
the most technical developments to appendices, and kept only the
important results in the main text. Here is a summary of the topics to be
covered in what follows and also of our main conclusions:

Sect.~2 will introduce the basics of the CCFM formalism, that the rest of
the work will rely on. In particular, in Sect.~\ref{sec:bfklrel} we shall
clarify the relation between the phase--space of the CCFM evolution and
that of the BFKL evolution, thus recovering previous results in
Ref.~\citep{Salam:1999ft}, but from a different perspective. In
Sect.~\ref{sec:angord} we shall present the standard version of the CCFM
evolution as an integral equation, and on this occasion we shall explain
the approximations which are involved in this rewriting and which are
often kept implicit in the literature. In Sect.~\ref{sec:virtformfac} and
Appendix A we discuss some subtle aspects, and correct some mistakes in
the literature, concerning the virtual (`non--Sudakov') form factors
which express the probability for not emitting gluon in the course of
the evolution. The careful derivation of these form factors, as outlined
in Sects~\ref{sec:angord} and \ref{sec:virtformfac} and in Appendix A,
will also allow us to better understand their physical origin and thus
propose some improved expressions for them, which treat more accurately
the kinematics (by including effects of recoils in the energy). The new
form factors, to be presented in appendix B, differ from the standard
ones by terms which are formally of higher order, but which may be
numerically important\footnote{We shall not include these new form
factors in our present numerical analysis since their structure is such
that they can be efficiently implemented only within Monte Carlo
simulations.} and thus interesting for the phenomenology.  Another aspect
of the form factor is its appropriate form in the formal high energy limit, which 
is relevant again for the comparison to BFKL and this will be discussed in appendix
C. 


Sections \ref{sec:inclccfm} and \ref{sec:ccfmsat} are the key sections in
this paper. In Sect~\ref{sec:inclccfm} we successively simplify the CCFM
formalism and reduce it to a set of simpler, integral and differential,
equations, which are more suitable for numerical simulations. The most
general equation, the integral equation \eqref{eq:ccfminteq1}, explicitely
preserves all the hard and soft gluon emissions from the $t$-channel 
propagators. 
This is the
equation at the basis of the CASCADE Monte Carlo event generator
\citep{Jung:2000hk}, and it can be generalized to include saturation
effects, as we shall explain in Sect.~\ref{sec:saturation}. But for the
present purposes, it is preferable to work with simpler versions of the
CCFM evolution, namely Eqs.~\eqref{eq:ccfmdiffeq3} and
\eqref{eq:ccfmdiffeq4}, which are more `inclusive' --- in the sense that
some of the virtual form factors are used to cancel the `soft' gluon
emissions. Such cancellations are not exact, but rather require some
additional kinematical approximations, which are however in the spirit of
the CCFM formalism. These approximations are also similar to those
underlying the `Linked Dipole Chain' (LDC) model
\citep{Andersson:1995ju}. And indeed, our \eqnum{eq:ccfmdiffeq3} appears
to be equivalent to the master equation in Ref.~\citep{Andersson:1995ju},
although our respective approaches are quite different. Intriguingly, it
turns out that if one performs the same type of approximations starting
with the BFKL formalism, one is again led to the same two equations
\eqref{eq:ccfmdiffeq3} and \eqref{eq:ccfmdiffeq4} (see the discussion in
Sect.~\ref{sec:ccfmcfbfkl}). In our opinion, this points out towards a
deep similarity between the BFKL and CCFM formalisms, which when applied
to inclusive observables differ only in the accuracy to which they treat
the kinematics.

In Sect.~\ref{sec:ccfmsat} we introduce the absorptive boundary method
which effectively implements saturation within a linear evolution
equation, so like BFKL or CCFM. First, in Sect.~\ref{sec:saturation}, we
describe this method on the example of the BFKL equation, where the
comparison with the non--linear BK equation will allow us to demonstrate
the success of the method. Then, in Sect.~\ref{sec:anmdim}, we
analytically study the high--energy behaviour of the approximate CCFM
equations \eqref{eq:ccfmdiffeq3} and \eqref{eq:ccfmdiffeq4} and thus
determine the energy--dependence of the associated saturation momenta, in
the case of a fixed coupling. Our analysis shows that the CCFM evolution
is somewhat faster than the BFKL one: the respective saturation exponent
is slightly larger. On the other hand, the characteristic functions which
determine the momentum--dependent anomalous dimension, are very similar
to the BFKL one, for both Eq.~\eqref{eq:ccfmdiffeq3} and
\eqnum{eq:ccfmdiffeq4}. In Appendix D, we relax some of the
approximations used in deriving \eqnum{eq:ccfmdiffeq3} and thus obtain a
more accurate equation, whose high--energy predictions are even closer to
those of the BFKL equation.

Finally, Section \ref{sec:results} presents a systematic numerical
analysis of the various evolution equations --- the BFKL equation and the
simplified versions, Eqs.~\eqref{eq:ccfmdiffeq3} and
\eqref{eq:ccfmdiffeq4}, of the CCFM evolution --- with the purpose to
illustrate the role of the saturation boundary and also of the running
coupling effects. We first demonstrate that the respective linear
equations (no saturation boundary) lead to rather different evolutions,
which are moreover infrared unstable in the case of a running coupling.
Then we show that the infrared instability is cured after including the
saturation boundary (the saturation momentum effectively acts as a hard
infrared cutoff which increases with the energy) and moreover the
respective predictions of the various equations remain very close to each
other, up to astronomically high energies. Hence, in so far as the
unintegrated gluon distribution is concerned, the BFKL and CCFM
evolutions properly corrected for saturation and endowed with a running
coupling are rather similar to each other.

\section{Review of CCFM}
\setcounter{equation}{0}

Our aim in this section is to review the CCFM approach, mainly the work
in \citep{Catani:1989sg}. This will prepare us in understanding our
subsequent strategy for simplifying this formalism and most efficiently
complete it with a saturation boundary. The original formulation in
\citep{Catani:1989sg} is very careful and complete, but also quite
technical and not always transparent. We shall therefore try to mostly
give a geometrical representation of the equations to be presented here.
In this process, we will derive to some interesting results that we were
not aware of, and also clarify some points which are often confusing in
the literature, such as the precise form of the ``non--Sudakov'' form
factor $\Delta_{ns}$.

\subsection{Kinematics and Basics}
\label{subsec:kin}

\begin{figure}[t]{\centerline{
    \includegraphics[angle=0, scale=0.6]{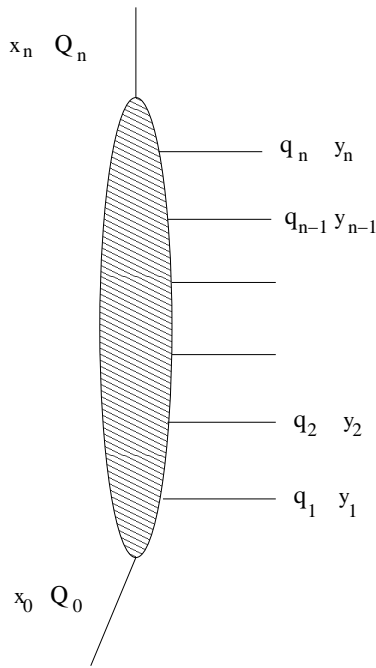}}
\caption {\sl \label{fig:ladder} Kinematics of the gluon radiation. The last
  t-channel propagator is denoted by $Q_n$, while the real $s$-channel gluons
  are represented by the horizontal lines. }
}
\end{figure}
We use Fig.~\ref{fig:ladder} to define the kinematics and schematically
introduce the physical picture. This figure represents a gluon ladder as
produced by the CCFM evolution; we denote by $q_i$ the transverse momenta
of the real, $s$--channel, gluons, and with $Q_i$ the transverse momenta
of the space--like, $t$--channel, propagators, which are not explicitly
shown in the figure. The incoming virtual gluon has zero transverse
momentum $Q_0=0$ (that is, this gluon is taken to be collinear with the
parent hadron, not shown here), and hence one has $Q_i = -\sum_{k=1}^i
q_k$. We will use $y_i$ and $x_i$ to denote the energy fractions of the
$s$--channel and $t$--channel gluons, respectively, measured with respect
to the energy $E$ of the incoming proton. In Fig.~\ref{fig:ladder}, the
$s$--channel (or `real') gluons are enumerated according to their energy:
\begin{eqnarray}
x=x_n \ll y_n \ll y_{n-1} \ll \dots \ll y_2 \ll y_1 \approx 1\,,
\label{eq:energyord}
\end{eqnarray}
but this ordering is not necessarily the same as that of the gluon
emissions along the ladder (i.e., it is not assumed that the gluon with
energy fraction $y_i$ is emitted out right after that with fraction
$y_{i-1}$, etc.). Rather, as we shall shortly see, the real gluon emissions in
the CCFM ladder are ordered according to their angles $\xi_i \equiv
q_i^2/(y_i^2E^2)$. This ordering issue is potentially confusing, since
e.g. the relation $Q_i = -\sum_k^i q_k$ would be strictly true if the
labels $i$ attached to gluons in Fig.~\ref{fig:ladder} were also
indicating their order of emission, {\em i.e.}, if the real gluons
emissions were ordered according to their energy (which they are not).
The resolution of this puzzle, to be explained in detail later on, is
that the actual emissions which do not obey energy ordering are also soft
in the sense of carrying little transverse momenta
(`$k_\perp$--conserving'), and hence they do not affect the momenta $Q_i$
of the $t$--channel gluons: the latter are fully determined by the `hard'
emissions which are simultaneously ordered in angle and in energy. (See
Sect.~\ref{sec:angord} for details.)

The (integrated) CCFM gluon structure function can be written as
\begin{eqnarray}
\mathcal{A}(x,\bar{\xi}) = \sum_n \int \prod_{i=1}^n \left( \abar \frac{\rmd \xi_i}{\xi_i}
\theta(\bar{\xi} - \xi_i)
\frac{\rmd y_i}{y_i} \theta(y_i - y_{i+1})
\sum_{perm} \theta(\xi_{l_i} - \xi_{l_{i-1}}) \right) \nonumber \\
\times  \frac{1}{x_n}\delta(x-x_n)
S_{eik}^2(y_1,\bar{\xi})S^2_{ne}(12\dots n)
\label{eq:structurefunc}
\end{eqnarray}
where $y_{n+1} \equiv 0$, and $S_{eik}$ and $S_{ne}$ are the virtual
corrections associated with the eikonal and the non-eikonal vertices
respectively. The theta function $\theta(\xi_{l_i} - \xi_{l_{i-1}})$ is
a consequence of the quantum coherence between successive emission which
implies that the emission angle must increase when moving upwards along
the ladder (\emph{i.e.}, towards the hard scattering). Notice that one
can have any angular ordering for the given energy ordering. (We shall
later relabel the real gluons according to the angular ordering.) The
angle $\bar{\xi}$ in the argument of $\mathcal{A}$ is the maximum angle
allowed by coherence and is determined by the kinematics of the hard
scattering; roughly, $\bar{\xi}\simeq Q^2/x^2E^2$ with $Q^2$ the
virtuality of the incoming photon. In this case, the structure function
\eqref{eq:structurefunc} gives the gluon distribution,
$\mathcal{A}(x,\bar{\xi}) = xg(x,Q^2)$ \citep{Marchesini:1994wr}.

\comment{If one also defines $z_i$ by $x_i\equiv z_{i}x_{i-1}$ (and
therefore $y_i=x_{i-1}(1-z_i)$), then
\begin{eqnarray}
\frac{1}{x}\prod_i \frac{\rmd y_i}{y_i}=\prod_i \frac{\rmd z_i}{z_i(1-z_i)} =
\prod_i  dz_i \biggl (\frac{1}{z_i}
+\frac{1}{1-z_i}\biggr )
\label{eq:splitfunc}
\end{eqnarray}
which is the more familiar term for the splitting function. Thus in the
CCFM ladder there are two vertices contributing to the splitting $Q_{i-1}
\to q_i + Q_i$. The one corresponding to the small--$z_i$ pole $1/z_i$ is
dubbed the ``non-eikonal'' vertex as it comes from the piece of the three
gluon vertex in which the polarization of the parent gluon is inherited
by the real gluon $q_i$. The opposite $1/(1-z_i)$ pole is dubbed the
``eikonal'' vertex, and in this case the polarization, together with most
of the energy, is inherited by the $t$-channel propagator $Q_i$. In
contrast, in a BFKL ladder, only the $1/z_i$ pole would be present, and
the energy ordering would be the same as the actual emission ordering
along the cascade.
 }

The virtual form factors in \eqref{eq:structurefunc} are given by
\begin{eqnarray}
S_{eik}(y_1,\bar{\xi}) &=& \exp\left( -\frac{1}{2}\,\abar
\int^{y_1} \frac{\rmd y}{y} \int^{\bar{\xi}} \frac{\rmd \xi}{\xi} \right) \label{eq:eik} \\
S_{ne}(12\dots n) &=& \prod_{k=1}^n \exp \left( \frac{1}{2}\,\abar \int_{y_{k+1}}^{y_k} \frac{\rmd y}{y}
\int_{\xi(Q_k)}^{\bar{\xi}} \frac{\rmd \xi}{\xi} \right) \equiv \prod_{k=1}^n S_{ne}(k)
\label{eq:noneik}
\end{eqnarray}
where $\xi(Q_k) \equiv Q_k^2/(y^2E^2)$ and the integral over the
transverse momentum $q^2$ has been written in terms of the associated
angular variable $\xi^2\equiv q^2/y^2E^2$. Notice that the exponent in
the non--eikonal form factor is positive. The apparent divergence in
$S_{eik}$ and in the real emission density $d\xi/\xi$ is regulated by a
collinear momentum cut $q_0$ (which also regulates the $dy/y$ divergence
in $S_{eik}$).

One should be aware that \eqref{eq:structurefunc} does not correspond to
an exclusive final state. In \eqref{eq:structurefunc} one has already
inclusively summed over all subsequent emissions from the outgoing gluons
$q_i$. Each such gluon can further radiate within a cone of half opening
angle $\xi_i$. These final emissions are such that the real emission
probability is exactly compensated by virtual corrections of the type
$S_{eik}$, and they are therefore not visible in the expression
\eqref{eq:structurefunc} for the (inclusive) gluon distribution. For the
study of exclusive final states, however, one needs to include all the
emissions.

\subsection{The phase space and relation to BFKL}
\label{sec:bfklrel}

In this subsection we shall describe in detail the phase--space for real
and virtual gluon emissions within the CCFM ladder and then argue that,
up to subleading effects, this is essentially the same as the
phase--space for the BFKL evolution. This will allow us to conclude that
the CCFM and BFKL evolutions become identical with each other in the high
energy limit. (See also Ref. \citep{Salam:1999ft} for similar
considerations.)

\begin{figure}[t] { \centerline{
    \includegraphics[angle=0, scale=0.5]{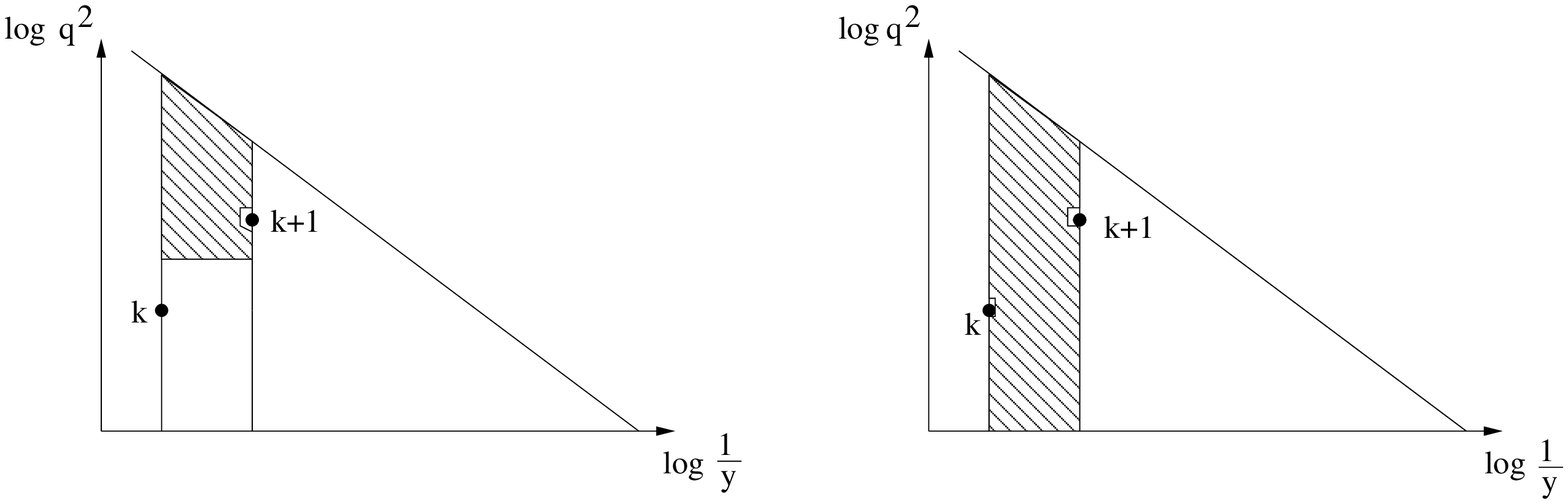}}
  \caption {\sl \label{fig:ccfm2emiss} Geometrical representation of
    two emissions (fat dots) in CCFM shown in the $(\ln(1/y), \ln q^2)$-phase
    space. The total phase space is bounded by the diagonal line which indicates
    the maximal angle allowed by coherence. The shaded regions show the phase regions
    over which $S_{ne}(k)$ (left) and $S_{eik}$ (right) are integrated over. }
}
\end{figure}

Let us start with the virtual emissions and rewrite the eikonal form
factor in \eqref{eq:eik} as
\begin{eqnarray}
S_{eik}(y_1,\bar{\xi}) &=& \prod_{k=1}^n \exp\left( -\frac{1}{2}
\bar{\alpha}_s  \int_{y_{k+1}}^{y_k}
\frac{\rmd y}{y} \int^{\bar{\xi}} \frac{\rmd \xi}{\xi} \right) \nonumber \\
\!\! &\equiv&\!\!\prod_{k=1}^n S_{eik}(k)
\end{eqnarray}
with $y_{n+1} \equiv 0$. Let us now consider the two emissions in
Fig.~\ref{fig:ccfm2emiss}. In this figure the horizontal
axis\footnote{These type of diagrams have been widely used for example in
\citep{Andersson:1995ju}, but notice that there the horizontal is taken
as $\ln(\xi)$ instead of $\ln(1/y)$. Our choice is the same as the one in
\citep{Salam:1999ft}.} is ln$(1/y)$ while the vertical axis is
ln$(q/q_0^2)$ (from now on we shall for simplicity omit using
$q_0$). A black dot in such figures denotes a real gluon with the
respective values for the energy ($y$) and the transverse momentum
($q^2$). Since $\ln \xi = \ln ( q^2/y^2E^2) = \ln (q^2/E^2) +
2\ln(1/y)$, the emission angle will be constant along diagonal lines in
the figure. The diagonal line shown in figure \ref{fig:ccfm2emiss} denotes the
maximum angle, determined by $\bar{\xi}$. The diagonal lines parallel to
this line and which pass through the gluons will indicate the angle of
the gluons (see e.g. Fig.~\ref{fig:ccfmemissions}).

\begin{figure}[t]{\centerline{
    \includegraphics[angle=0, scale=0.55]{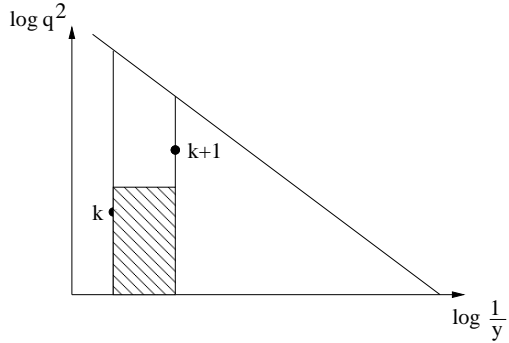}}
\caption {\sl\label{fig:bfkl2emiss} The shaded region is the region left
  over where there is a net contribution from the virtual form factors. 
  It is bounded from above by $Q_k$. The real emissions are represented 
  by fat dots.  }
}
\end{figure}

The shaded regions in the two figures indicate the phase space over which
the non--eikonal and eikonal form factors are integrated over. The
rightmost figure, representing $S_{eik}(k)$, is easier to understand.
Here the $y$ integral is bounded by $y_{k+1}$ and $y_k$, that is we
integrate over the region between the two vertical lines, and the $\xi$
integral goes up to $\bar{\xi}$, as can be seen in the figure.  To
understand the leftmost figure, note that in $S_{ne}(k)$ defined in
\eqref{eq:noneik}, the $y$--integral is the same as in $S_{eik}(k)$ while
the $\xi$--integral (or, equivalently, that over $q^2=y^2E^2\xi^2$) is
integrated from $Q_k^2$ up. The horizontal line bounding the shaded
region in that figure corresponds to $\ln Q_k^2$. Since the exponent is
negative in $S_{eik}(k)$ and is positive in $S_{ne}(k)$, we get a
complete cancelation in the region of overlap. It is then easy to see
from Fig.~\ref{fig:ccfm2emiss} that the region which is left is the one
shown in Fig.~\ref{fig:bfkl2emiss}, and that the net exponent is
negative.

Now, remember that in the BFKL evolution, the virtual corrections are
contained in the ``non--Sudakov'' (or ``non--eikonal'') factor
$\Delta^{BFKL}_{ne}(k)$ defined by
\begin{eqnarray}
\Delta_{ne}^{(BFKL)}(k) = \exp \left( -\abar
\int_{y_{k+1}}^{y_k} \frac{\rmd y}{y} \int^{Q_k^2}_{q_0^2}
 \frac{\rmd q^2}{q^2} \right ),
\label{eq:bfklnoneik}
\end{eqnarray}
and we see that this corresponds to the shaded area in
Fig.~\ref{fig:bfkl2emiss}. Thus we find
\begin{eqnarray}
S^2_{ne}(k) \cdot S^2_{eik}(k) = \Delta^{BFKL}_{ne}(k).
\label{eq:bfklccfm}
\end{eqnarray}

\begin{figure}[t]{\centerline{
    \includegraphics[angle=0, scale=0.6]{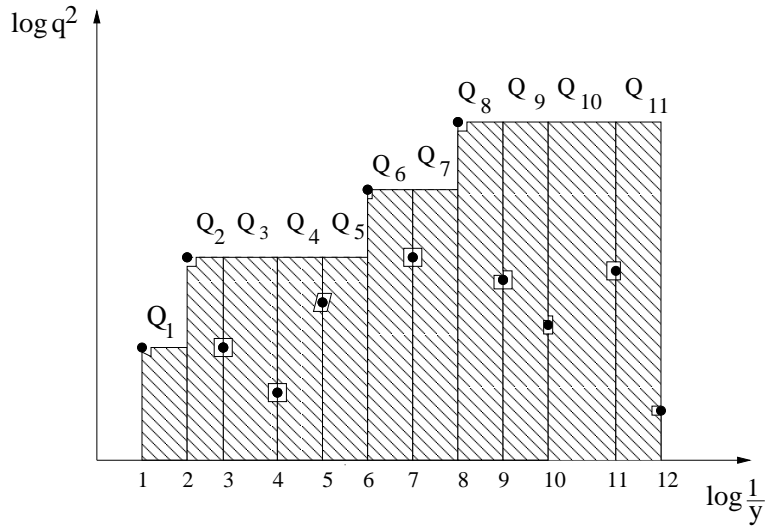}}
\caption {\sl\label{fig:ccfminitial} A complete set of emissions in the
  CCFM phase space. Real emissions are represented by fat dots.
  The shaded regions are the regions over which
  we have contributions from the virtual form factors. We have here
  ignored
  the constraint from the maximal angle. }
}
\end{figure}

Consider now a complete set of emissions in the initial chain as shown in
Fig.~\ref{fig:ccfminitial}. In the figure the gluons are obviously
enumerated according to their energy. The explanation of the phase space
in figure is the following. The $Q_i$ are determined by the relation $Q_i
= -\sum_k^i q_k$ . For example, since $Q_0=0$, we have $Q_1=q_1$ (in the
following all the momenta denote the norm of the transverse components).
Since $q_2 \gg q_1$ we have $Q_2\simeq q_2$,  where the small recoil is
neglected in the figure. Then the subsequent real gluons have small
momenta so that $Q_{k+1} \approx Q_k$ up to gluon 6 which has large
momentum, and therefore $Q_6\simeq q_6$ and so on.  In the end we get an
integral over the total shaded region in Fig.~\ref{fig:ccfminitial}, and
again this is exactly the same that we would have in BFKL.

Of course in CCFM the total phase space is determined by $\bar{\xi}$ so
in Fig.~\ref{fig:ccfminitial} we have assumed the constraint $\xi <
\bar{\xi}$ not to cut the shaded regions. Hence, a difference between the
CCFM and BFKL ladder appears towards the end of the chain. This
difference however is not enhanced by $\ln(1/x)$, and hence it is
subleading from the viewpoint of the BFKL resummation. Moreover, the CCFM
real--gluon emissions are ordered according to their angle, and not to
the energy. However, in the ensuing gluon distribution
\eqref{eq:structurefunc} there is a sum over all the possible angular
orderings for a given energy ordering. Therefore the  phase space for
real emissions is also the same, up to the subleading difference
mentioned above.

In general, however, the energy weighting of the real and virtual
emissions is different for the BFKL and CCFM evolutions. For the latter,
this is encoded in the splitting functions in
Eqs.~\eqref{eq:structurefunc} and \eqref{eq:eik}--\eqref{eq:noneik},
which show that the (real and virtual) gluon emissions are distributed
logarithmically in $y$ --- the rapidity of the $s$--channel gluons. The
BFKL evolution, on the other hand, retains those diagrams which resum all
orders in $\alpha_s\ln(1/x)$ : these are gluon ladders in which the
$t$--channel gluons are strongly ordered in longitudinal momentum
($x=x_n\ll x_{n-1}\ll\cdots\ll x_1\ll x_0$) and distributed with the
logarithmic weight ${dx_i}/{x_i}$. Clearly, strong ordering in the
$t$--channel implies a similar ordering in the $s$--channel --- from
$x_i\ll x_{i-1}$, it follows that $y_i\simeq x_{i-1}\ll y_{i-1} \simeq
x_{i-2}$ ---,  so that the positions (in energy) of the t-channel
propagators uniquely determine the positions of the real gluons. This is
true in the strict high energy limit, where $s\to\infty$ at fixed $Q^2$,
and  $\abar$ is very small. Beyond this, however, it is important that
the CCFM configurations are generated according to the rule of quantum
coherence, and hence they represent realistic final states (at least, up
to further emissions from the real CCFM gluon, as explained as the end of
Sect.~\ref{subsec:kin}). Thus, although the two types of evolution
provide identical results for the (inclusive) gluon distribution in the
formal high--energy limit the CCFM evolution is more appropriate for
describing actual final states (see the discussion at the end of section
\ref{sec:angord}). Moreover, this {\em formal} high--energy limit becomes
meaningless in the presence of saturation, as we shall later explain, and
when this limit is properly taken (see in Sect.~\ref{sec:ccfmsat}),
differences are expected to appear already in inclusive quantities, so
like the gluon distribution.

\subsection{The structure of the angular ordered cascade}
\label{sec:angord}

\begin{figure}[t]{\centerline{
    \includegraphics[angle=0, scale=0.6]{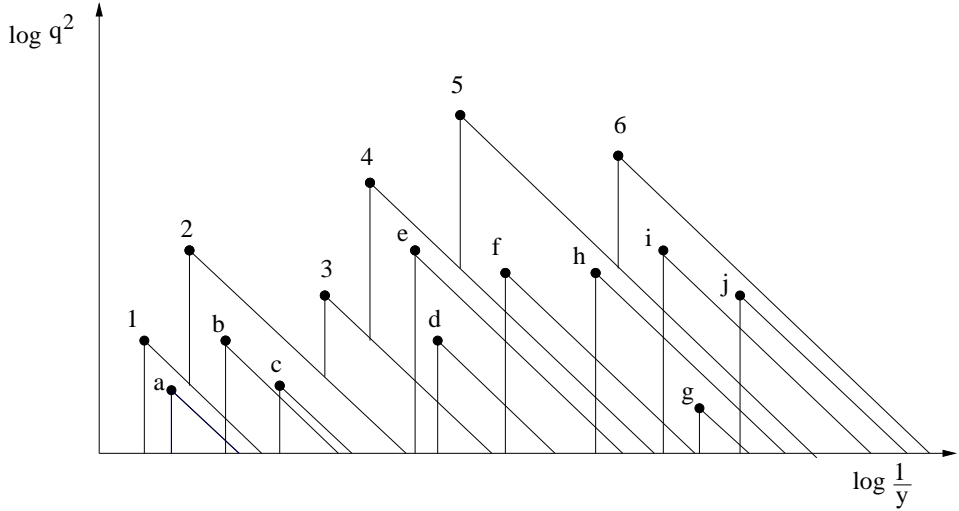}}
\caption {\sl\label{fig:ccfmemissions} Dividing the CCFM radiation
  into two subsets: hard emissions (enumerated emissions) and soft emissions
  (indexed by small letters) as explained in the text. All real emissions
  are represented by fat dots. }
}
\end{figure}

Returning to Eq.~\eqref{eq:structurefunc}, this can be further simplified
to obtain a more familiar expression for the gluon distribution. To that
aim, we shall divide the initial state radiation into two classes,
``soft'' and ``hard'' (or ``fast'') gluons \citep{Catani:1989sg}. This is
done as follows. Consider the set of initial gluons shown in
Fig.~\ref{fig:ccfmemissions}. The ``hard'' subset of gluons are those
which are not in angle followed by a gluon with more energy, \emph{i.e.}
with higher $y$. All other gluons are defined as being ``soft'' gluons.
In Fig.~\ref{fig:ccfmemissions}, the gluons marked by 1, 2, 3, 4, 5 and 6
are hard gluons, since as compared to them, there are no other gluons
with larger angle and higher energy. (Recall that a larger angle would
mean a gluon above the respective diagonal line, while a higher energy
means a gluon to the left of the vertical line through the gluon.) The
gluons marked by $a, b, c, d, e, f, g, h, i$ and $j$ are soft gluons. For
example, $a$ is in angle followed by 1 and 1 has larger energy than $a$
since it is located to the left of it. Gluon $d$ is followed by $e$ which
has larger energy, which itself is followed by 4 in angle which has also
larger energy than $e$. The hard gluons are thus ordered \emph{both} in
angle and in energy.

\begin{figure}[t]{\centerline{
    \includegraphics[angle=0, scale=0.6]{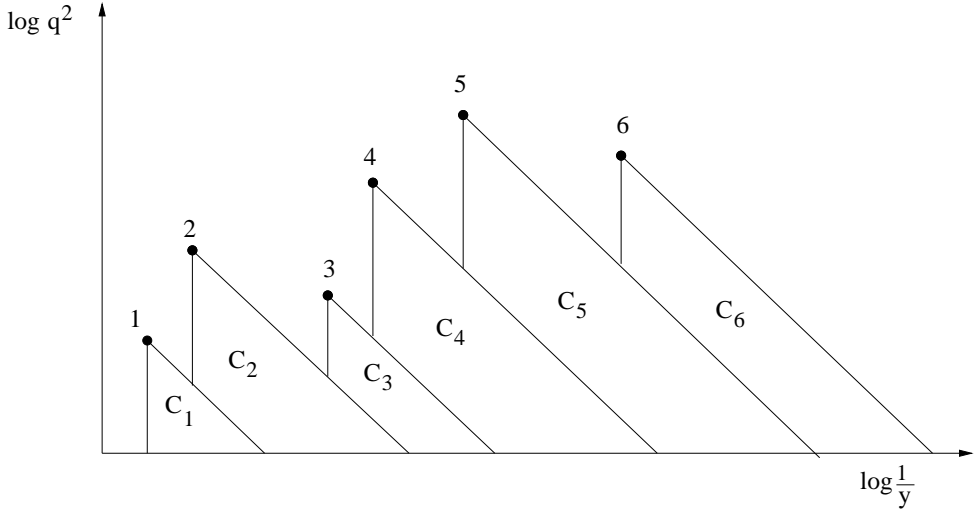}}
\caption {\sl\label{fig:ccfmphase} The representation of the hard
  emissions, enumerated fat dots, together with the clusters $C_i$. For each hard emission $i$,
  the cluster $C_i$ contains all the soft emissions associated with that
  hard emission. }
}
\end{figure}

The soft gluons can furthermore be divided into clusters. Define the
cluster $C_k$ as consisting of those soft gluons which have their angle
between $\xi_{k-1}$ and $\xi_{k}$, and which have energies less than
$y_k$. Thus in Fig.~\ref{fig:ccfmemissions}, $a$ belongs to cluster $C_1$
since $0 < \xi_a < \xi_1$ ($\xi_0 \equiv 0$) and $y_a < y_1$, $b$ and $c$
belong to $C_2$ since $\xi_1 < \xi_{b,c} < \xi_2$ and so on. In this
example there are no gluons in $C_3$. The phase spaces for the clusters
$C_i$ are shown in Fig.~\ref{fig:ccfmphase}. We now see that the ``soft''
gluons which belong to the cluster $C_k$ are indeed soft in the sense of
having lower energies than the ``hard'' gluon $k$ which defines the
cluster.

\begin{figure}[t]{\centerline{
    \includegraphics[angle=0, scale=0.6]{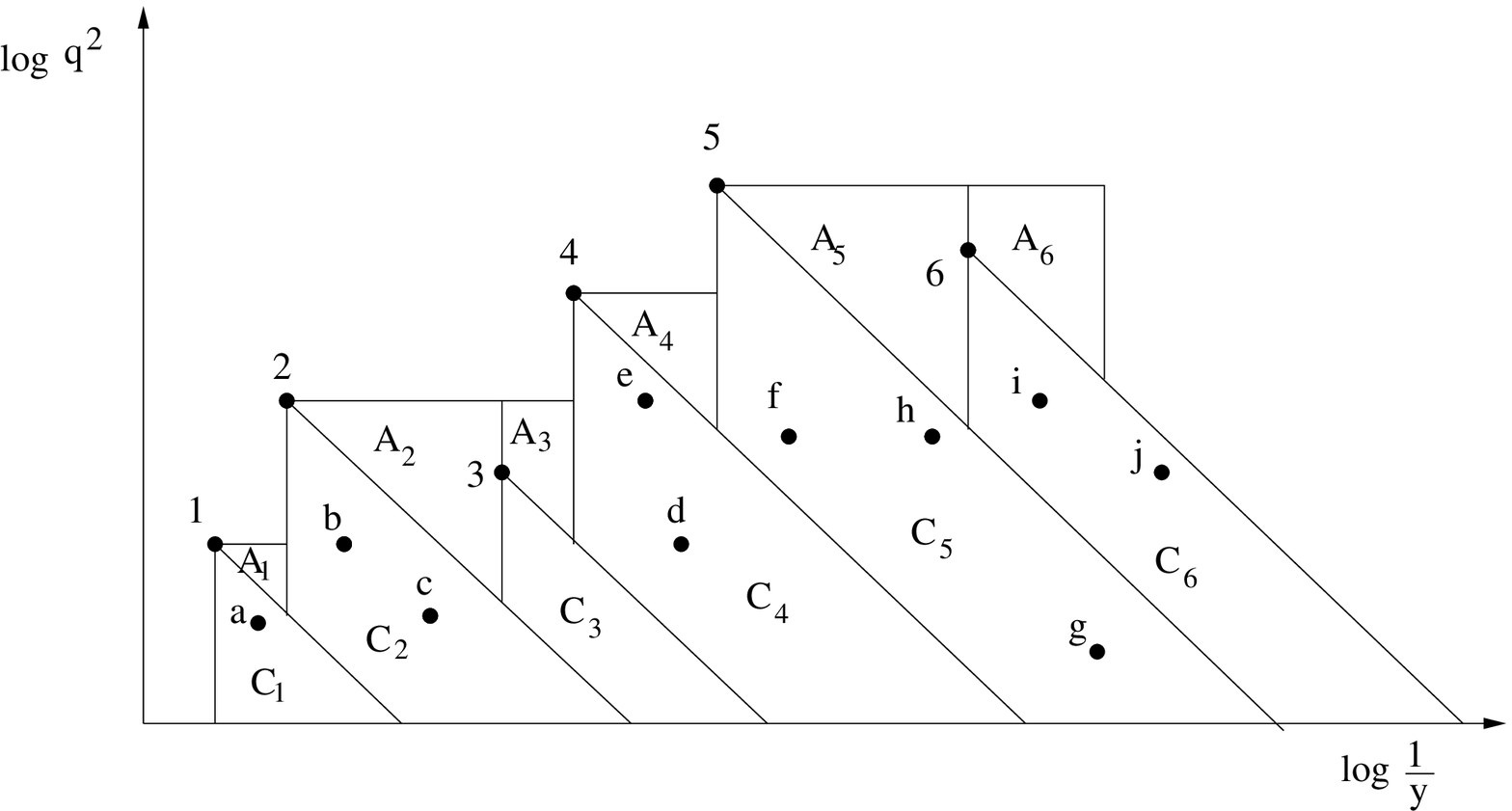}}
\caption {\sl\label{fig:ccfmhardnsoft} The hard and soft emissions,
represented
  by fat dots,  in CCFM
  together with the phase space regions for the ``non-Sudakov'' (regions $A_i$)
  and the ``Sudakov'' (regions $C_i$) form factor. The Sudakov cancels the real
  soft emissions. }
}
\end{figure}

The advantage of this separation is that it allows us to rewrite the
distribution \eqref{eq:structurefunc} in a simpler way. To that aim,
notice that the soft gluons in $C_k$ have all transverse momenta smaller
than $q_k$. This is obvious from the figures and more formally follows
from the fact that\footnote{We denote by $\xi_{s, k}$ and $y_{s, k}$ the
angle and energy fraction for a soft gluon belonging to cluster $C_k$.}
$\xi_{s,k} < \xi_k$ and $y_{s,k} < y_k$ implies $q_{s,k}^2 <
(y_{s,k}^2/y_k^2)q_k^2 < q_k^2$. Then one can write
\begin{eqnarray}
S^2_{ne} = \prod_{k \in \mathscr{A}} S^2_{ne}(k) \approx
\prod_{k \in \mathscr{H}} S^2_{ne}(k)
\label{eq:nehard}
\end{eqnarray}
where $\mathscr{A}$ denotes the set of all emissions while $\mathscr{H}$
denotes the subset of hard emissions, and the phase--space for a hard
gluon in $\mathscr{H}$ is the whole respective cluster  $C_k$, as shown
in Fig.~\ref{fig:ccfmphase}. The gluon distribution
\eqref{eq:structurefunc} can then be written as
\begin{eqnarray}
\mathcal{A}(x, \bar{\xi}) &=& \sum_{n=1}^\infty \int \prod_{k=1}^n
\left( \abar \frac{\rmd y_k}{y_k} \frac{\rmd \xi_k}{\xi_k}
S^2_{ne}(y_{k+1},y_k,Q_k)\theta(\xi_{k+1} - \xi_k)\theta(y_k - y_{k+1})
\right) \frac{1}{x_n} \delta(x-x_n) \nonumber \\
&\times& S_{eik}^2(y_1,\bar{\xi})\prod_{k=1}^{n+1} \left ( \sum_{m=0}^\infty \abar^m \int_{C_k}
\prod_{i=1}^m \frac{\rmd y_i}{y_i} \frac{\rmd \xi_i}{\xi_i}\theta(\xi_{i+1} - \xi_i)
\right ).
\label{eq:ccfmstruct}
\end{eqnarray}
where $\xi_{n+1} \equiv \bar{\xi}$, and we now index the gluons by using
the angular ordering. The meaning of this equation is simple: it says
that we can construct each chain by adding an arbitrary number of soft
gluons between each pair of hard emissions. We can now further simplify
this expression, by using cancelations between real and virtual
contributions to the emission of the soft gluons.

To that purpose, we refer to Fig.~\ref{fig:ccfmhardnsoft} where we have
identified the hard and soft emissions from Fig.~\ref{fig:ccfmemissions},
and the numerated emissions are the hard ones. For these we also indicate
the angles. The soft gluons are marked by small letters. We now define
the new ``non--Sudakov'' form factor $\Delta_{ns}$ by
 \begin{eqnarray}
\Delta_{ns} = \prod_{k} \Delta_{ns}(k) = \prod_{k} \exp \left( -\abar A_k\right ).
\label{eq:deltadef}
\end{eqnarray}
Similarly we define the ``Sudakov'' factors $\Delta_s(k)$ by
\begin{eqnarray}
\Delta_s = \prod_{k} \Delta_s(k) = \prod_{k} \exp \left( -\abar C_k \right).
\label{eq:sudakov}
\end{eqnarray}
Thus we have
\begin{eqnarray}
S^2_{eik}(y_1,\bar{\xi})\cdot \prod_k S^2_{ne}(k) =
\prod_k \Delta_{ns}(k)\cdot \Delta_s (k).
\label{eq:nesudakov}
\end{eqnarray}
Now, the summation over the real soft emissions in each cluster $C_k$ in
\eqref{eq:ccfmstruct} exponentiates,
\begin{eqnarray}
\sum_{m=0}^\infty \abar^m \int_{C_k}\prod_{i=1}^m
\frac{\rmd y_i}{y_i} \,\frac{\rmd \xi_i}{\xi_i}\,\theta(\xi_{i+1} - \xi_i)
  = \exp \left( \abar C_k \right)
 \equiv \Delta_{R}^{soft}(k).
\nonumber \\
\end{eqnarray}
By the definition of the Sudakov in \eqref{eq:sudakov} we thus have
\begin{eqnarray}
\Delta_s(k) \cdot \Delta_{R}^{soft}(k) =1,
\end{eqnarray}
\emph{i.e.} the real soft emissions are exactly compensated by the
Sudakov form factors. After thus inclusively summing over all soft
emissions, the structure function can be finally written as
\begin{eqnarray}
\mathcal{A}(x, \bar{\xi})= \sum_{n=1}^\infty \int \prod_{k=1}^n
\left( \frac{\rmd y_k}{y_k} \frac{\rmd \xi_k}{\xi_k}
\Delta_{ns}(y_{k+1},y_k,Q_k)\theta(\xi_{k+1} - \xi_k)
\theta(y_k - y_{k+1})\right) \frac{1}{x_n} \delta(x-x_n). \nonumber \\
\label{eq:ccfmstruct2}
\end{eqnarray}
This expression involves the {\em hard} gluons alone.

\comment{Eq.~\eqref{eq:ccfmstruct2} is often quoted when pointing to the
difference between CCFM and BFKL results for the production of $k$
initial gluons. This is, however, misleading since one has in
\eqref{eq:ccfmstruct2} already summed over the soft emissions, and this
formula therefore does not give the complete radiation pattern for the
initial state.}

Before leaving this section, two comments are in order:

\texttt{(i)} In Fig.~\ref{fig:ccfmhardnsoft} the horizontal lines between
the real emissions represent the momenta of the $t$--channel gluons. As
obvious from the figure, these momenta are determined solely by the hard
subset of emissions (\emph{c.f.} \eqnum{eq:nehard}) which are ordered
\emph{in both} energy and angle. This \emph{a posteriori} explains why
the condition $Q_i = -\sum_k^i q_k$ is approximately true irrespective
whether the labels $i,\,k$ refer to energy ordering, or to the angular
ordering (which is the actual order of the gluon emissions). This
argument shows that there is an implicit approximation in the CCFM
formalism --- the fact that soft gluon emissions are assumed not to
change the virtual transverse momenta. Hence, without further loss of
accuracy, we will later use similar approximations to simplify the
expression of the gluon distribution even further (see
Sect.~\ref{sec:inclccfm}).

The second important point is the relation to BFKL mentioned earlier and
which deserves some clarifications. To compare to BFKL, it is convenient
to define $z_k$ by $x_k\equiv z_{k}x_{k-1}$. This implies
$y_k=x_{k-1}(1-z_k)$), and therefore
\begin{eqnarray}
\frac{1}{x_n}\prod_k \frac{\rmd y_k}{y_k}=\prod_k \frac{\rmd z_k}{z_k(1-z_k)} =
\prod_k  dz_k \biggl (\frac{1}{z_k}
+\frac{1}{1-z_k}\biggr ).
\label{eq:splitfunc}
\end{eqnarray}
Thus in the CCFM ladder one can distinguish two vertices contributing to
the splitting $Q_{k-1} \to q_k + Q_k$. The one corresponding to the
small--$z_k$ pole $1/z_k$ is dubbed the ``non--eikonal'' vertex as it
comes from the piece of the three gluon vertex in which the polarization
of the parent gluon is inherited by the real gluon $q_k$. The opposite
$1/(1-z_k)$ pole is dubbed the ``eikonal'' vertex, and in this case the
polarization, together with most of the energy, is inherited by the
$t$--channel propagator $Q_k$.  The hard emissions previously identified
are associated with the $1/z_k$ pole in \eqnum{eq:splitfunc}, while the
soft emissions with the $1/(1-z_k)$ pole \citep{Catani:1989sg}. In
contrast, in a BFKL ladder, only the $1/z_k$ pole would be present, and
the energy ordering coincides with the actual sequence of emissions along
the cascade. In that case the typical gluons are such that $z_k\ll 1$ and
hence $y_k\simeq x_{k-1}$, as anticipated at the end of
Sect.~\ref{sec:bfklrel}. Thus in the corresponding phase--space
integrals, like \eqnum{eq:bfklnoneik}, one can replace the measure
$dy_k/y_k$ by $dz_k/z_k$. We are now prepared for our second comment:

\texttt{(ii)} Despite the formal equivalence between the CCFM and BFKL
evolutions (in the high energy limit), the latter cannot be used to
generate the final state, not only because it does not obey the condition
of angular ordering (as required by quantum coherence), but also because
the Regge kinematics $z_k\ll 1$ cannot be ensured in practice when trying
to generate a BFKL ladder. The reason is as follows: the BFKL emission
probabilities for real and, respectively, virtual gluons are separately
infrared divergent (see e.g. \eqnum{eq:bfklnoneik}) and thus require an
infrared regulator $q_0$. Although the dependence upon $q_0$ formally
cancels in the complete result, the introduction of this soft momentum
cutoff will falsify the condition that $z_k\ll 1$ in the intermediate
steps. Indeed, when a new value for $z_k$ is randomly generated with
probability law $\Delta_{ne}^{(BFKL)}(k)$, the typical value value is
such that
\beq
 \ln\frac{1}{z_k}\, \sim\, \frac{1}{\abar}\,\frac{1}{\ln (Q_k^2/q_0^2)}\,,
\eeq
which in principle should be of $\order{1/\abar}$ for the Regge
kinematics to apply, but in reality becomes of $\order{1}$ (meaning
$z_k\sim\order{1}$ as well) whenever $q_0$ is taken to be small enough.

Within the CCFM evolution, this problem is avoided due to the presence of
both types of poles, $1/z_k$ and $1/(1-z_k)$, and to the angular
ordering. In that context, the dependence on $q_0$ is present in the
Sudakov form factor, as the areas $C_k$ in Fig.~\ref{fig:ccfmhardnsoft}
are cut from below by $q_0$. In the $q_0 \to 0$ limit, successive
emissions will become very close to each other in angle. Indeed, an
emission typically occurs when the corresponding region $C_k$ (that we
now define separately for {\em each} emission, either hard, or soft; see
Fig.~\ref{fig:ccfmhnsemissions}) has an area of $\order{1}$,
\beq
\abar \ln \left ( \frac{q_k^2}{q_0^2}\sqrt{\frac{\xi_{k-1}}{\xi_k}}
\right ) \ln \sqrt{\frac{\xi_k}{\xi_{k-1}}} \sim 1,
\eeq
which implies $\xi_k \to \xi_{k-1}$ when $q_0 \to 0$.  In that case one
can identify two limiting cases: \texttt{(i)} $z \ll 1$ and $q$ more or
less similar to the momentum of the previous emission, or \texttt{(ii)}
$z \approx 1$ and $q$ either of the order of, or much smaller than, the
$q$ of the previous emission. These two possibilities are precisely the
type of emissions already present in CCFM and therefore the structure of
the cascade is not altered by $q_0$, unlike what happens in BFKL.

Of course, in the presence of saturation the dependence  on any soft
momentum disappears naturally, as the dynamically generated saturation
momentum, which grows rapidly in the course of the evolution, provides a
natural cutoff (see the discussion in Sect.~\ref{sec:saturation}).

\subsection{The virtual form factors}
\label{sec:virtformfac}

In this subsection we shall display some more explicit formul\ae{} for
the non--Sudakov and Sudakov form factors $\Delta_{ns}$ and $\Delta_s$,
whose detailed derivation is presented in appendix A. Although such
formul\ae{} were already presented in the original work
\citep{Catani:1989sg}, it turns out that they are often written in a
wrong way in the literature (see the discussion in appendix A). To avoid
such errors, and also in order to be able to generalize these form
factors by including non--leading effects --- a task that we address in
appendix B
---, it is essential to have a proper understanding of the derivation of
the corresponding formul\ae. This is briefly discussed here and then in
more detail in appendix A.

Let us first recall that for the hard emissions, $y_k = (1-z_k)x_{k-1}
\approx x_{k-1}$, while for the soft emissions $x_k = z_k x_{k-1} \approx
x_{k-1}$. Then one can as a first approximation set $z_k=0$ for the hard
emissions (except in the $1/z_k$ pole), and $z_k=1$ for the soft
emissions (except in the $1/(1-z_k)$ pole). Therefore the region $A_k$
has the transverse momentum bounded from below by $\xi_k$, and from above
by $Q_k$, while the respective $y$ integral is bounded between $y_k$ and
$y_{k+1}$. Thus we approximately have
\begin{eqnarray}
\Delta_{ns}(k) = \exp \left( -\abar \int_{x_{k}}^{x_{k-1}}\frac{\rmd y}{y}
\int_{\xi_k}^{Q_k^2/(y^2E^2)} \frac{\rmd \xi}{\xi} \right ).
\label{eq:cnoneikonal1}
\end{eqnarray}
Defining $y \equiv z\, x_{k-1}$, and switching  from $\xi$ to $q$ by
using $\xi=q^2/(y^2E^2)$ one gets
\begin{eqnarray}
\Delta_{ns}(k) = \exp \left( -\abar \int_{z_k}^{1}\frac{\rmd z}{z}
\int_{z^2q_k^2/(1-z_k)^2}^{Q_k^2} \frac{\rmd q^2}{q^2} \right ).
\label{eq:cnoneikonal2}
\end{eqnarray}
The reason we did not set $1-z_k=1$ in the lower limit of the $q$
integral is because the CCFM equation is usually written in terms of the
so--called rescaled momenta defined by $p_k\equiv q_k/(1-z_k)$ so that
the factor $(1-z_k)$ is absorbed into the definition of $p_k$. Then
\begin{eqnarray}
\Delta_{ns}(k) = \exp \left( -\abar \int_{z_k}^{1}\frac{\rmd z}{z}
\int_{z^2p_k^2}^{Q_k^2} \frac{\rmd q^2}{q^2} \right )
\label{eq:nonsud}
\end{eqnarray}
which is the form used in \citep{Catani:1989sg}. Equation \ref{eq:nonsud}
for $\Delta_{ns}$ is, however, usually written in a different way in the
literature (see \eqnum{eq:wrongnonsud} in the appendix). In appendix A,
we give a more careful derivation of the non-Sudakov form factor, and we
demonstrate that the correct form is indeed given by \ref{eq:nonsud}.

\begin{figure}[t]{\centerline{
    \includegraphics[angle=0, scale=0.8]{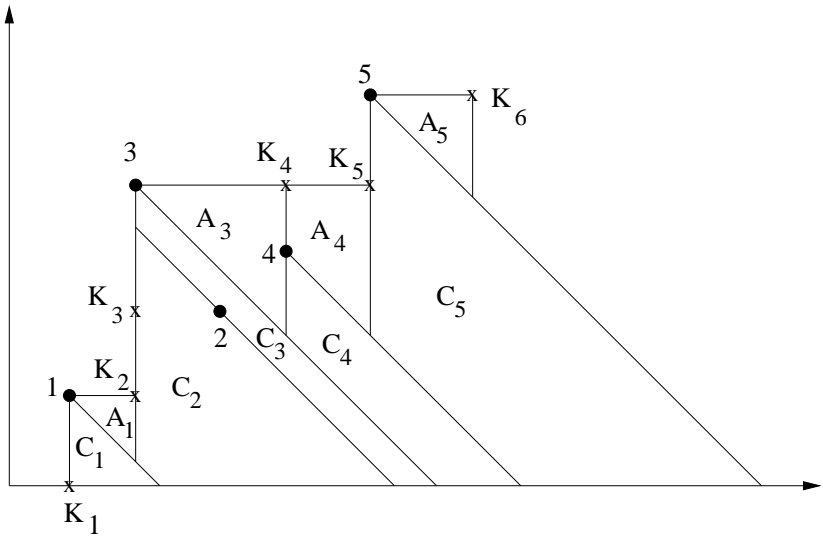}}
\caption {\sl\label{fig:ccfmhnsemissions} Hard and soft emissions, with the
Sudakov (regions
  $C_i$) defined for each individual emission. To leading order there is no
  non-Sudakov (regions $A_i$) associated with the soft emissions. Real emissions
  are again represented by fat dots, while the $k_\perp$ and energy fraction of the
  virtual propagators are denoted by crosses. In this example, emission number 2
  is a soft emission. }
}
\end{figure}

To write down the Sudakov one first needs to define it for each, hard and
soft, individual emission. In Fig.~\ref{fig:ccfmhnsemissions} we show an
explicit chain of hard and soft emissions, where the real emissions are
indexed according to their angular ordering, and where we also
explicitely show the virtual t-channel propagators by crosses. The
individual Sudakov form factors are then defined as the integrals over
the regions $C_k$ in the figure (note that these are not the same regions
$C_k$ as before, now we define such a region for each emission, not just
for the hard ones).

We see that the region $C_k$ is to the left bounded by the energy of
$Q_{k-1}$, while there is no lower limit for the
energy\footnote{Eventually there will be a limit coming from the soft
momentum cut, so that the region does not extend to infinite size}. In
momenta it is bounded between the angles of the real gluons $q_{k-1}$ and
$q_k$. Therefore we may write the Sudakov as
\begin{eqnarray}
\Delta_s(k) = \exp \left ( -\abar
\int_{\xi_{k-1}}^{\xi_k} \frac{\rmd  \xi}{\xi}\int_{\epsilon}^{x_{k-1}} \frac{\rmd y}{y}  \right )
\label{eq:sud1}
\end{eqnarray}
where $\epsilon$ represents the soft cutoff. Since we must have $q > q_0$
we get $\xi = q^2/(y^2E^2) > q_0^2/(y^2E^2)$ and $y > q_0/(\sqrt{\xi}E) =
\epsilon$. Defining $p = \sqrt{\xi}x_{k-1}E$, we have $\epsilon =
x_{k-1}q_0/p$ and
\begin{eqnarray}
\Delta_s(k) = \exp \left ( -\abar
\int_{z_{k-1}^2p_{k-1}^2}^{p_{k}^2}\frac{\rmd p^2}{p^2} \int_{\epsilon'}^1 \frac{\rmd y}{y}  \right ).
\label{eq:sud}
\end{eqnarray}
where $\epsilon' = \epsilon/x_{k-1}=q_0/p$. Now if one wishes one can let
$y=1-z$ and then the usual form for the Sudakov form factor is obtained.

\section{More inclusive versions of CCFM}
\setcounter{equation}{0} \label{sec:inclccfm}

In section \ref{sec:angord} we have shown that one can use the Sudakov
form factors to cancel the real soft emissions, and this resulted in a
simplified expression for the gluon distribution, \eqnum{eq:ccfmstruct2}.
This distribution is more ``inclusive'' than the original one,
\eqnum{eq:structurefunc}, which explicitly includes all soft emissions,
yet it is equivalent to it for the calculation of the gluon distribution.
In this section, we shall construct other, even more inclusive, versions
of the CCFM evolution, which are better adapted for numerical
calculations. In these constructions, we shall exploit the flexibility
which exists in defining the CCFM evolution, as associated with the
various approximations involved in its derivation. Note that ``getting
more inclusive'' is not the only possibility for deriving different
versions of CCFM. In appendix B we shall derive yet another version by
including non--leading effects related to recoils. That version could be
implemented in a Monte Carlo simulation, so like CASCADE.

\subsection{Integral equations}

From now on we shall work with the `unintegrated' gluon distribution,
i.e., the number of gluons with a given longitudinal momentum fraction
$x$ and a given transverse momentum $k$, which is obtained by undoing the
integral over the last angular variable $\xi_n$ in
\eqnum{eq:structurefunc} (or \eqref{eq:ccfmstruct}) and replacing
$\xi_n\to k^2/x^2E^2$. It will be also convenient to replace the maximal
angle $\bar\xi$ by a corresponding momentum variable $\bar q$, via the
substitution $\bar{\xi} = \bar{q}^2/(x^2E^2)$; as explained after
\eqnum{eq:structurefunc}, one has roughly $\bar q^2\simeq Q^2$ (the
virtuality of the space--like photon exchanged in DIS). The integral
equation satisfied by $\mathcal{A}(x, k, \bar{q})$ is easy to derive from
\eqnum{eq:ccfmstruct}, and reads
\begin{eqnarray}
\mathcal{A}(x, k, \bar{p}) = \abar \int_x^1 dz
\int \frac{\rmd ^2p}{\pi p^2} \,
\theta (\bar{p} - zp) \Delta_s(\bar{p},zp)
\left ( \frac{\Delta_{ns}(k,z,p)}{z} + \frac{1}{1-z} \right )
\nonumber \\
\times \mathcal{A}\left(\frac{x}{z}, |k+(1-z)p|, p\right),
\label{eq:ccfminteq1}
\end{eqnarray}
where we are using rescaled momenta within the integrand: $\bar{p} =
\bar{q}/(1-x)$ and $p= {q}/(1-z)$. The third argument of the gluon
distribution inside the integrand, \emph{i.e.} $p$, truly means that the
maximal angle corresponding to this distribution is the angle $\xi$ of
the emitted real gluon, that is, ${\xi} = {q}^2/(y^2E^2)=z^2p^2/x^2E^2$,
with $y=(1-z)(x/z)$. Hence, this equation can be read as follows: the
final $t$--channel gluon with energy fraction $x$ and transverse momentum
$k$ (and for a maximum emission angle measured by $\bar{q}$) is generated
via the splitting $k' \to k+q$ of a previous $t$--channel gluon with
energy fraction $x/z > x$ and transverse momentum $|k'|= |k+q|$ (and for
a maximum emission angle measured by $p$ and $z$). This is the most
exclusive version of the integral equation and includes all the hard and
soft emissions. This equation is implemented in the CASCADE event
generator \citep{Jung:2000hk}.

In the more inclusive version we can sum over all soft emissions (in the
regions $C_i$ in Fig.~\ref{fig:ccfmphase}) so that the Sudakov factors
disappear and we are left with the gluon distribution in
\eqnum{eq:ccfmstruct2}. This version gives rise to the following integral
equation
\begin{eqnarray}
\mathcal{A}(x, k, \bar{p}) = \abar \int_x^1 \frac{\rmd z}{z}
\int \frac{\rmd ^2p}{\pi p^2}\, \theta (\bar{p} - zp)\Delta_{ns}(k,z,p)
\mathcal{A}\left(\frac{x}{z}, |k+(1-z)p|, p\right).
\label{eq:ccfminteq2}
\end{eqnarray}
This equation is simpler to solve than \eqref{eq:ccfminteq1} but is still
not very easy to deal with, not even numerically. Notice that one might
as well use \eqnum{eq:ccfminteq2} as the basis for an event generator,
but then all soft emissions must later be included as final state
radiation. In what follows, however, we shall be concentrating on the
small--$x$ part of the gluon distribution, where the small--$z$ values
($z\ll 1$) are dominating. In that case one can replace all rescaled
momenta with regular momenta, and rewrite $|k+(1-z)p|\to |k+q|$ in the
argument of $\mathcal{A}$ within the integrand; also the energy ordering
becomes automatic. Hence the equation becomes
 \begin{eqnarray}
\mathcal{A}(x, k, \bar{q}) = \abar \int_x^1 \frac{\rmd  z}{z}
\int \frac{\rmd ^2 {q}}{\pi q^2} \,\theta (\bar{q} - zq)\Delta_{ns}(k,z,q)
\mathcal{A}\left(\frac{x}{z}, |{k}+{q}|, p\right)\,,
\label{eq:ccfminteq21}
 \end{eqnarray}

\subsection{Geometrical representation of the real vs. virtual cancelations}

Starting with \eqnum{eq:ccfminteq21}, we shall later derive the most
inclusive, and also the simplest, version of the CCFM equation. To that
aim it is important to understand in depth the structure of the virtual
corrections encoded in the `non--Sudakov' form factor $\Delta_{ns}$,
\eqnum{eq:deltadef}. Note first that, despite its name, this form factor
is essentially not different from a genuine Sudakov one, since it also
represents the negative exponent of an area in the phase space, namely
the areas $A_i$ in figures \ref{fig:ccfmhardnsoft} and
\ref{fig:ccfmhnsemissions}. It can therefore be used to cancel the real
emissions confined to this phase space, as was first noted in
\citep{Andersson:1995ju}.  As we review in appendix
\ref{sec:correctnonsud} however, this is strictly true only if the
additional kinematical constraint $k^2> zq^2$, which ensures that the
squared 4--momenta of the $t$--channel propagators are dominated by their
transverse part \citep{Catani:1989sg, Kwiecinski:1996td}, is enforced
within \eqref{eq:ccfminteq21}.

\begin{figure}[t]{\centerline{
    \includegraphics[angle=0, scale=0.6]{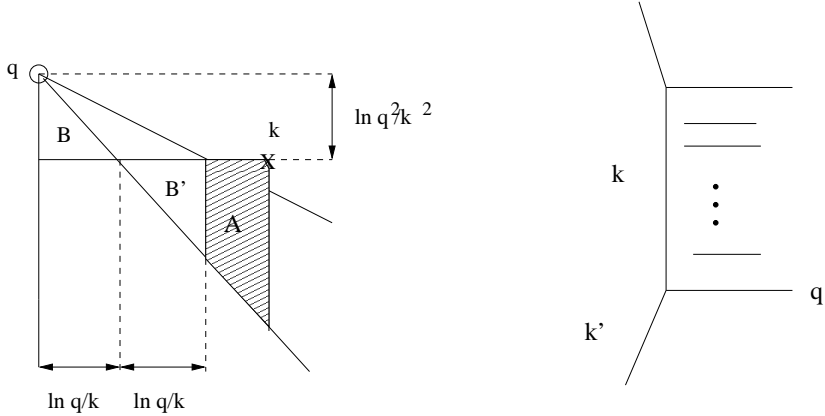}}
\caption {\sl\label{fig:nonsudpickc} The phase space for $\Delta_{ns}$. The
region
  $B$ contributes with negative weight, and it is seen that $B=B'$. The
  shaded region $A$ is thus the region left over, which is also the region
  to which the (real) $k_\perp$-conserving emissions are confined. To the right
  we show symbolically the insertion of $k_\perp$-conserving emissions between
  a pair of non-$k_\perp$-conserving emissions. The propagator $k$ is unchanged
  by this insertion.}
}
\end{figure}


Assuming $k^2> zq^2$ from now on, we can distinguish between two cases:
$k \geq q$ and $k < q$. When $k \geq q$, the regions $A_i$ in figures
\ref{fig:ccfmhardnsoft} and \ref{fig:ccfmhnsemissions} correspond to real
emissions with $q \ll k \approx k'$ and $\Delta_{ns}$ is guaranteed to be
smaller than 1. Consider now the situation in Fig.~\ref{fig:nonsudpickc}
where $k < q$. As we show in appendix \ref{sec:correctnonsud}, we have
\beq\label{DNS}
\Delta_{ns} = \exp(-\abar(A + B' - B))
\eeq
where the regions $A$ (shaded region), $B$ and $B'$ are shown in
Fig.~\ref{fig:nonsudpickc}. The first observation is that the triangular
regions $B$ and $B'$ have the same area, and hence they cancel in the
exponent of \eqnum{DNS}. The upper diagonal line in
Fig.~\ref{fig:nonsudpickc} indicates the line through which the
kinematical constraint limit $k^2=zq^2$ holds, while the lower diagonal
line indicates the angle of $q$.  Now, the emissions lying below $k$ are,
in the spirit of the approximations made in section \ref{sec:angord} ,
$k_\perp$- conserving. When the kinematical constraint is included, these
emissions are confined to the shaded region (region $A$) in figure
\ref{fig:nonsudpickc}. This can be understood as follows: due to angular
ordering and the fact that we are looking at $k_\perp$-conserving
emissions, all subsequent radiation must lie in the region $B'$ or in
$A$. However, if we had a real emission in region $B'$, then because that
emission is $k_\perp$-conserving, the $t$-channel propagator emitting
this gluon would have transverse momentum approximately equal to $k$ and
it would necessarily have bigger energy than the real gluon. Therefore we
see that it must be located under the upper diagonal line in
Fig.~\ref{fig:nonsudpickc}. This, however, would violate the kinematical
constraint and such an emission is therefore not possible. Thus we are
left with the fact that all real $k_\perp$-conserving emissions are
confined to region $A$.  Therefore the inclusive summation over all the
real emissions, which are inserted in between two
non--$k_\perp$-conserving emissions, leads to a factor
\beq
\Delta_R\, =\, \exp(\abar \,A).
\eeq
To conclude (recall that $B'=B$)
\beq
\Delta_R \cdot \Delta_{ns} = \exp(\abar ( A - A + B -B))= 1
\eeq
which shows that the ``non-Sudakov'' factor cancels the
$k_\perp$-conserving emissions. This was first noticed in
\citep{Andersson:1995ju}, and it was later used in \citep{Salam:1999ft}
as well\footnote{ In \citep{Salam:1999ft}, however, only the
possibilities $k' \approx k \gg q$ and $k \approx q \gg k'$ were
considered. In that case there is obviously no need for the kinematical
constraint.}. One is then left with a much simpler formula for
$\mathcal{A}$ which can be derived from \eqnum{eq:ccfminteq2} after
including the constraint $k^2> zq^2$, and the fact that we are left only
with the non--$k_\perp$-conserving emissions. This last constraint,
however, can be enforced in various ways, which are all consistent with
each other within the present approximations. Therefore there is no
unique equation that one can derive. In what follows we shall consider
two different possibilities and then study the ensuing equations.

\subsection{Deriving the differential equations}
\label{sec:derivediffeq}

In Ref.~\citep{Avsar:2009pv} we have the restriction to
non--$k_\perp$-conserving emissions by introducing the theta function
$\theta(q^2- \mathrm{min}(k^2, k'^2))$ into the r.h.s. of the integral
\eqnum{eq:ccfminteq21}. This was also the prescription originally used in
Ref.~\citep{Andersson:1995ju}, and the equations derived in
\citep{Avsar:2009pv} and \citep{Andersson:1995ju} are indeed equivalent.
After inserting this constraint together with the `kinematical' one $k^2>
zq^2$ and removing $\Delta_{ns}$, \eqnum{eq:ccfminteq21} becomes
\begin{eqnarray}
\mathcal{A}(x, k, \bar{q}) = \abar \int_x^1 \frac{\rmd z}{z}
\int \frac{\rmd^2q}{\pi q^2} \, \theta (\bar{q} - zq)\theta (k^2 - zq^2)
\theta(q^2- \mathrm{min}(k^2, k'^2))
\mathcal{A}\left(\frac{x}{z}, k', q\right).\nn
\label{eq:firststep}
\end{eqnarray}
Since $\bar{q} \geq k$ for all cases of physical interest (recall that
$\bar q^2\simeq Q^2$ in DIS), we further have $\bar{q}^2 \geq k^2 \geq
zq^2 \geq z^2q^2$. Therefore the angular ordering is automatic and
$\theta (\bar{q} - zq)$ can be neglected. This means that the dependence
on the third variable $\bar{q}$ drops out, at least in the l.h.s. But a
similar argument holds also for the function for $\mcal{A}(x/z, k', q)$
under the integral, because we have $q \geq k'$ (indeed, we are left only
with emissions satisfying $k' \approx q \gg k$ or $k \approx q \gg k'$).
Dropping then the dependence of $\mathcal{A}$ upon its third variable, we
obtain
 \begin{eqnarray} \mathcal{A}(x, k) = \abar \int_x^1 \frac{\rmd
z}{z} \int \frac{\rmd^2q}{\pi q^2} \, \theta (k^2 - zq^2) \theta(q^2-
\mathrm{min}(k^2, k'^2)) \mathcal{A}\left(\frac{x}{z}, k'\right)\,.
\label{eq:secondstep}
 \end{eqnarray}
The next step is to perform the  integration over the azimuthal angle
$\phi$. To that aim, it is convenient to replace $\theta(k^2-zq^2)$ by
$\theta(k^2-zk'^2)$, which is allowed within the current
approximations\footnote{Indeed, we have either $k' \approx q \gg k$, in
which case the replacement is obviously correct, or $k\approx q \gg k'$,
in which case both the first and the second theta function can be
replaced by 1.}, and then switch the integration variables from $q$ to
$k'$ and, respectively, from $z$ to $x/z$ (which we rename as $z$).
Making this replacement and doing the $\phi$ integral one then gets
\begin{eqnarray}
\mathcal{A}(x, k) = \abar \int_x^1 \frac{\rmd z}{z}
\int \frac{\rmd k'^2}{\vert k'^2-k^2\vert} \, \theta (z- xk'^2/k^2)\,
h(\kappa )\,
\mathcal{A}(z, k')\,,
\label{eq:ccfminteq3}
\end{eqnarray}
where
\begin{eqnarray}
h(\kappa) =  1 -\frac{2}{\pi}\arctan\left(\frac{1+\sqrt{\kappa}}{1-\sqrt{\kappa}}\sqrt{
\frac{2\sqrt{\kappa}-1}{2\sqrt{\kappa}+1}} \right ) \theta(\kappa-1/4).
\label{eq:h}
\end{eqnarray}
and $\kappa \equiv \mathrm{min}(k^2, k'^2)/\mathrm{max}(k^2, k'^2)$.
Differentiating w.r.t to $Y$ we finally deduce the following differential
equation for the unintegrated gluon distribution in the CCFM formalism
  \begin{eqnarray}
  \partial_Y\mathcal{A}(Y, k) = \abar \int \frac{\rmd k'^2}{\vert k^2-k'^2\vert} \,
h(\kappa) \biggl( \theta(k^2-k'^2)\mathcal{A}(Y,k') \biggr. \nonumber \\
+  \biggl. \theta(k'^2-k^2)\theta(Y-\ln(k'^2/k^2))\mathcal{A}(Y- \ln(k'^2/k^2),k')
\biggr).
\label{eq:ccfmdiffeq3}
\end{eqnarray}
As mentioned above, \eqref{eq:ccfmdiffeq3} is equivalent to the master
equation in the LDC formalism  \citep{Andersson:1995ju} (up to some
trivial redefinitions: the distribution in \citep{Andersson:1995ju}
corresponds to our $k^2\mcal{A}(Y,k)$ times the proton radius and some
constants).

\comment{ However, note that in \citep{Andersson:1995ju} the gluon
distribution is defined via
\begin{eqnarray}
xg(x,Q^2) = \int^{Q^2} \frac{\rmd ^2k}{k^2}\mcal{A}(x,k),
\end{eqnarray}
and thus the distribution in \citep{Andersson:1995ju} corresponds to our
$k^2\mcal{A}(Y,k)$ (times the proton radius and some constants). If one
uses this in \eqref{eq:ccfmdiffeq3} we indeed get the same equation as in
\citep{Andersson:1995ju}. }

\comment{Notice that the first constraint is less restrictive, as it
automatically follows from the second one. It turns out that the latter
constraint gives a larger Pomeron intercept. One could also think of
using $q \geq k$ or $q \geq k'$ as explicit constraints, and the
equations then derived are very similar to \eqref{eq:ccfmdiffeq3}.
However, in that case the evolution is not symmetric in $k$ and $k'$
anymore, unlike BFKL and the other equations we derive. We therefore do
not implement these last two constraints. We now implement the constraint
$q^2 = \mathrm{max}(k^2, k'^2)$.}

Yet another way to implement the restriction to non--$k_\perp$-conserving
emissions is to switch the integration variable in \eqnum{eq:ccfminteq21}
from $q$ to $k'=|q+k|$ and then replace $q^2$ with $\mathrm{max}(k^2,
k'^2)$. Note that this constraint is more restrictive that the theta
function in \eqnum{eq:firststep}. In this case the angular integration in
\eqref{eq:firststep} becomes trivial. Also the replacement $\theta(k^2-
zq^2) \to \theta(k^2-zk'^2)$ is now exact, and so is also the requirement
$q \geq k'$ (which, we recall, allows one to ignore the dependence of
$\mathcal{A}(x, k, \bar{q})$ upon its third, `maximal angle', variable).
We thus deduce
\begin{eqnarray}
  \partial_Y\mathcal{A}(Y, k) = \abar \int \frac{\rmd k'^2}
  {\mathrm{max}(k^2, k'^2)} \,
\biggl( \theta(k^2-k'^2)\mathcal{A}(Y,k') \biggr. \nonumber \\
+  \biggl. \theta(k'^2-k^2)\theta(Y-\ln(k'^2/k^2))
\mathcal{A}(Y- \ln(k'^2/k^2),k') \biggr).
 \label{eq:ccfmdiffeq4}
 \end{eqnarray}

\subsection{More on the relation to BFKL}
\label{sec:ccfmcfbfkl}

Before moving on to discuss the issues of unitarity and saturation, we
would like discuss the effects of the kinematic constraint on the BFKL
equation,  and we would like to show that the same approximations used
above in deriving the more inclusive equations \eqref{eq:ccfmdiffeq3} and
\eqref{eq:ccfmdiffeq4} applied on BFKL, with the kinematical constraint,
leads exactly to the same equations. In section \ref{sec:results} we will
present numerical results of the BFKL equation with the kinematic
constraint included, with and without the saturation boundary to be
described in the next section.

\begin{figure}[t] {\centerline{
    \includegraphics[angle=0, scale=0.6]{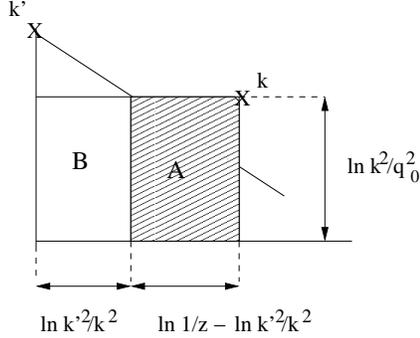}}
\caption {\sl\label{fig:bfklnonsudpic} The phase space for the modified BFKL
  non-Sudakov form factor when the kinematical constraint is taken into account.
  Region $B$ which was previously included is now excluded. Real $k_\perp$-conserving
  emissions are confined to region $A$. }
}
\end{figure}

The BFKL equation can be written
\begin{eqnarray}
\mathcal{A}(Y,k) =  \abar \int_0^Y dy \int \frac{\rmd ^2q}{\pi q^2}\,
\theta (Y - y + \ln(k^2/k'^2)) \Delta'(Y-y, k)\, \mathcal{A}(y, k')
\label{eq:bfklinteq}
\end{eqnarray}
where we have included the kinematical constraint $k^2 > z k'^2$ (with $z
= e^{y-Y}$). The function $\Delta'$ is the non-Sudakov form factor
modified to include the kinematical constraint. The correct form for
$\Delta'$ can be found by requiring that once more it can be used to
exactly compensate the $k_\perp$-conserving emissions which are now also
modified by the kinematical constraint. The phase space for $\Delta'$ is
illustrated in Fig.~\ref{fig:bfklnonsudpic}. The region we are looking
for is the shaded one, region $A$. Region $B$ which was allowed before is
now excluded due to the kinematical constraint. Indeed the real
$k_\perp$-conserving emissions are confined to region $A$, since just
like for CCFM, any real emission in $B$ would create a $t$-channel
propagator below the diagonal line, violating the kinematical constraint.
It is then seen that (see also \citep{Kwiecinski:1996td})
\begin{eqnarray}
\Delta'(Y-y, k) = \theta(k - k')\Delta(Y-y, k) \nonumber \\ +
\,\,\,\, \theta(k'-k)
\exp\left (- \abar \ln(k^2/q_0^2)(\ln(k^2/k'^2) + Y -y) \right )
\label{eq:bfkldiffeq}
\end{eqnarray}
where $\Delta(Y-y, k)$ is the usual non-Sudakov, and the second factor is
just $\exp(- \abar A)$ (again if $k > k'$ the kinematical constraint is
automatic).

Next we inclusively sum over the $k_\perp$-conserving emissions in
\eqref{eq:bfklinteq}, canceling $\Delta'$. If we use the explicit
constraint $q > \mathrm{min}(k, k')$, we get
\begin{align}
\mathcal{A}(Y,k) =  \abar \int_0^Y dy \int \frac{\rmd ^2k'}{\pi \vert k - k'\vert^2}\,
\theta (Y - y + \ln(k^2/k'^2)) \theta ( \vert k-k'\vert^2 - \mathrm{min}(
k^2, k'^2) )\mathcal{A}(Y,k'),\nonumber \\
\label{eq:bfklkincons}
\end{align}
and  this leads to an equation which is \emph{exactly} identical to
\eqref{eq:ccfminteq3} and \eqref{eq:ccfmdiffeq3}. We can also use the
constraint $q=\mathrm{max}(k, k')$ in which case we obtain
\eqref{eq:ccfmdiffeq4}. We should also mention that we could repeat the
arguments in appendix \ref{sec:betteraccuracy} for BFKL with the
kinematical constraint. In that case the region left over by the
real-virtual cancellations is exactly equal to the region $C'$ in
\eqref{eq:Cprime}.

\section{CCFM evolution with saturation boundary}
\setcounter{equation}{0} \label{sec:ccfmsat}

The CCFM evolution, so like any other linear evolution in perturbative
QCD, predicts an unlimited growth of the gluon distribution with
increasing $Y$, thus leading to unitarity violations in the high energy
limit. This is so since the linear evolution equations miss the
non--linear phenomena responsible for unitarization, which are gluon
saturation and multiple scattering. It turns out, however, that the
phenomenon of gluon saturation merely acts as a kind of cutoff, which
limits the growth of the gluon distribution, but does not modify the
mechanism responsible for this growth. This makes it possible to mimic
the effects of saturation by appropriately implementing this cutoff on
the linear evolution equations, without a detailed understanding of the
underlying non--linear phenomena. In this section we shall motivate and
describe the implementation of this cutoff --- actually, an {\em
absorptive boundary condition} ---, which then will be used, in
Sect.~\ref{sec:results}, within numerical simulations of the CCFM and
BFKL evolutions in the presence of unitarity corrections.

\subsection{Unitarity and Saturation momentum}
\label{sec:saturation}

In what follows we shall explain our method for effectively implementing
saturation on the example of the BFKL equation \citep{BFKL}. This is
interesting since the corresponding non--linear generalization which
obeys unitarity is also known --- this is the Balitsky--Kovchegov (BK)
equation \citep{Balitsky:1995ub,Kovchegov:1999yj} ---, and thus it can be
used to test our method. The derivation of the BK equation has been
recently pushed to next--to--leading order accuracy
\citep{Balitsky:2006wa,Kovchegov:2006vj,Balitsky:2008zz,Balitsky:2009xg},
but here we shall limit ourselves to its leading--order version, which is
also the accuracy of the CCFM formalism. However, this LO version will be
eventually extended to include a running coupling (both for BFKL and for
CCFM), since the running coupling effects modify in an essential way the
high energy evolution --- these are the only NLO corrections which remain
important for asymptotically high energy.

The (leading--order) BK equation for the unintegrated gluon distribution
reads
\beq\label{eq:BK}
\partial_Y \mcal{A}(Y,k) =  \abar \int \frac{\rmd k'^2}{k'^2} \left
\{ \frac{k'^2 \mcal{A}(Y,k') - k^2\mcal{A}(Y,k)}{\vert k^2-k'^2\vert} +
\frac{k^2\mcal{A}(Y,k)}{\sqrt{4k'^4+k^4} }\right \}
-{\bar\alpha_s}\big(\mathcal{A}(Y,k)\big)^2 \,.
  \eeq
The terms linear in $\mcal{A}(Y,k)$ in this equation represent the BFKL
equation, whereas the last, quadratic, term is responsible for gluon
saturation. One can roughly think about this last term as describing the
recombination of two gluons into one, but this picture is quite crude:
the actual non--linear phenomena responsible for gluon saturation are
much more complex and should be rather viewed as the blocking of new
gluon emissions by strong color fields \citep{CGCreviews}. Since this
equation is non--linear, we should be more specific about the
normalization of the function $\mcal{A}(Y,k)$. Our conventions are such
that the standard, `integrated' gluon distribution is computed as
 \beq\label{GDF}
x g(x,Q^2)\,=\,\frac{4N_c^2}{\pi^2\abar} \int^{Q^2}\! \frac{\rmd^2
{k}}{(2\pi)^2}\int
 \rmd^2{b}\ \mathcal{A}(Y,{k},{b})\,,\eeq
where $b$ denotes the 2--dimensional impact parameter in the transverse
space. Note that, strictly speaking, the non--linear effects are
non--local in $b$. The simple form for the non--linear term shown in
\eqnum{eq:BK} is obtained under the further assumption that the hadron is
homogeneous in $b$ (a `large nucleus'). With these conventions, the {\em
gluon occupation number} --- i.e., the number of gluons of a given color
per unit rapidity per unit volume in transverse phase--space
--- is not exactly $\mathcal{A}(Y,k)$, but rather
$\mathcal{A}(Y,k)/\abar$ (up to a numerical factor).

So long as $\mathcal{A}(Y,k)\ll 1$, the non--linear terms in
\eqnum{eq:BK} can be neglected, and then this equation predicts the
rapid, BFKL, growth of the gluon distribution, which is exponential in
$Y$, together with diffusion in transverse space (see below). When
$\mathcal{A}(Y,k)\sim {1}$ (corresponding to a physical occupation number
$\order{1/\abar}$), the non--linear effects become important and tame
this growth. For a given rapidity $Y$, this happens at a specific value
of the transverse momentum $k=Q_s(Y)$, called the {\em saturation
momentum}, which grows rapidly with $Y$ (see below). One can show that at
low momenta $k\lesssim Q_s(Y)$, the occupation number essentially {\em
saturates}, in the sense that is shows only a very slow
increase\footnote{The dependence of $Q_s(Y)$ upon $Y$ is such that the
physical occupation number at $k\lesssim Q_s(Y)$ grows linearly with $Y$;
that is, what saturates is the {\em rate} for gluon emission
\citep{CGCreviews}.} with $Y$: $\mathcal{A}(Y,k)\simeq \ln[Q_s(Y)/k]$.
Hence, for a given $Y$, the gluon distribution $\mathcal{A}(Y,k)$
produced by the BK equation looks like a {\em front}, which interpolates
between the dilute ($\mathcal{A}(Y,k)\ll 1$) BFKL tail\footnote{In
particular, for extremely high $k\ggg Q_s(Y)$, the BFKL distribution
approaches the bremsstrahlung spectrum, $\mathcal{A}(Y,k)\sim 1/k^2$,
whereas at moderate $k$ this is modified by the BFKL `anomalous
dimension'.} at high transverse momenta $k\gg Q_s(Y)$ and a saturation
region at low transverse momenta $k\lesssim Q_s(Y)$, where
$\mathcal{A}(Y,k)\sim \ln[Q_s(Y)/k]$. The transition between these two
region occurs around $k=Q_s(Y)$, where $\mathcal{A}(Y,k)\sim {1}$. With
increasing $Y$, this transition value $Q_s(Y)$ is rapidly increasing,
i.e., the front moves up to higher values of $k$.

What is remarkable about this dynamics is that the progression of the
front with increasing energy and also its shape at high $k$ are fully
determined by the BFKL evolution of the dilute tail at $k\gg Q_s(Y)$, and
thus can be inferred from the linear, BFKL, equation alone. One says that
the saturation front is ``pulled by its tail''. This property is central
to our analysis: it implies that some essential features of the dynamics
in the presence of saturation, like the energy dependence of the
saturation momentum, can be studied without a detailed knowledge of the
non--linear effects responsible for saturation. Hence, a similar study
can be performed on the basis of other equations, so like CCFM, whose
non--linear generalizations are not known.

The pulled--front property property is highly non--trivial ---  there are
many examples of non--linear equations which develop a {\em pushed
front}, i.e., a front whose progression is driven by the growth and
accumulation of `matter' behind the front \citep{Saar} --- and so far it
has not been rigorously demonstrated for the general case in QCD. In the
case of a {\em fixed coupling} and for {\em sufficiently high energy},
this property follows from the identification \citep{Munier:2003vc}
between the asymptotic form of the BK equation at high
energy\footnote{This is obtained via the gradient expansion of the
non--locality in Eq.~(\ref{eq:BK}) to second order in $\del/\del\rho$,
with $\rho=\ln k^2$ (`diffusion approximation').} and the FKPP equation
(from Fisher Kolmogorov, Petrovsky, and Piscounov) of statistical
physics. (The FKPP equation describes a `reaction--diffusion process' in
the mean field approximation corresponding to very large occupation
numbers at saturation; see e.g. the review paper \citep{Saar}.) However,
this identification does not extend to a {\em running} coupling, and it
was in fact shown \citep{Dumitru:2007ew} that the high--energy evolution
with running coupling is not in the same universality class as the
reaction--diffusion. Yet, the pulled--front property appears to hold for
that case too, and also for relatively small rapidities (\emph{i.e.}, for
the early stages of the evolution), as most convincingly demonstrated so
far by the numerical simulations in Ref.~\citep{Avsar:2009pv}, where the
solutions to the BK equation with running coupling have been
systematically compared to those of the BFKL equation supplemented with a
saturation boundary condition.

\begin{figure}[]
\begin{center}
\includegraphics[scale=.8,angle=270]{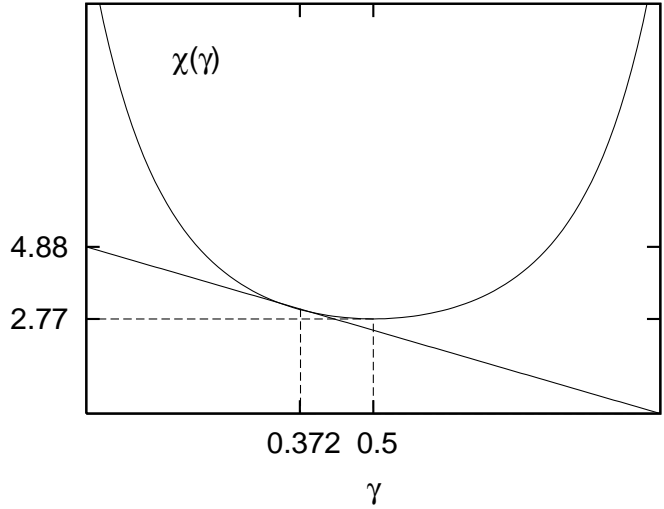}
    \caption{\label{CHI}
    The BFKL eigenvalue $\chi(\gamma)$ and the graphical
    solution to the saturation problem.
    The value $\gamma_s
    = 0.327$ corresponds to the saturation saddle point and the value
    $\chi(\gamma_s)/\gamma_s = 4.88$ determines the asymptotic energy
    dependence of the saturation momentum (for comparison, the ``hard
    pomeron'' saddle point $\gamma_{\mathbb P}  =1/2$ and its
    intercept $\omega_{\mathbb P} = 4 \ln 2 = 2.77$ are shown).}
\vspace*{.5cm}
\end{center}
\end{figure}

In order to gain some analytic insight into the role and the form of this
boundary condition, it is useful to briefly review the computation of the
saturation momentum from the BFKL equation
\citep{Iancu:2002tr,Mueller:2002zm} --- in the fixed coupling case, for
simplicity (the corresponding developments for the case of a running
coupling can be found in Refs.
\citep{Mueller:2002zm,Triantafyllopoulos:2002nz}). We start with the
Mellin representation of the BFKL solution, that is
 \beq\label{TBFKL}
\mcal{A}(Y, k) = \int\limits_{C} \frac{\rmd \gamma}{2\pi i} \ \rme^{
\abar\chi(\gamma)Y - (1-\gamma) \rho}\tilde{\mcal{A}}(\gamma),
\eeq
where $\rho \equiv \ln(k^2/Q^2_0)$ with $Q_0$ an arbitrary reference
scale, $\tilde{\mcal{A}}(\gamma)$ is the initial condition at $Y=0$, and
$\chi(\gamma)$ is the BFKL characteristic function, \emph{i.e.} the
eigenvalue of the BFKL kernel in Mellin space:
 \beq\label{chigamma}
\chi(\gamma)= 2\psi(1)-\psi(\gamma)-\psi(1-\gamma),\qquad
 \psi(\gamma)\equiv \rmd \ln \Gamma(\gamma)/\rmd\gamma\,.\eeq
The integration contour $C$ runs parallel to the imaginary axis with $0<
{\rm Re}(\gamma) < 1$. For real values of $\gamma$ in between 0 and 1
(the relevant range for computing the saddle point; see below), the
function $\chi(\gamma)$ is displayed in Fig.~\ref{CHI}.

\eqnum{TBFKL} is expected to be correct so long as $\mcal{A}(Y, k)\ll 1$,
\emph{i.e.} for high enough momenta $k\gg Q_s(Y)$. Here we shall assume
that this expression can be also used to approach the saturation line
`from the above' (\emph{i.e.} from momenta $k$ larger than $Q_s$), and
hence to approximately determine the latter from the condition that
$\mathcal{A}(Y,k)\sim {1}$ when $k=Q_s(Y)$. To that aim, we shall also
rely on the saddle point approximation, which is appropriate for high
enough $Y$ --- namely, such that $\abar Y\gg 1$. By combining the saddle
point condition
 \beq
 \abar Y \chi'(\gamma_s) = -\rho_s\,,
  \eeq
with the condition that $\mathcal{A}(Y,k)\sim {1}$ along the saturation
line:
 \beq
 \abar Y \chi(\gamma_s) - (1-\gamma_s)\rho_s\,=0,
 \eeq
one obtains $\gamma_s$ and the saturation line $\rho_s(Y)\equiv \ln
[Q_s^2(Y)/k_0^2]$ as follows \citep{Iancu:2002tr}
 \beq\label{gammasBFKL}
 \frac{\chi'(\gamma_s)}{\chi(\gamma_s)}&\,=\,&-\frac{1}{1-\gamma_s}
 \quad\Longrightarrow\quad \gamma_s\approx 0.372,\nn
 \rho_s(Y)&\,\simeq\,& \lambda_s Y
  \quad\mbox{with}\quad
 \lambda_s\,\equiv\,\abar\,\frac{\chi(\gamma_s)}{1-\gamma_s}
 \,\approx 4.883\abar\,.\eeq
A geometrical interpretation of the above equation for $\gamma_s$ is
shown in Fig.~\ref{CHI}.

For $\rho$ larger than $\rho_s$, but not {\em much} larger, one can
estimate the gluon distribution by expanding the integrand in
Eq.~(\ref{TBFKL}) around $\gamma_s$ : writing $\gamma=\gamma_s-i\nu$, and
expanding to second order in $\nu$, one finds
 \beq\label{expsat}
 \abar Y \chi(\gamma) -(1-\gamma)\rho\,
  \simeq\, - (1-\gamma_s)(\rho-\rho_s) - i\nu
 (\rho-\rho_s) - D_sY\nu^2,\eeq
where $D_s=\abar\chi''(\gamma_s)/2 \approx 24.26\abar$. This expansion is
valid so long as $1<\rho-\rho_s\ll  D_sY$. After also performing the
Gaussian integration over $\nu$, one finds \citep{Iancu:2002tr}
 \beq\label{TSAT}
 \mathcal{A}(Y,k)\,\simeq\,
 \frac{1}{\sqrt{\pi D_s Y}}\,\left(\frac{Q_s^{2}}{k^2}\right)^{1-\gamma_s}
  \exp\left\{-\frac{\ln^2(k^2 /Q_s^2)}
 {4D_sY} \right\},\eeq
with $Q_s^2(Y)=Q_0^2\rme^{\lambda_sY}$. \eqnum{TSAT} exhibits a
power--like spectrum in $k$ with anomalous dimension (\emph{i.e.},
deviation from the bremsstrahlung spectrum) $\gamma_s$, which is further
modified by BFKL diffusion (the last factor, which is a Gaussian in
$\ln(k^2 /Q_s^2)=\rho-\rho_s$).

Although obtained solely from the linear, BFKL, equation, the above
results for the behaviour near saturation are essentially correct: they
coincide with the respective predictions of the BK equation for
asymptotically high energy. It turns out that one can do even better:
still without resorting on the actual non--linear \eqnum{eq:BK}, one can
also determine the sub--asymptotic behaviour of $Q_s^2(Y)$ at large $Y$,
and refine the approach of $\mathcal{A}(Y,k)$ towards saturation. The
main observation, which lies also at the heart of the subsequent
developments in this work, is that the saturation region acts like an
{\em absorptive boundary}  \citep{Mueller:2002zm}, which not only tames
the growth of the gluon distribution, but also prevents the BFKL
diffusion towards lower momenta $k\lesssim Q_s(Y)$.

After also implementing this absorptive boundary, in a way to be shortly
specified, the previous formul\ae{} are modified as follows
\citep{Mueller:2002zm}: the gluon distribution becomes
 \beq\label{Arho} \mathcal{A}(Y,\rho) \, \sim \,(\rho-\rho_s+\delta)\,
 {\rm e}^{-(1-\gamma_s) (\rho-\rho_s)}\,
 \exp\left\{-\frac{(\rho-\rho_s)^2} {2D_s Y}
 \right\},\eeq
where $\delta\sim 1$ is some unknown constant and $\rho_s(Y)$ is now
under control up to next--to--leading order in the asymptotic expansion
at high--energy:
 \beq
 \rho_s(Y)\,\simeq\, \lambda_s Y\, -
 \,\frac{3}{2\gamma_s}\,\mathrm{ln}Y\,.
 \label{satmom}
 \eeq
These results coincide indeed with the corresponding predictions of the
non--linear, BK, equation, as deduced via the correspondence with the
FKPP equation \citep{Munier:2003vc}. Notice in particular the property of
\emph{geometric scaling} \citep{Iancu:2002tr,Mueller:2002zm}: within the
region $\rho-\rho_s\ll \sqrt{2D_s Y}$, whose width is increasing with
$Y$, the Gaussian factor can be ignored in \eqnum{Arho}, and then this
expression reduces to a function of $\rho-\rho_s(Y)$ alone, \emph{i.e.}
of the scaling variable $Q^2/Q_s^2(Y)$. This property provides a natural
explanation for an important regularity observed in the HERA data for DIS
at small--$x$ \citep{Stasto:2000er}.

The success of the absorptive boundary method for the BFKL evolution
makes it compelling to try and use it as a systematic method for
enforcing saturation within an arbitrary linear evolution, so like CCFM.
Let us now explain in detail our practical implementation of the
absorptive boundary, as it will be used within numerical simulations. We
first introduce a line of constant gluon occupancy $\rho=\rho_c(Y)$ via
the condition
\beq\label{cline}
  \mathcal{A}(Y,\rho=\rho_c(Y))\,=\,c\,,
 \eeq
where the number $c$ is smaller than one, but not {\em much} smaller. The
saturation line $\rho_s(Y)$ would correspond to $c\sim 1$, so clearly
$\rho_s(Y)$ is smaller than $\rho_c(Y)$, but relatively close to it. (For
the BK evolution with fixed coupling, these two lines would be parallel
lines, with slope $\lambda_s$, separated from each other by
$\rho_c(Y)-\rho_s(Y)\sim \ln(1/c)$.) For $\rho <\rho_c(Y)$ and
sufficiently high energy, the solution $\mathcal{A}_{\rm BFKL}(Y,\rho)$
to the BFKL equation would become larger than one --- in fact,
arbitrarily large. If this equation is to be solved numerically, one may
think about enforcing saturation by hand, in the following way: at each
step in $Y$, one first identifies the corresponding point $\rho_c(Y)$
from the condition \eqref{cline}, and then one requires
$\mathcal{A}(Y,\rho)$ to remain finite and of $\order{1}$ for any $\rho$
sufficiently far below $\rho_c(Y)$ --- say, for $\rho\le
\rho_c(Y)-\Delta$ with $\Delta\simeq \ln(1/c)$. When decreasing $\rho$
below $\rho_c(Y)$, the solution $\mathcal{A}(Y,\rho)$ will typically
start by rising, then reach a maximum of $\order{1}$ and eventually
decrease to zero. We shall conventionally identify the saturation scale
$\rho_s(Y)$ with the position of this maximum. In this procedure, the
numbers $c$ and $\Delta$ are to be viewed as free parameters, which are
however correlated with each other, since $\Delta\sim \ln(1/c)$.

By construction, the value of $\mathcal{A}(Y,\rho)$ behind the saturation
front is not under control, as this is fixed by hand at a constant value.
In practice we shall choose this constant value to be zero:
$\mathcal{A}(Y,\rho)=0$ for $\rho\le \rho_c(Y)-\Delta$ (but other values
of $\order{1}$ will be also used, to test the sensitivity of the method
to this particular choice). Thus, clearly, our procedure cannot be used
for those physical problems which are sensitive to the details of the
saturation region, like deep inelastic scattering at low $Q^2\lesssim
Q_s^2(Y)$, or particle production at low transverse momenta. On the other
hand, this procedure accurately describes the dynamics of the front, in
that it provides the same results as the non--linear BK equation for the
energy dependence of the saturation momentum and for the gluon
distribution $\mathcal{A}(Y,k)$ at momenta $k
> Q_s(Y)$, for both fixed and running coupling, and for all values of
$Y$.

Let us illustrate the efficiency of this method with a few numerical
results for the case of a running coupling. We include running coupling
by pulling the $\abar$ factor inside the $k'$--integral in \eqnum{eq:BK}
and using the one--loop expression for the running coupling with scale
$Q^2={\rm max}(k^2, k'^2)$ and $\Lam= 200$~MeV. This simple prescription
is in agreement with the recently constructed running--coupling version
of the BK equation \citep{Balitsky:2006wa,Kovchegov:2006vj} . To avoid
the infrared divergence of the coupling at $Q^2=\Lam^2$, we shall replace
$\alpha_s(Q^2) \to \alpha_s(Q^2+\mu^2)$ for some parameter $\mu$. Our
default choice will be $\mu^2 = 0.5$ GeV$^2$, but we shall study the
sensitivity of our results to variations in $\mu$. Our initial condition
$\mathcal{A}(Y=0,k)$ is given by the bremsstrahlung spectrum for $k >
1$~GeV (with maximal height $\mathcal{A}=0.5$) and it vanishes for $k <
1$~GeV.

\begin{figure}[t] {\centerline{
  \includegraphics[angle=270, scale=0.6]{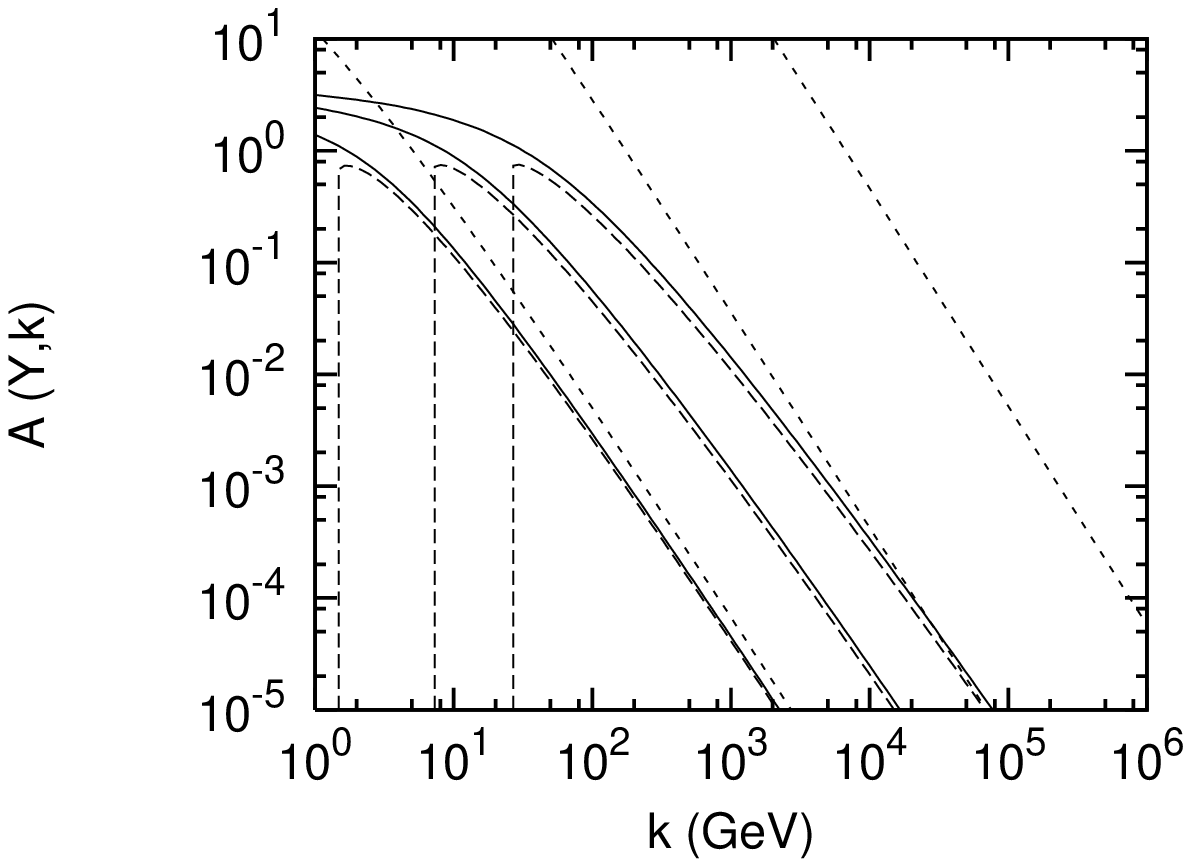}
  \includegraphics[angle=270, scale=0.6]{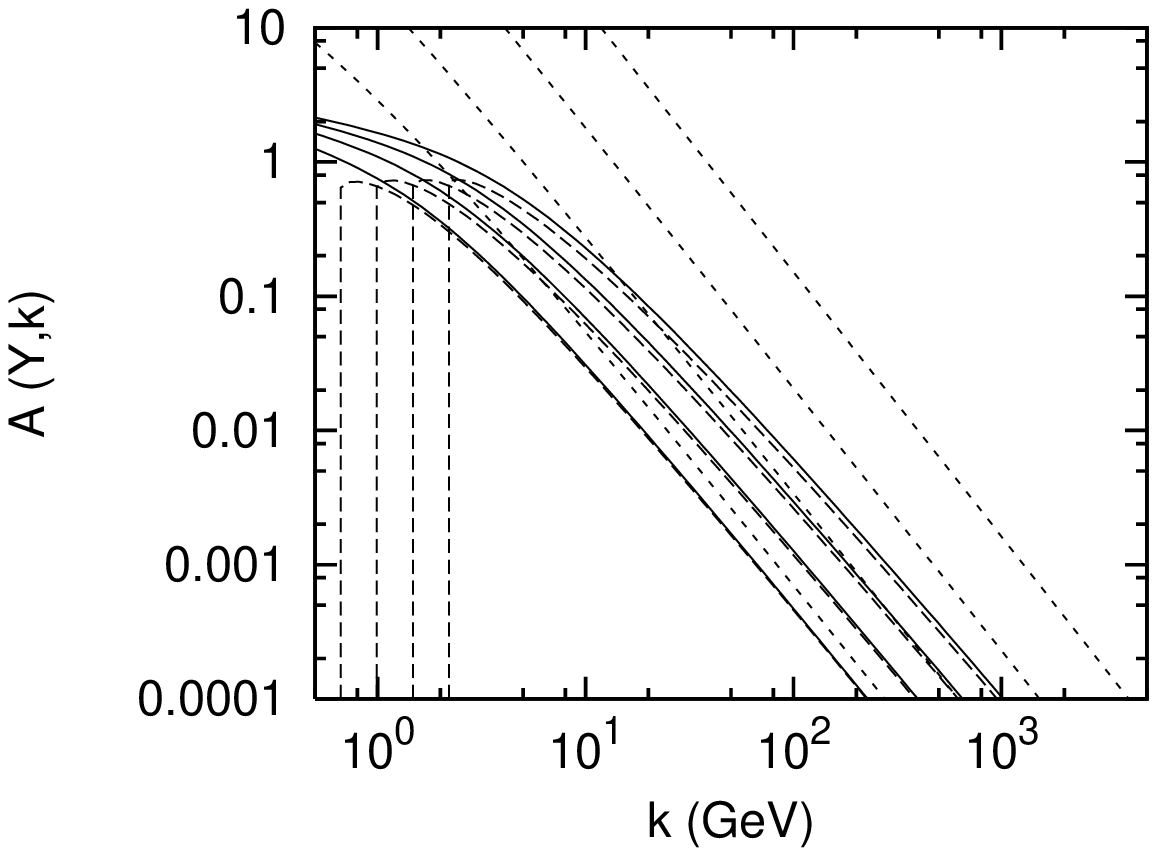}
}
\caption {\sl The running coupling results for: BK (solid curves), BFKL
with absorptive boundary (long dashed curves) and pure BFKL (short dashed
curves) for (left) $Y= 10, 20, 30$ and $40$, and (right) $Y=6, 8, 10$ and
12.
 For the absorptive boundary we used $c=0.4$ and $\Delta=2.0$.
\label{fig:bfklrunres1} }}
\end{figure}

The pure BFKL evolution with running coupling is known to be infrared
unstable: the rise of the gluon distribution is much faster at small
values of $k$ (where the coupling is larger) and, besides, this rapid
accumulation of gluons in the infrared is also feeding the growth at
higher $k$. Therefore the linear evolution behaves quite differently
compared to the non--linear one, even at high $k$. To illustrate that, we
have also included in Fig.~\ref{fig:bfklrunres1} the results of the
strict BFKL evolution, which indeed show a much faster progression
towards high $k$ as compared to the BK equation. On the other hand, one
observes a perfect matching between the saturation fronts provided by BK
and, respectively, BFKL with saturation boundary. This shows that the
front remains of the pulled type even with running coupling and at the
same time demonstrates the success of our method for effectively
implementing saturation. We also note that in the corresponding figures
in \citep{Avsar:2009pv}, the values $c=0.1$ and $\Delta=5.0$ were used
instead of $c=0.4$ and $\Delta=2.0$, thus showing that the results do not
depend on the specific values of these parameters (as long as they are
correlated as $\Delta \sim \ln(1/c)$).

\begin{figure}[t] {\centerline{
  \includegraphics[angle=270, scale=0.6]{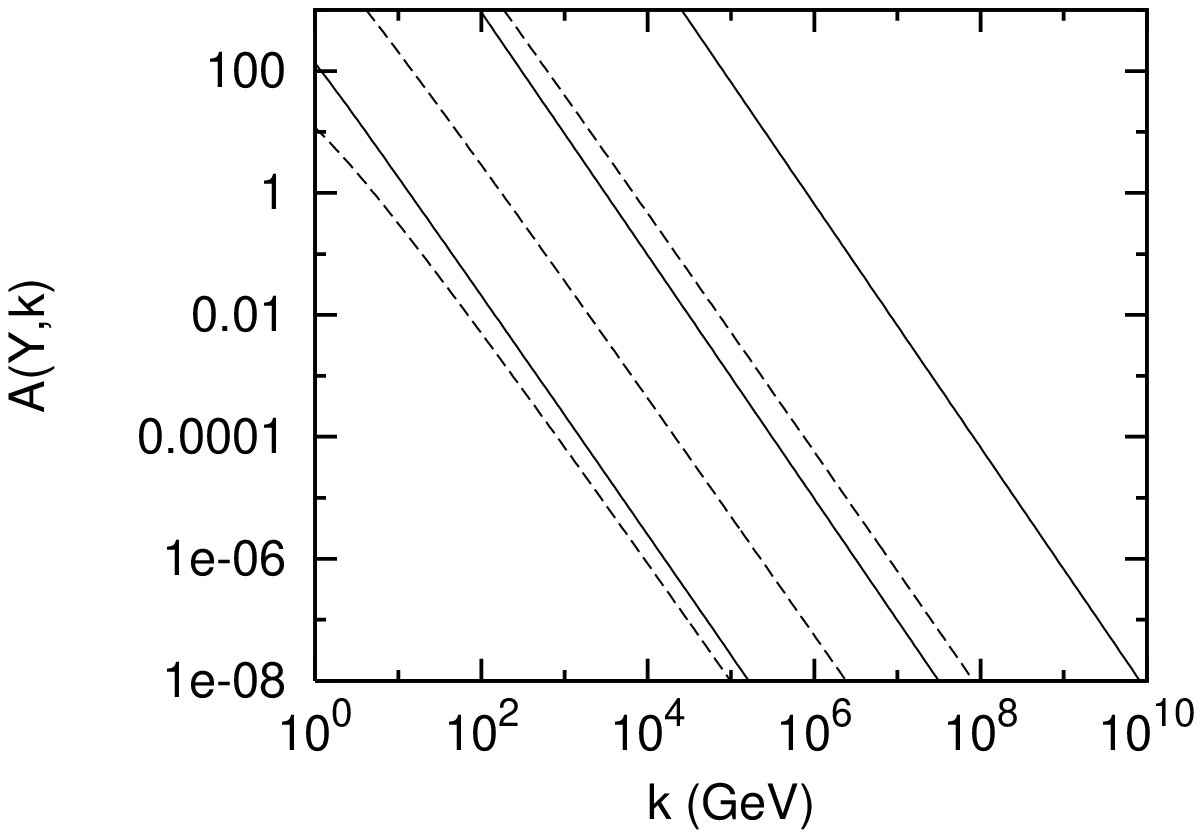}
  \includegraphics[angle=270, scale=0.6]{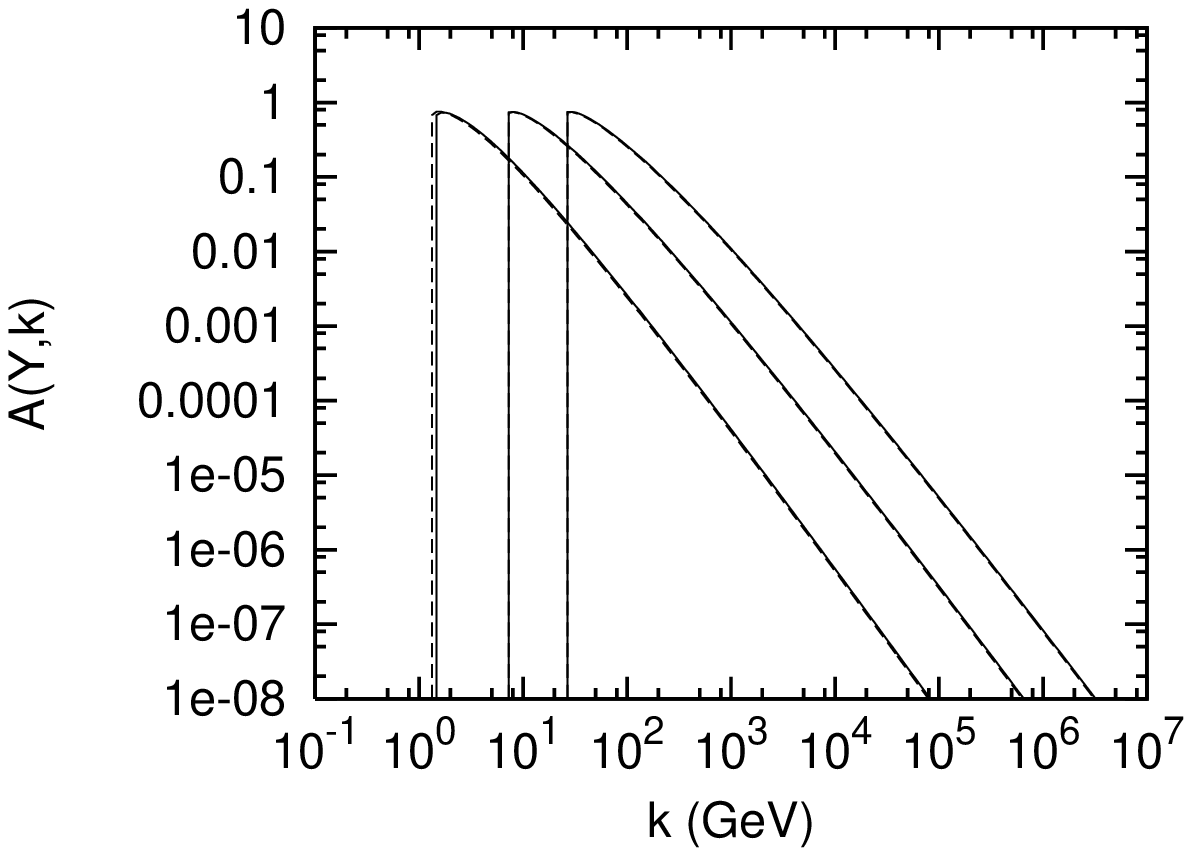}}
\caption {\sl Testing the sensitivity of the BFKL evolution to the IR cutoff
$\mu$. Left: The pure BFKL evolution with $\mu^2 = 0.5$ GeV$^2$ (solid
lines) and $\mu^2 = 2.0$ GeV$^2$ (dashed lines), for $Y=10, \,20$ and 30. Right: The
BFKL evolution with absorptive boundary for $\mu^2 = 0.5$ GeV$^2$ (solid
lines), $\mu^2 = 2.0$ GeV$^2$ (dashed lines) and for the same values for $Y$
as before. \label{fig:bfklrunres2} }}
\end{figure}

In particular, the infrared problem is cured by saturation: the
saturation scale effectively acts as an infrared cutoff, which becomes
`hard' ($Q_s^2(Y)\gg\Lam^2$) for sufficiently high energy. To better
illustrate this, we exhibit in Fig.~\ref{fig:bfklrunres2} results
obtained for different values of the IR cutoff $\mu^2$ inserted in the
running coupling. Unlike the pure BFKL results (left figure), which are
extremely sensitive to a change in $\mu$, the results corresponding to
the saturation boundary condition (right figure) show no sensitivity
whatsoever.

From the curves in Fig.~\ref{fig:bfklrunres1}, it is also possible to
extract the $Y$--dependence of the saturation momentum $\rho_s(Y)$ for
running coupling. We find that the squared--root law $\rho_s \simeq
\lambda_r \sqrt{Y}$ predicted by the theory
\citep{Iancu:2002tr,Mueller:2002zm,Triantafyllopoulos:2002nz} for
asymptotically high energies provides a good fit to our numerical results
for $Y\ge 10$, with a fitted value $\lambda_r\simeq 2.9$ which agrees
reasonably well with the (asymptotic) theoretical
expectation\footnote{For asymptotically large $Y$, the running--coupling
BFKL evolution yields \citep{Iancu:2002tr,Mueller:2002zm} :
$\rho_s(Y)\simeq \sqrt{2\lambda_0 b_0Y}$ where $\lambda_0\simeq 4.88$ is
the same number as in \eqnum{satmom} and $b_0\equiv {12N_c}/({11N_c
-2N_f})$ is the coefficient in the one--loop running coupling:
$\bar\alpha(Q^2)={b_0}/{\ln (Q^2/\Lam^2)}$. In our simulations, we use
$N_f=0$, hence we expect $\lambda_r\equiv \sqrt{2\lambda_0 b_0}\simeq
3.26$, which is indeed consistent with the fit to the curves in
Fig.~\ref{fig:bfklrunres1}.} $\lambda_r\simeq 3.2$.

Now, from the phenomenological point of view, we are more interested in
values of $Y$ which are not that large, say $Y\le 14$ (corresponding to
$x \gtrsim 10^{-6}$), as relevant for forward jet production at LHC. With
that in mind, we also show in Fig.~\ref{fig:bfklrunres1} (right) the
results for lower values of $Y$, between 6 and 12 units; one can thus see
that the absorptive boundary method works equally well also for such
lower rapidities.

\subsection{CCFM evolution in presence of saturation: analytic results}
\label{sec:anmdim}

Within numerical simulations, it is straightforward to implement the
saturation boundary condition in any of the previous versions of the CCFM
equation: the complete, integral, equation \eqref{eq:ccfminteq1}, or one
of its simpler, differential, equations deduced in
Sect.~\ref{sec:derivediffeq}. Given the approximations made in deriving
the latter, it is clear that the corresponding results will not be
exactly the same. In particular, the saturation momentum deduced from the
full CCFM equation \eqref{eq:ccfminteq1} will also depend upon the
maximal angle variable $\bar\xi$ introduced by the kinematics of the
external scattering; in the case of DIS, this means that $Q_s$ will
depend upon the virtuality scale $Q^2$ at which one measures the
structure functions. It would be very interesting to study this
dependence, and also the energy dependence of $Q_s$, by numerically
solving the integral equation \eqref{eq:ccfminteq1} in the presence of
saturation. This could be done via Monte Carlo simulations (e.g., by
correspondingly extending the CASCADE event generator
\citep{Jung:2000hk}), which would also allow one to study the effects of
saturation on the structure of the final state. Such numerical
calculations would be however tedious in practice, and it would be very
difficult to reach high values of $Y$ in this way. Our interest here is
mostly conceptual: we would like to understand how saturation modifies
the CCFM evolution, and how the respective predictions compare to those
based on the BFKL evolution (with saturation once again). To that aim, we
shall focus on the differential versions of the CCFM equation, for which
we shall present some analytic estimates in this section, and then a
systematic numerical study in Sect.~\ref{sec:results}.

The analytic study is based on the Mellin transform, as already
introduced in \eqnum{TBFKL}. In what follows we will generically denote
the kernel eigenfunction (the quantity denoted as $\abar \chi(\gamma)$ in
\eqref{TBFKL}) as $\omega(\abar, \gamma)$, keeping the notation
$\chi(\gamma)$ only for the BFKL eigenfunction \eqref{chigamma}. In the
case of \eqnum{eq:ccfmdiffeq4}, the function $\omega$ can be analytically
computed, while for \eqnum{eq:ccfmdiffeq3} it must be constructed
numerically. Namely, for \eqnum{eq:ccfmdiffeq4}, we get
\beq
\omega = \abar\left ( \frac{1}{\gamma} + \frac{1}{1-\gamma + \omega}
\right ),
\label{omegaeq}
\eeq
which is easily solved to give
\beq
\omega = - \frac{1}{2}\left ( 1-\gamma - \frac{\abar}{\gamma} \right ) +
\sqrt{\frac{1}{4}\left (1- \gamma - \frac{\abar}{\gamma} \right )^2 +
\frac{\abar}{\gamma}}.
\eeq
For $\abar \to 0$ this expression reduces to
\beq
\omega = \abar\left ( \frac{1}{\gamma} + \frac{1}{1-\gamma} \right )
\label{eq:simplechi}
\eeq
where $1/\gamma$ is recognized as the collinear piece (the one which
dominates when $k \gg k'$), while $1/(1-\gamma)$ is the anti--collinear
one ($k' \gg k$). The high--energy saddle point coincides with the BFKL
one, $\gamma = 1/2$, but the intercept is given by 4, instead of $2.77$
for BFKL.  The saturation saddle point, $\gamma_s$, is found according to
\eqnum{gammasBFKL} which now becomes
\beq
\omega'(\gamma_s) = \frac{\omega(\gamma_s)}{1-\gamma_s}\,.
\label{eq:satsaddlepoint}
\eeq
This gives $\gamma_s = 1/3\simeq 0.33$, which is quite close to the
respective BFKL value $\gamma_s \simeq 0.37$. The saturation exponent
$\lambda_s$ controlling the growth of the saturation momentum is again
obtained from \eqref{gammasBFKL} which in this case gives 6.75, and thus
is larger than the respective BFKL value 4.88.

Although the eigenfunction \eqref{eq:simplechi} gives a too high
intercept and a too large speed of the saturation front, it is well known
that it approximates the general shape for the BFKL eigenfunction,
$\chi(\gamma)$, very well. (This explains why the respective saturation
saddle points are so close to each other.) In fact if one just subtracts
from \eqref{eq:simplechi} the difference of the two intercepts, $4 - 4\ln
2$, then an almost perfect approximation of $\chi(\gamma)$ is obtained.
Note that a constant shift in the eigenfunction corresponds in momentum
space to the inclusion of a local term $\propto \mcal{A}(Y,k)$ in the
right--hand side of the differential equation.

\begin{figure}[t]{\centerline{
    \includegraphics[angle=270, scale=0.55]{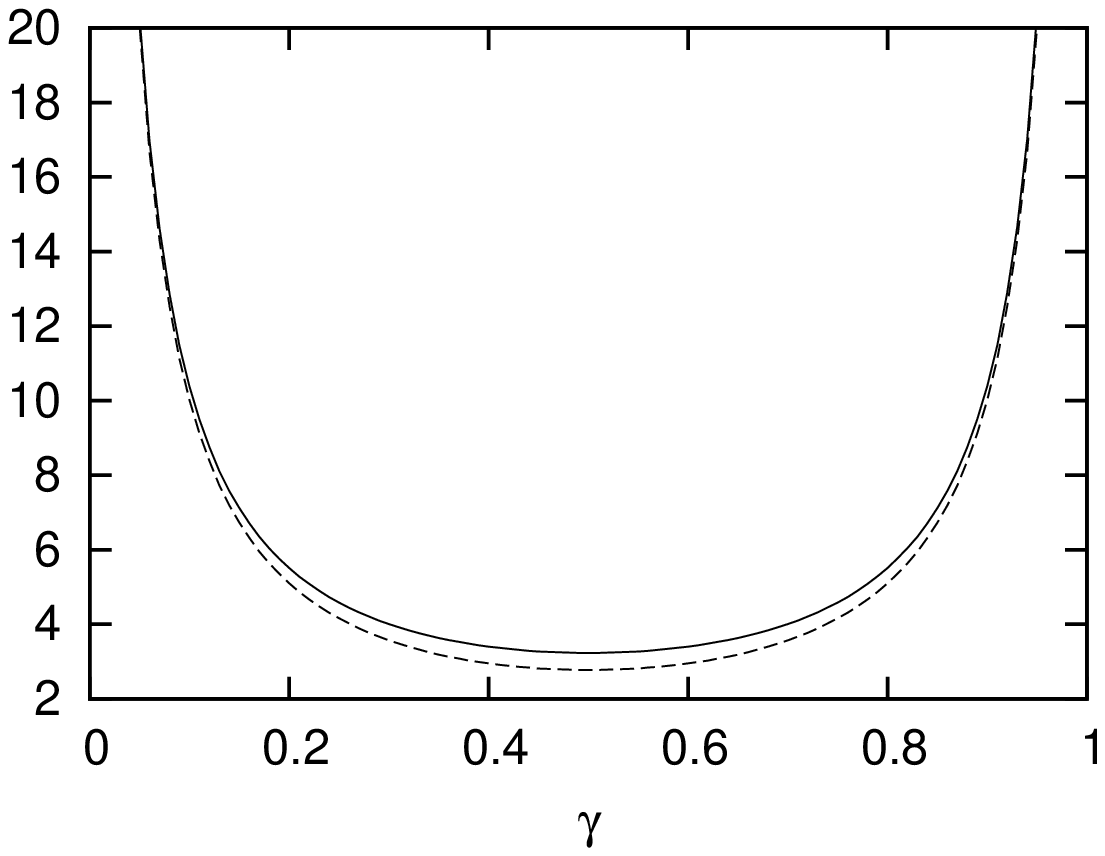}
    \includegraphics[angle=270, scale=0.55]{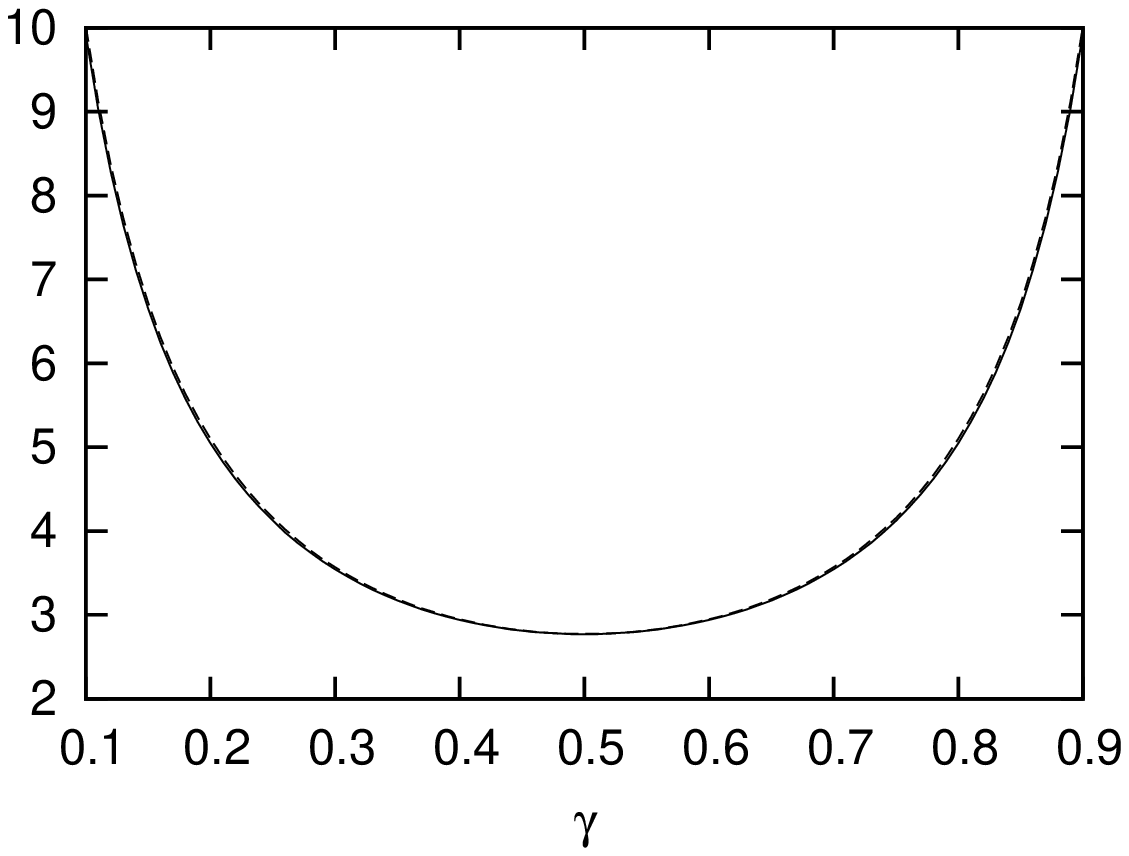}}
    \caption {\sl\label{fig:chiplot} Left: The Mellin space eigenfunction
      of the asymptotic approximation of \eqnum{eq:ccfmdiffeq3} (solid line)
      compared to the BFKL eigenfunction $\chi(\gamma)$ (dashed line). Right:
      The same as in left but with a constant 0.45 subtracted from the solid line. }
  }
\end{figure}

For \eqnum{eq:ccfmdiffeq3}, we write $\omega = \abar\tilde{\chi}(\gamma,
\omega)$, where $\tilde{\chi}$ can be numerically constructed. In the
weak coupling limit ($\abar \to 0$) we have
\beq
\omega = \abar\tilde{\chi}(\gamma) = \abar \int_0^1 \rmd t\
\frac{h(t)}{1-t} (t^{-\gamma} + t^{-(1-\gamma)}).
\eeq
To find the behaviour at the endpoints, $\gamma = 0$ and $\gamma =1$, we
concentrate on the region $t \approx 0$ and use that $h(t) =1$ for $t <
1/4$. This yields
\begin{align}
\int_0^{1/4} \rmd t\
 \frac{1}{1-t}(t^{-\gamma} + t^{-(1-\gamma)})=\nonumber \\
= \frac{(1/4)^{1-\gamma}}{1-\gamma}F(1-\gamma, 1,
2-\gamma, 1/4) + \frac{(1/4)^{\gamma}}{\gamma}F(\gamma, 1, 1+\gamma,
1/4),
\end{align}
where $F$ is the hypergeometric function. Since at the endpoints we have
$F \to 1$ for each term, we find the same endpoint behaviour as
\eqref{eq:simplechi}, and hence as in BFKL. For the complete interval we
plot $\tilde{\chi}$, together with $\chi$, in Fig.~\ref{fig:chiplot}. We
see that the minimum of $\tilde{\chi}$, \emph{i.e.} the intercept, is a
bit above the BFKL one
--- this is numerically found as 3.23 ---, and therefore so is also the
respective saturation exponent, found as $\lambda_s\simeq 5.56$. On the
other hand, the shape of the eigenfunction is again very similar to the
BFKL one: in the right figure \ref{fig:chiplot} we have subtracted the
difference between the two intercepts, \emph{i.e.} 0.45, from
$\tilde{\chi}$, and then the two curves almost coincide. So, no
surprisingly, the saturation anomalous dimension $\gamma_s\simeq 0.35$ is
close to the respective BFKL value. As previously noted, such a constant
subtraction is tantamount to a contribution proportional to
$\mcal{A}(Y,k)$ in the differential equation, which in turn can be
associated with additional virtual corrections. In appendix
\ref{sec:betteraccuracy} we will see
that such virtual terms naturally appears in \eqnum{eq:ccfmdiffeq3} when
the real--virtual cancellations are treated with a better accuracy.

\begin{figure}[t]{\centerline{
    \includegraphics[angle=270, scale=0.75]{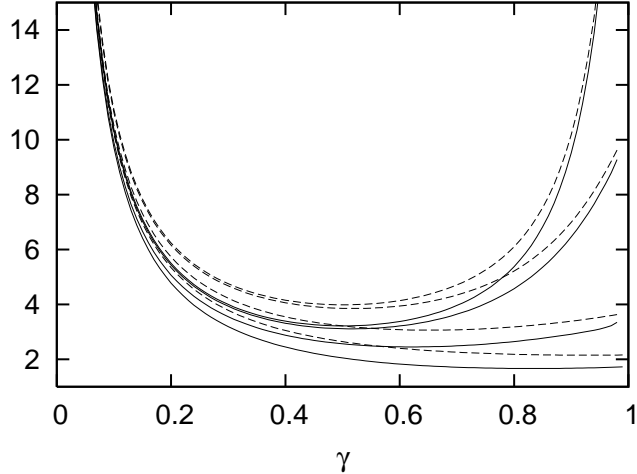}}
    \caption {\sl\label{fig:chiplot2} The eigenfunctions $\omega/\abar$
      for equations \eqref{eq:ccfmdiffeq3} (solid line) and
      \eqref{eq:ccfmdiffeq4} (dashed line), for 4 values of the coupling (from up
      to down) : $\abar = 0.001\,,0.01,\,0.1\,,$
      and 0.4. The very small values of $\abar$ are included to illustrate
      the emergence of the anti--collinear pole at $\gamma=1$ in the limit
      $\abar\to 0$.}
}
\end{figure}

Before leaving this section, we demonstrate  the behaviour
of the eigenfunctions for any $\abar$. For finite $\abar$, the
anti-collinear pole is screened by $\omega$ appearing in the r.h.s.
of \eqref{omegaeq} and the corresponding equation for \eqref{eq:ccfmdiffeq3}.
The pole appearing in \eqref{eq:simplechi}  is really valid only
when $\abar = 0$.
In Fig.~\ref{fig:chiplot2} we plot the eigenfunctions, $\omega/\abar$, for
\eqref{eq:ccfmdiffeq3} and  \eqref{eq:ccfmdiffeq4} for various
values of $\abar$. The figure clearly demonstrates the emergence of  the anti-collinear pole
 as $\abar \to 0$.

\section{Numerical results} \label{sec:results}
\setcounter{equation}{0}

In this section, we shall present numerical solutions for the two
differential versions of the CCFM evolution derived in section
\ref{sec:derivediffeq}, namely Eqs.~\eqref{eq:ccfmdiffeq3} and
\eqref{eq:ccfmdiffeq4}, and also for the BFKL equation including the
kinematical constraint, cf. Eqs.~\eqref{eq:bfklinteq} and
\eqref{eq:bfkldiffeq}. We shall mostly focus on the unitarity--preserving
versions of these equations, as obtained after implementing the
saturation boundary described in Sect.~\ref{sec:saturation}. But before
doing that, it is also instructive to compare the respective linear
evolutions, in order to illustrate the significant differences between
them (in agreement with the discussion in Sect.~\ref{sec:saturation}),
and thus emphasize the ambiguity in our current understanding of
high--energy evolution within perturbative QCD. Remarkably, it will turn
out that this ambiguity is drastically reduced after including the
saturation effects in the form of the absorptive boundary. Namely, the
main conclusion that will emerge from our analysis is that all the
equations under consideration --- the CCFM equations
\eqref{eq:ccfmdiffeq3} and \eqref{eq:ccfmdiffeq4}, and the BFKL equation
with and without the kinematical constraint ---, which differ
significantly from each other in the linear regime, lead nevertheless to
very similar predictions for the energy dependence of the saturation
momentum after adding the saturation boundary condition and with a
running coupling. Note that the inclusion of the running coupling is
essential in that sense: with a fixed coupling, important differences
persist even after including saturation, as it should be expected from
the discussion in Sect.~\ref{sec:saturation}.

For the purposes of the numerics, it is convenient to rewrite
\eqnum{eq:bfklinteq} in differential form and perform the azimuthal
integration; this yields:
\beq
\partial_Y \mathcal{A}(Y,k) \!\! &=& \!\!\!\! \abar \!\! \int dk'^2 \left
( \frac{\theta( k - k')\mathcal{A}(Y,k') +
\theta( k' - k)\theta(Y - \ln(k'^2/k^2))\mathcal{A}(Y - \ln(k'^2/k^2),k')
}{\vert k^2 - k'^2\vert}
\right. \nonumber \\
&-&  \left. \frac{k^2}{k'^2} \left( \frac{\mathcal{A}(Y,k)}{\vert k^2 - k'^2\vert} -
 \frac{\mathcal{A}(Y,k)}{\sqrt{4k'^4+k^4} }\right ) \right ).
\label{eq:bfkldiffeq2}
\eeq

\begin{figure}[t] {\centerline{
  \includegraphics[angle=270, scale=0.57]{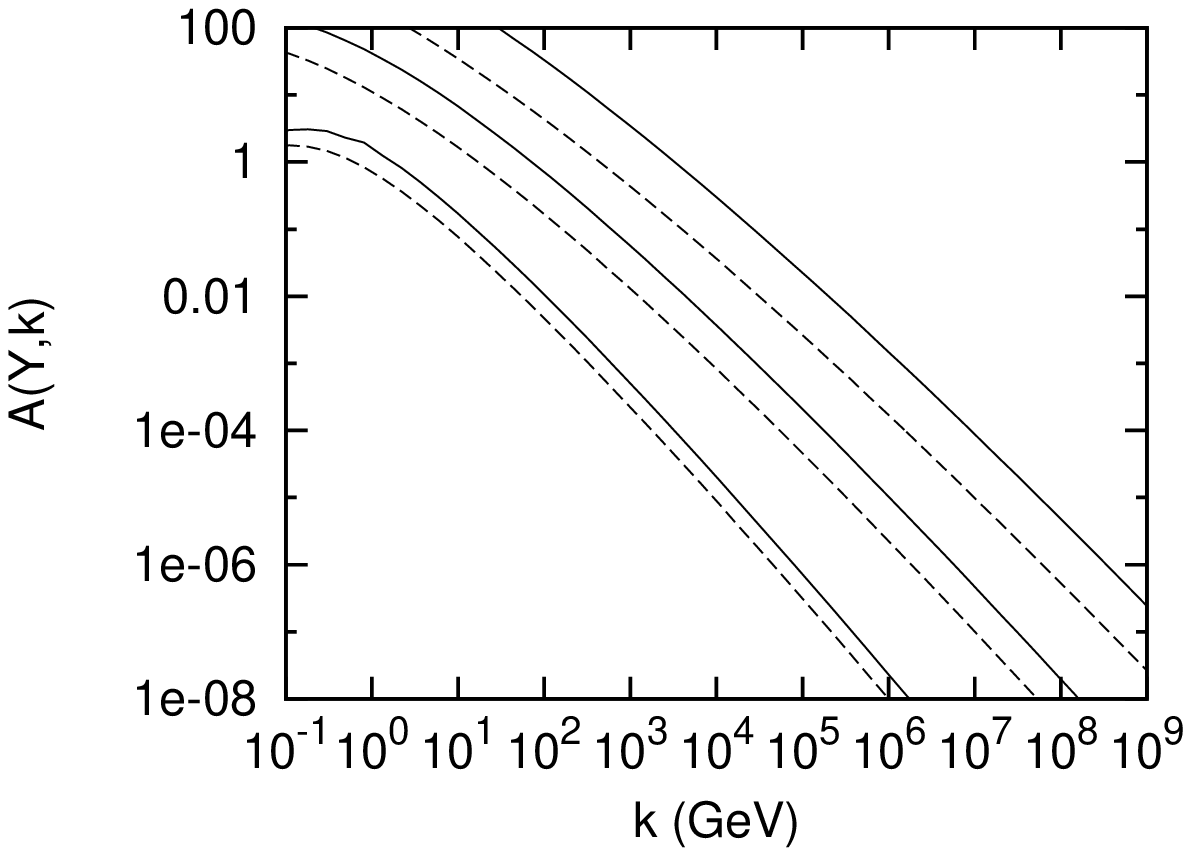}
  \includegraphics[angle=270, scale=0.57]{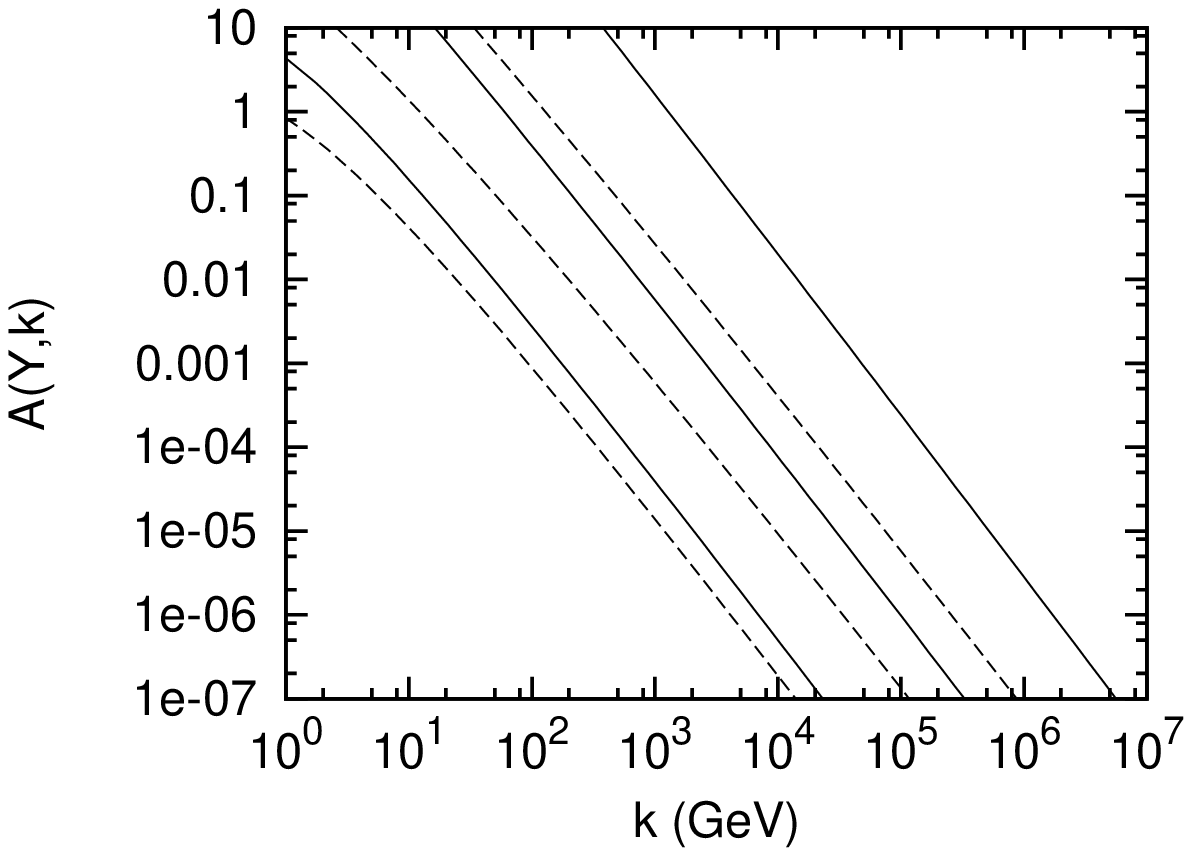}}
\caption {\sl The solutions to \eqref{eq:ccfmdiffeq3} (solid lines), and
the BFKL equation including the kinematical constraint
\eqref{eq:bfkldiffeq2} (dashed lines) for $Y=10, 20$ and 30. Left: fixed
coupling $\alpha_s=0.2$. Right: running coupling. \label{fig:bfklkcres1}
}}
\end{figure}

In Fig.~\ref{fig:bfklkcres1} we show the solutions to the linear
equations \eqref{eq:ccfmdiffeq3} and \eqref{eq:bfkldiffeq2}. As expected
from the discussion in Sect.~\ref{sec:saturation}, one sees that the CCFM
solution is somewhat faster. The difference is more pronounced for a
running coupling, due to the infrared instability of the linear
evolution. But even with a running coupling, the present equations
\eqref{eq:bfkldiffeq2} and \eqref{eq:ccfmdiffeq3}, which include the
kinematical constraint, predict a growth for the gluon distribution which
is strongly reduced as compared to the strict BFKL equation (without
kinematical constraint). This was shown in Ref.~\citep{Avsar:2009pv}
where the solutions of \eqref{eq:ccfmdiffeq3} have been compared to the
solutions of the pure BFKL equation.


\begin{figure}[t] {\centerline{
   \includegraphics[angle=270, scale=0.57]{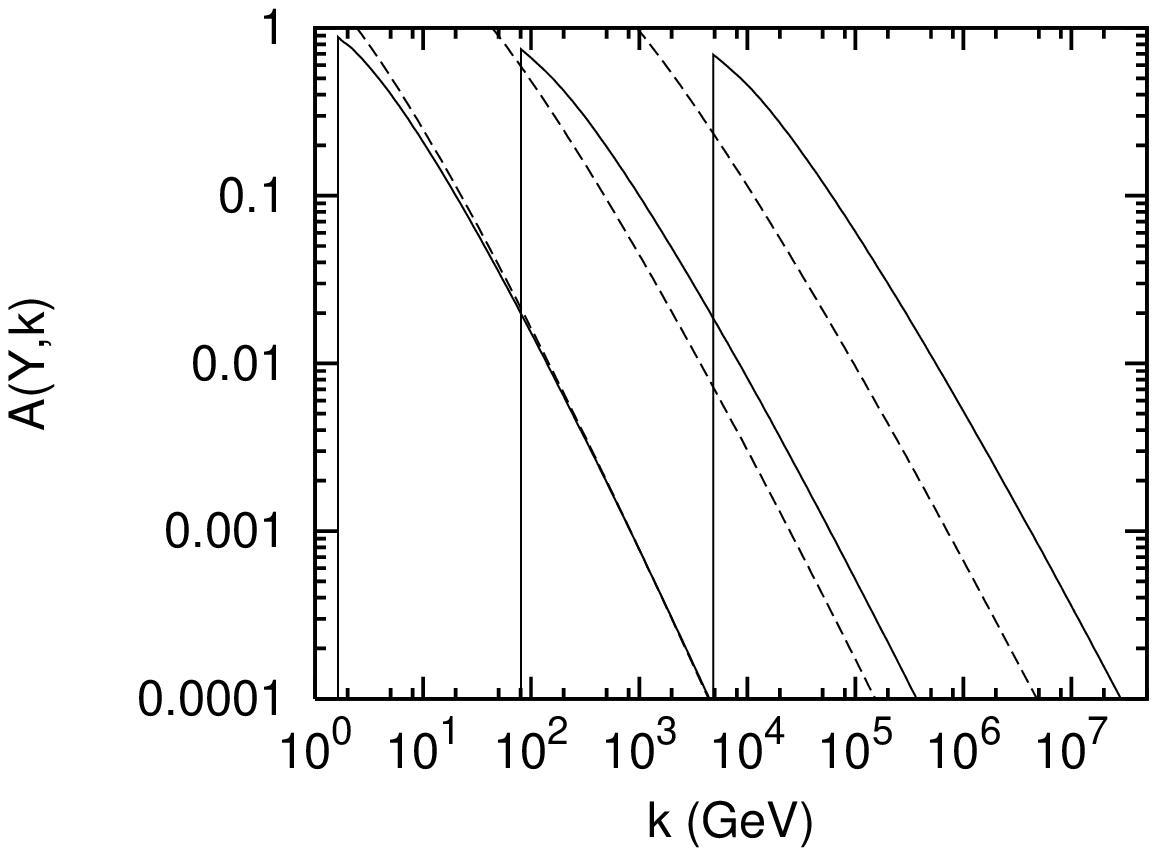}
  \includegraphics[angle=270, scale=0.57]{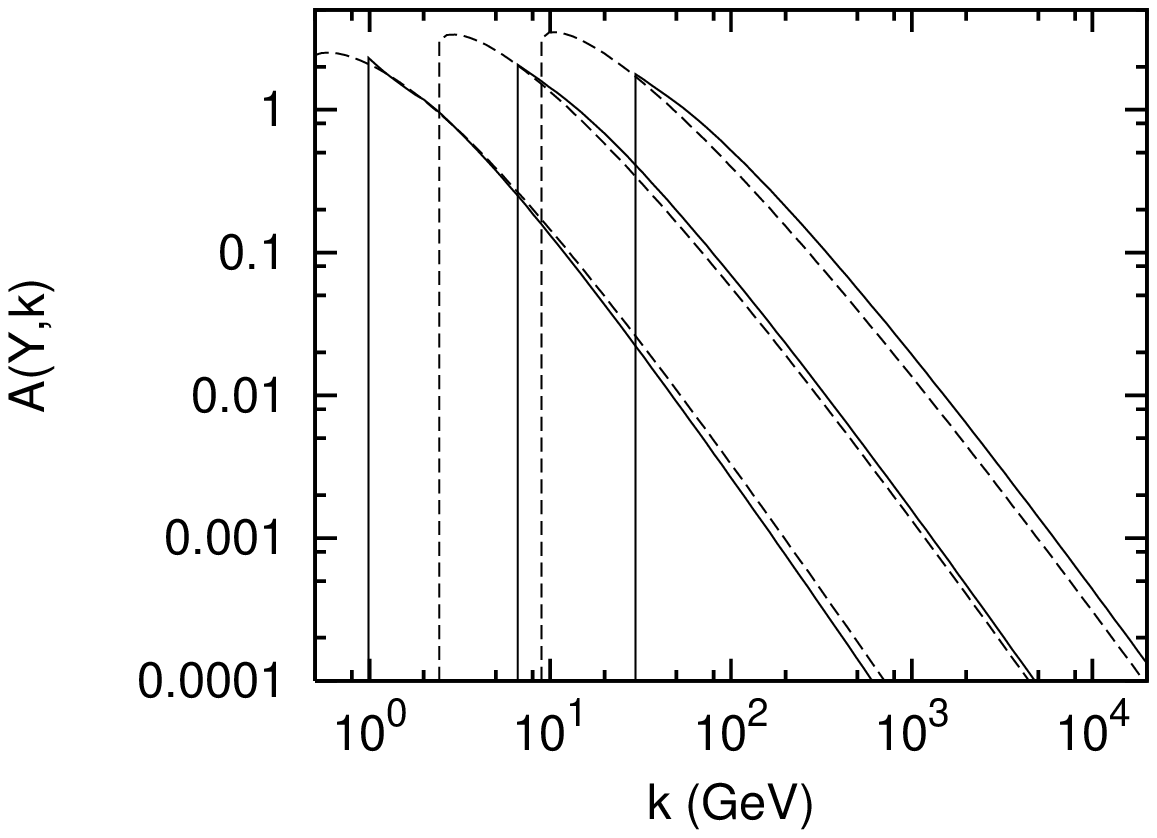}
 }
  \caption { \sl The solutions to equations \eqref{eq:ccfmdiffeq3} (solid lines)
    and \eqref{eq:bfkldiffeq2} (dashed lines) including the saturation boundary,
    for $Y=10, 20, 30$. The solutions to \eqref{eq:bfkldiffeq2} have been
    scaled so that the solutions match at $Y=10$. Left: fixed coupling $\alpha_s=0.2$.
    Right: running coupling.\label{fig:ccfmeqplot} }
}
\end{figure}

Let us now perform a similar comparison after adding the saturation
boundary. (Unless otherwise stated, we use the values $c=0.1$ and $\Delta
= 5.0$ for all calculations including the saturation boundary. A
different set of values will be used later on, to illustrate the
robustness of our results.) For the same $Y$ values as in Fig. \ref{fig:bfklkcres1},
we compare in Fig.~\ref{fig:ccfmeqplot}
the respective predictions of Eqs.~\eqref{eq:ccfmdiffeq3} and
\eqref{eq:bfkldiffeq2} for both fixed and running coupling.
To make the comparison between the solutions easier, we have scaled
the solutions to \eqref{eq:bfkldiffeq2} so that the solutions more or
less coincide at $Y=10$. For the fixed coupling case (left plot) the scaling
factor is 3.9 while for the running coupling case (right plot) it is
3.6. The difference between the fixed and the running coupling cases
is striking. For fixed coupling, the saturation front generated by
\eqref{eq:ccfmdiffeq3} is
seen to progress significantly faster than the one generated by \eqref{eq:bfkldiffeq2}.
For the running coupling case on the other hand,
it is clear that the respective fronts progress with similar
speeds. Thus in this case
the two equations have similar predictions for the energy
dependence of the saturation momentum, so that the entire difference between them
can (almost) be removed by either a  rescaling of the saturation momentum
$Q_s$, or as in the figure, a rescaling of $\mcal{A}$. This is clearly
not the case for a fixed coupling. Note also that the difference in
the fixed coupling case is visibly large even though the scale
of the figure is much larger than corresponding running coupling
plot.

\begin{figure}[t] {\centerline{
  \includegraphics[angle=270, scale=0.57]{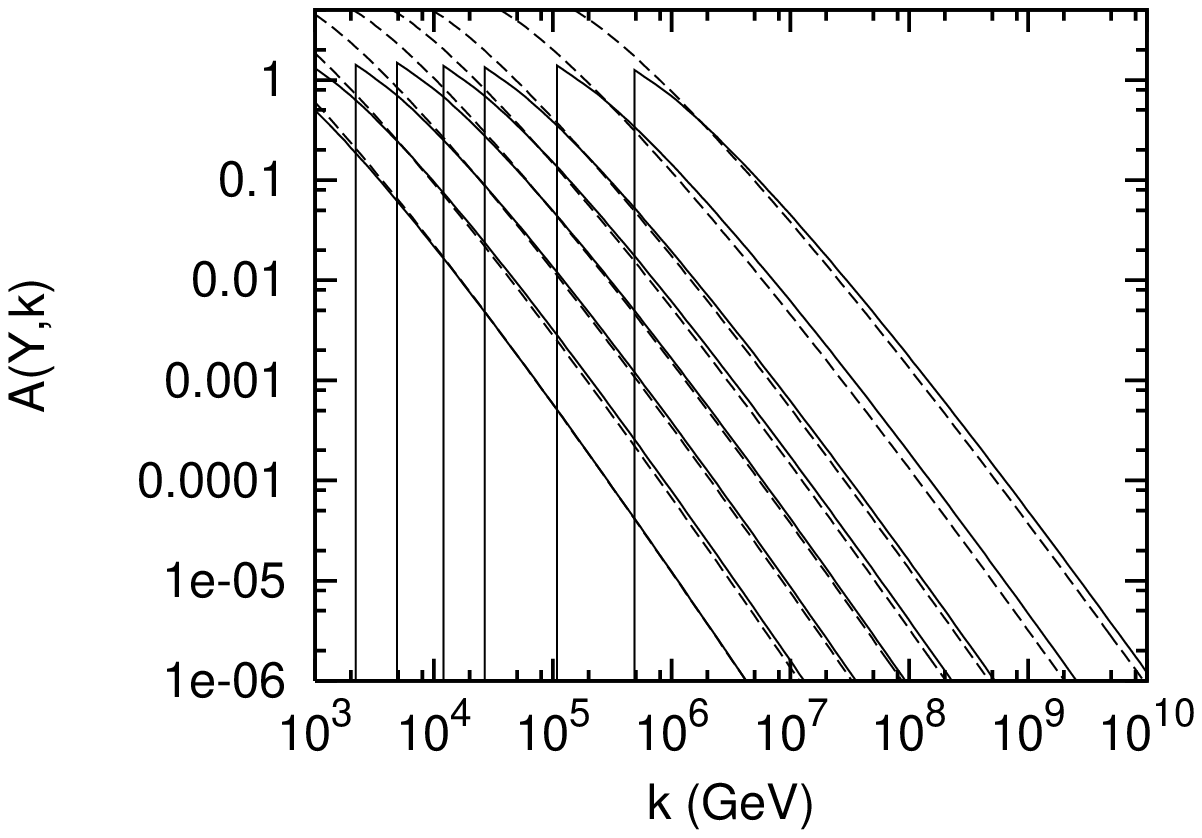}
  \includegraphics[angle=270, scale=0.57]{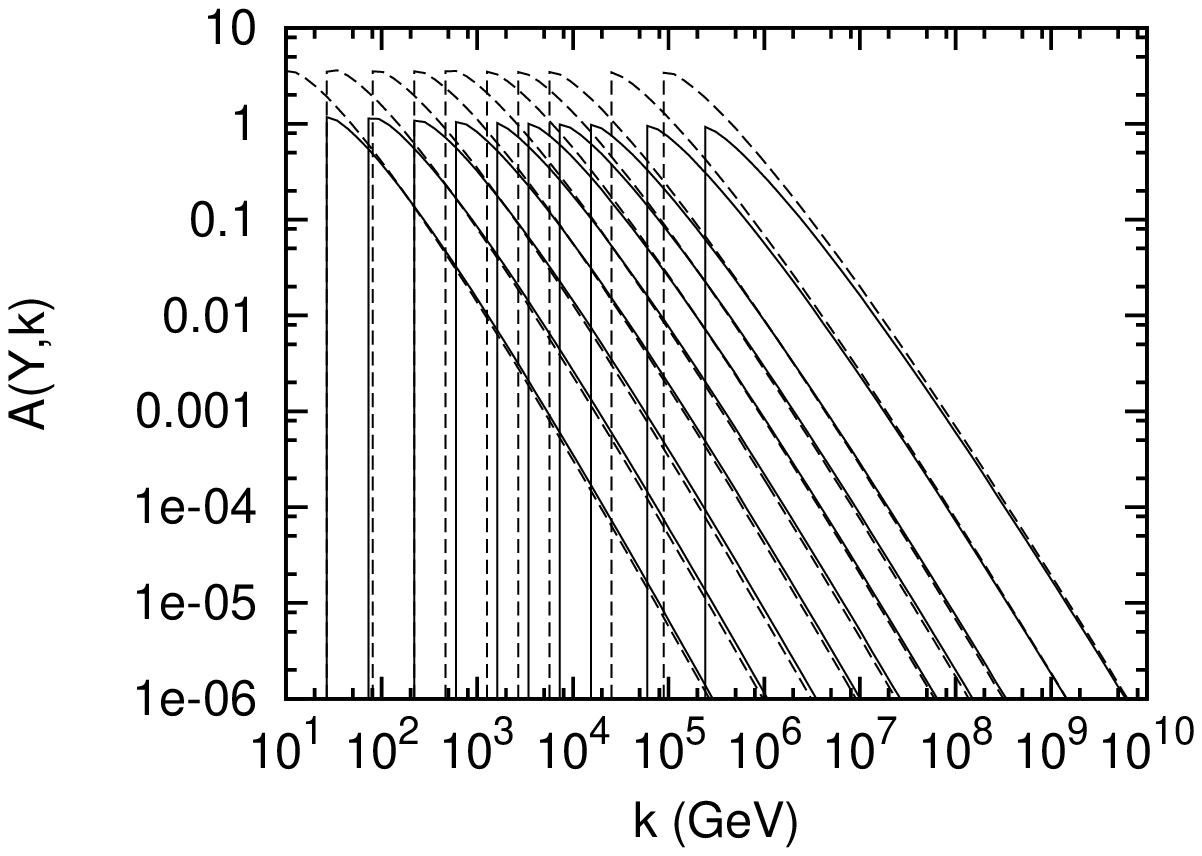}}
\caption { \sl Solutions to  \eqref{eq:ccfmdiffeq3},
\eqref{eq:bfkldiffeq2} and BFKL including the saturation boundary for
asymptotic $Y$ between 50 and 140 units, and for a running coupling.
Left: \eqref{eq:ccfmdiffeq3} (dashed lines) and \eqref{eq:bfkldiffeq2}
(solid lines). The solution to \eqref{eq:bfkldiffeq2} has been scaled by
a factor 6.5.  Right:  BFKL (dashed lines) and \eqref{eq:bfkldiffeq2}
(solid lines). The solution to \eqref{eq:bfkldiffeq2} has been scaled by
a factor 4. \label{fig:bfklkcres2} }}
\end{figure}

In Fig.~\ref{fig:bfklkcres2} (left) we show the respective results for
the running coupling case for
much higher values of $Y$. Also, in the right figure there, we compare
the absorptive boundary results for \eqnum{eq:bfkldiffeq2} and,
respectively, the standard BFKL equation without the kinematical
constraint. One clearly sees now that, as anticipated, all the equations
under consideration give approximately the same speed for the saturation
front, at least up to $Y=140$. To make this similarity more manifest, we
have rescaled the gluon distributions by appropriate factors, in such a
way to superpose the different sets of fronts. These factors account for
the difference in the overall normalization of the saturation momenta
provided by the different equations.

We have also checked that the energy dependence of the saturation
momentum is consistent with the form expected at running coupling,
\emph{i.e.}, $Q_s^2=Q_0^2\,\rme^{\lambda_r \sqrt{Y}}$. Fitting this form
to our results we find $\lambda_r \approx 3.1$ for
\eqnum{eq:ccfmdiffeq3}, and $\lambda_r \approx 2.9$ for
\eqnum{eq:bfkldiffeq2}. The normalization factor $Q_0$ is found to be a
factor 1.6 higher for \eqref{eq:ccfmdiffeq3} then for
\eqref{eq:bfkldiffeq2}. For BFKL without the kinematical constraint we
again find $\lambda_r \approx 2.9$.

\comment{Note that the values reported here, which have been obtained
using points $Y \geq 10$, should not yet be seen as the asymptotic
($Y\to\infty$) ones. It should be clear by now that the true asymptotic
comparisons between BFKL and CCFM really involve the $\abar \to 0$ limit,
which with a running coupling is not completely under control, though
with saturation the typical $\abar$ is indeed pushed towards increasingly
smaller values. On the practical side, determining accurately the
asymptotic value requires one to have a fine meshing of the $k_\perp$
space, as the very large values of $Q_s$ have a big effect on the fit.
 }

\begin{figure}[t] {\centerline{
  \includegraphics[angle=270, scale=0.7]{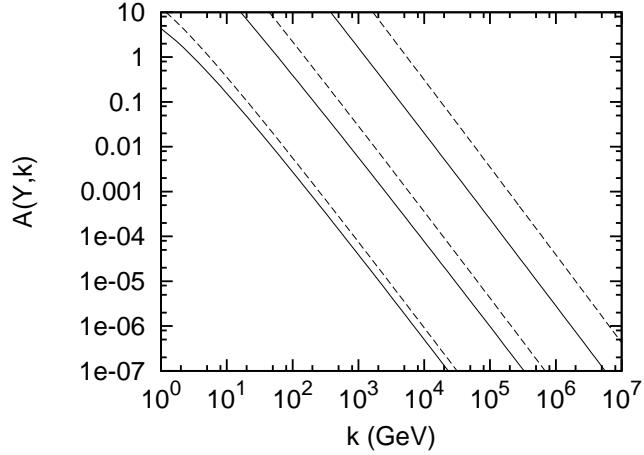}}
  \caption {\sl The solutions to equations \eqref{eq:ccfmdiffeq3} (solid lines)
    and \eqref{eq:ccfmdiffeq4} (dashed lines) in the linear case with
    a running coupling for $Y=10, 20$ and $30$.\label{fig:ccfmeqplot3} }
}
\end{figure}

 \begin{figure}[t] {\centerline{
  \includegraphics[angle=270, scale=0.57]{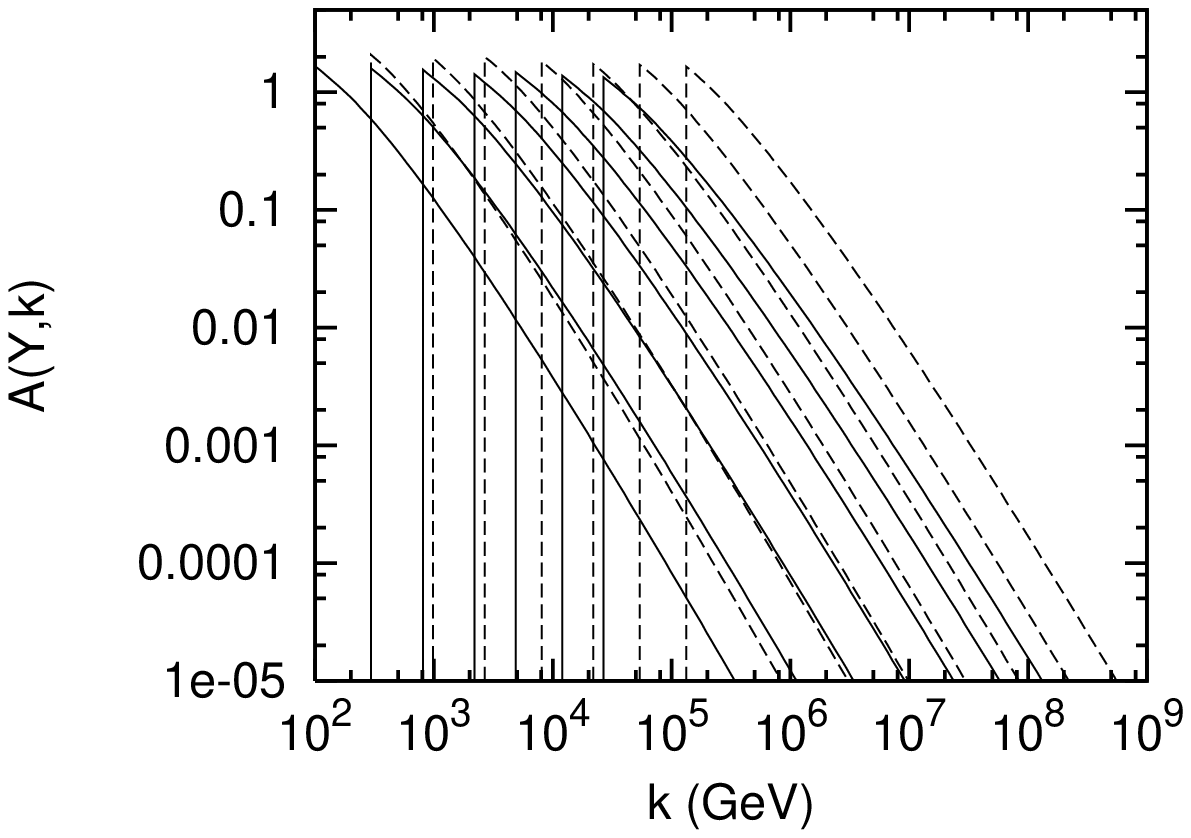}
  \includegraphics[angle=270, scale=0.57]{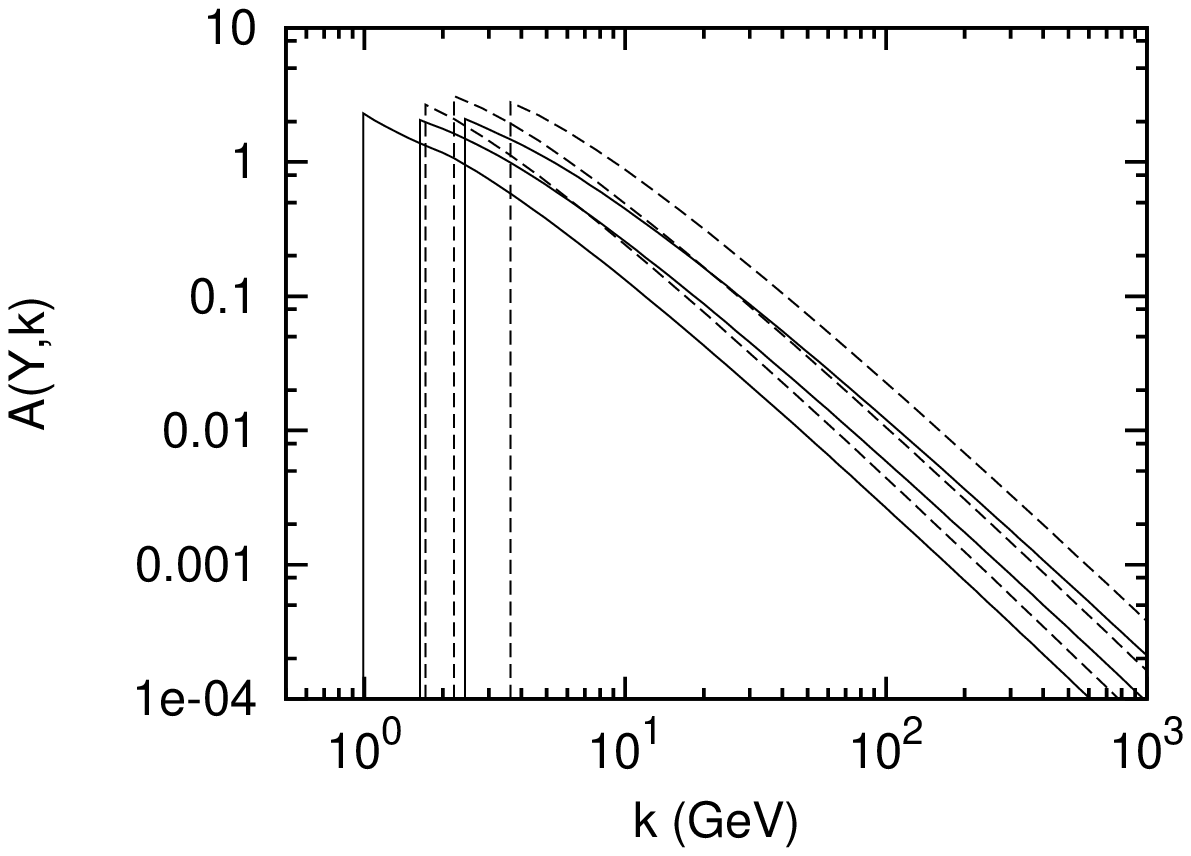}
}
  \caption { \sl The solutions to equations \eqref{eq:ccfmdiffeq3} (solid lines)
    and \eqref{eq:ccfmdiffeq4} (dashed lines) including the saturation boundary
    and for a running coupling for Left: Very high $Y$ between 40 and 100.
    Right: The phenomenologically
    relevant values of $Y=10, 12, 14$.\label{fig:ccfmeqplot2} }
}
\end{figure}

Figure \ref{fig:ccfmeqplot3} shows the comparison between the two
versions of the CCFM evolution, Eqs.~\eqref{eq:ccfmdiffeq3} and
\eqref{eq:ccfmdiffeq4}, in the linear case (no saturation boundary) and
for a running coupling. We see that \eqref{eq:ccfmdiffeq4} gives a faster
growth, but the shapes of the curves are similar; this was to be expected
in view of the discussion in Sect.~\ref{sec:anmdim}. The corresponding
results after adding the saturation boundary are shown in
Fig.~\ref{fig:ccfmeqplot2}. It is then find that the speeds of the
saturation fronts are quite similar, with \eqnum{eq:ccfmdiffeq4} giving a
slightly larger speed, meaning a higher value for the saturation
exponent. Up to the $Y$ values shown in the figure, we find the value
$\lambda_r \approx 3.3$ for \eqref{eq:ccfmdiffeq4} compared to the value
$\lambda_r \approx 3.1$ for \eqref{eq:ccfmdiffeq3} mentioned before. The
difference in the normalization factor $Q_0$ is a factor 1.4, with
\eqref{eq:ccfmdiffeq4} giving the larger value.

So far all these results have been obtained for the fixed set of values
$c=0.1$ and $\Delta=5.0$. Theoretically, the different values of $c$ and
$\Delta$ should only affect the absolute normalization of the gluon
distribution and of $Q_s$, while leaving  unchanged their shape and,
respectively, energy dependence. To check these expectations and
illustrate the robustness of our results we now repeat some of the
calculations with the set $c=0.4$ and $\Delta=2.0$. (Recall that these
two parameters are correlated as $\Delta \sim \ln 1/c$.)

\begin{figure}[t] {\centerline{
  \includegraphics[angle=270, scale=0.57]{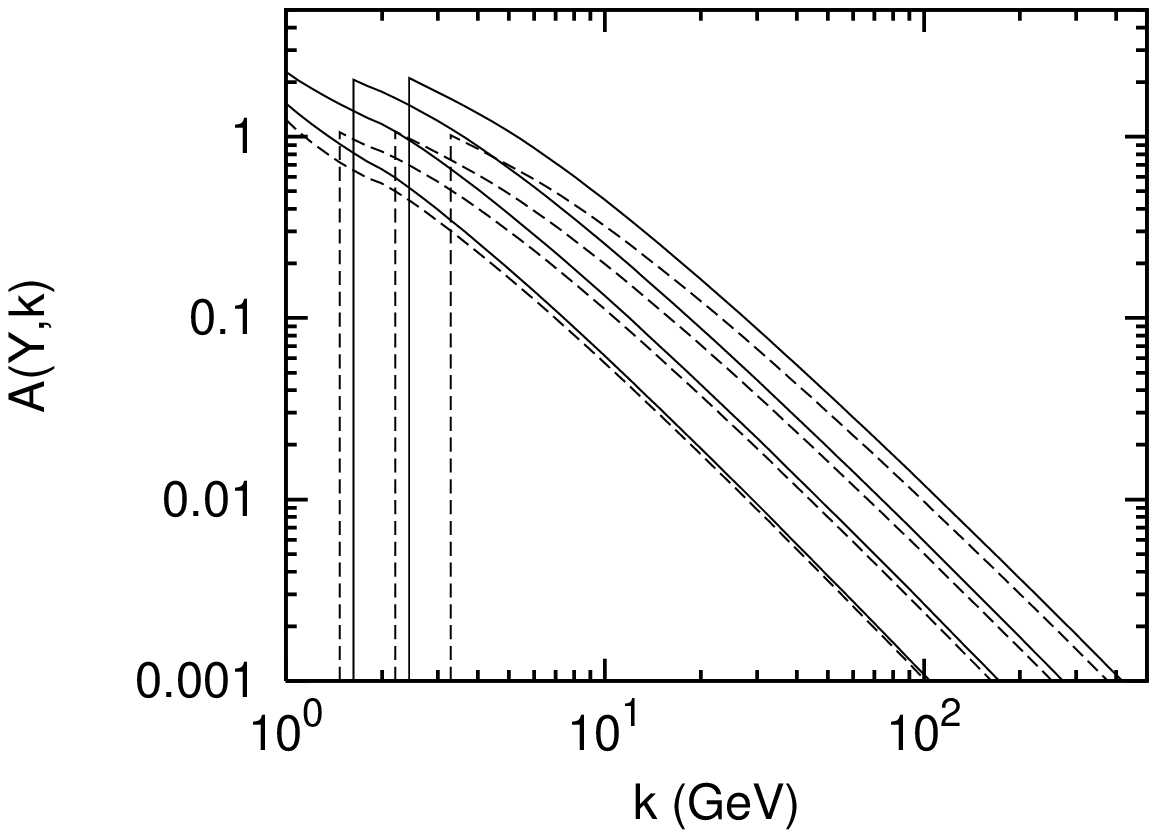}
   \includegraphics[angle=270, scale=0.57]{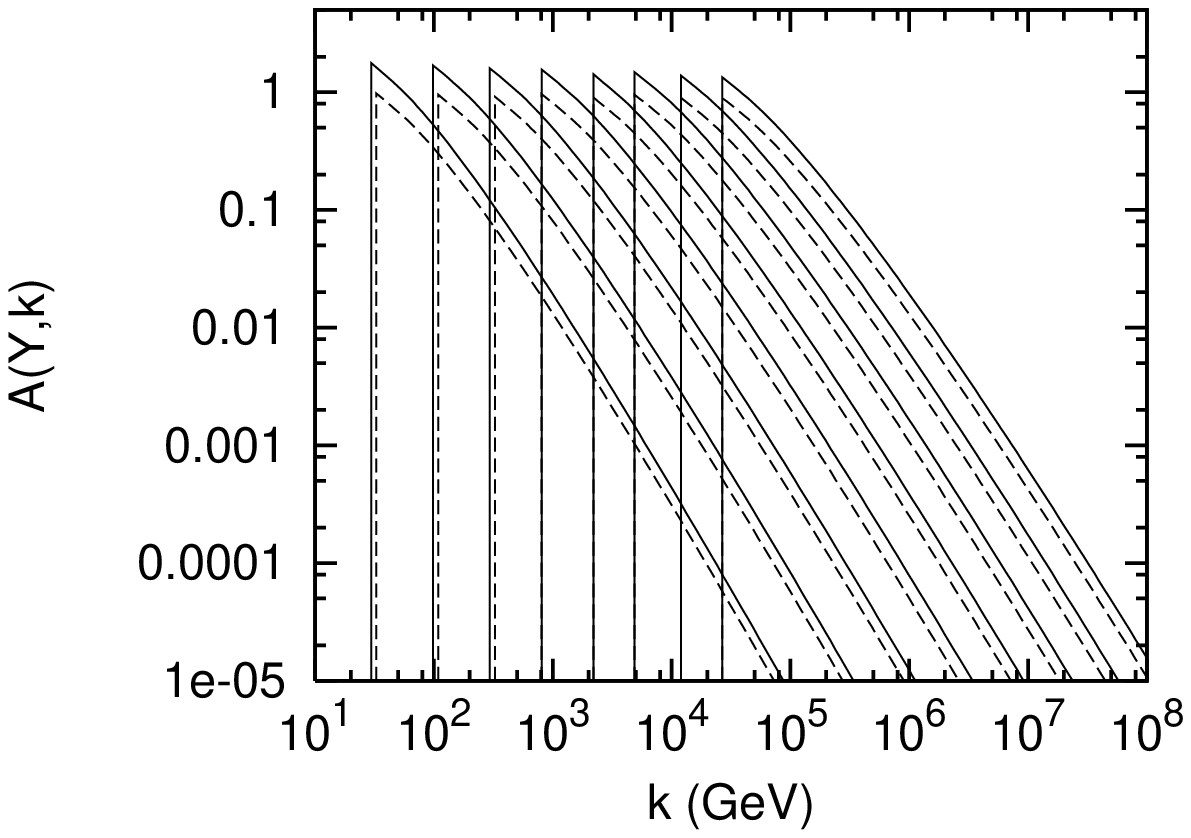}
}
\caption { \sl  Comparing the solutions to \eqnum{eq:ccfmdiffeq3}
(with saturation boundary and running coupling) using two
set of values for the saturation parameters: $c=0.1, \Delta=5.0$ (solid lines),
$c=0.4, \Delta=2.0$ (dashed lines). Left: $Y=8, 10, 12, 14$. Right:
$Y$ between 30 and 100.
\label{fig:ccfmeqplot8} }}
\end{figure}

\begin{figure}[t] {\centerline{
   \includegraphics[angle=270, scale=0.7]{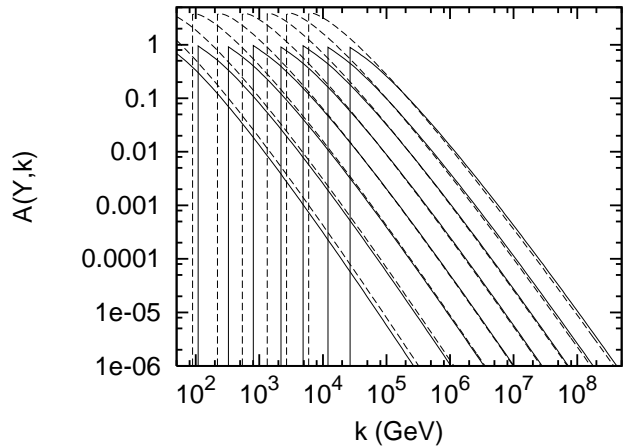}}
\caption { \sl  Comparing the solutions of Eqs.~\eqref{eq:ccfmdiffeq3}
and  \eqref{eq:bfkldiffeq2} for  $c=0.4, \Delta=2.0$, and large $Y$
between 30 and 100 units.  The solutions to  \eqref{eq:bfkldiffeq2}  have been
scaled up by a factor 5.5.
\label{fig:ccfmeqplot7} }}
\end{figure}

In Fig.~\ref{fig:ccfmeqplot8} we compare the results obtained with the
two sets of values for the case of \eqnum{eq:ccfmdiffeq3}. For the
phenomenologically relevant values between $Y=8$ and 14, it appears that
the curves are moving with slightly different speeds, and it seems that
there is a slight difference in the slope of the gluon distribution which
cannot be removed by a pure rescaling. The differences are, however, tiny
and are not significant within the numerical certainty. For a much larger
interval $Y$ where the differences should become more visible, we in fact
see that the shapes of the curves and their speed are very similar to
each other. The difference in normalization is around a factor 1.5, with
the solutions with  $c=0.4$ and $\Delta=2.0$ lying below the default
ones. Furthermore, in Fig.~\ref{fig:ccfmeqplot7} we show a comparison
between Eqs.~\eqref{eq:ccfmdiffeq3} and \eqref{eq:bfkldiffeq2} using this
new set of parameters.  We have here again scaled up the solutions to
\eqref{eq:bfkldiffeq2}, but this time by a factor of 5.5, compared to 6.5
in Fig.~\ref{fig:bfklkcres2}. Once again, the two sets of fronts appear
to have similar speeds and shapes, with the CCFM solution progressing
slightly faster than the BFKL one. We have checked that the speeds of the
fronts come out the same as in the default case of
Fig.~\ref{fig:bfklkcres2}.

\begin{figure}[t] {\centerline{
  \includegraphics[angle=270, scale=0.57]{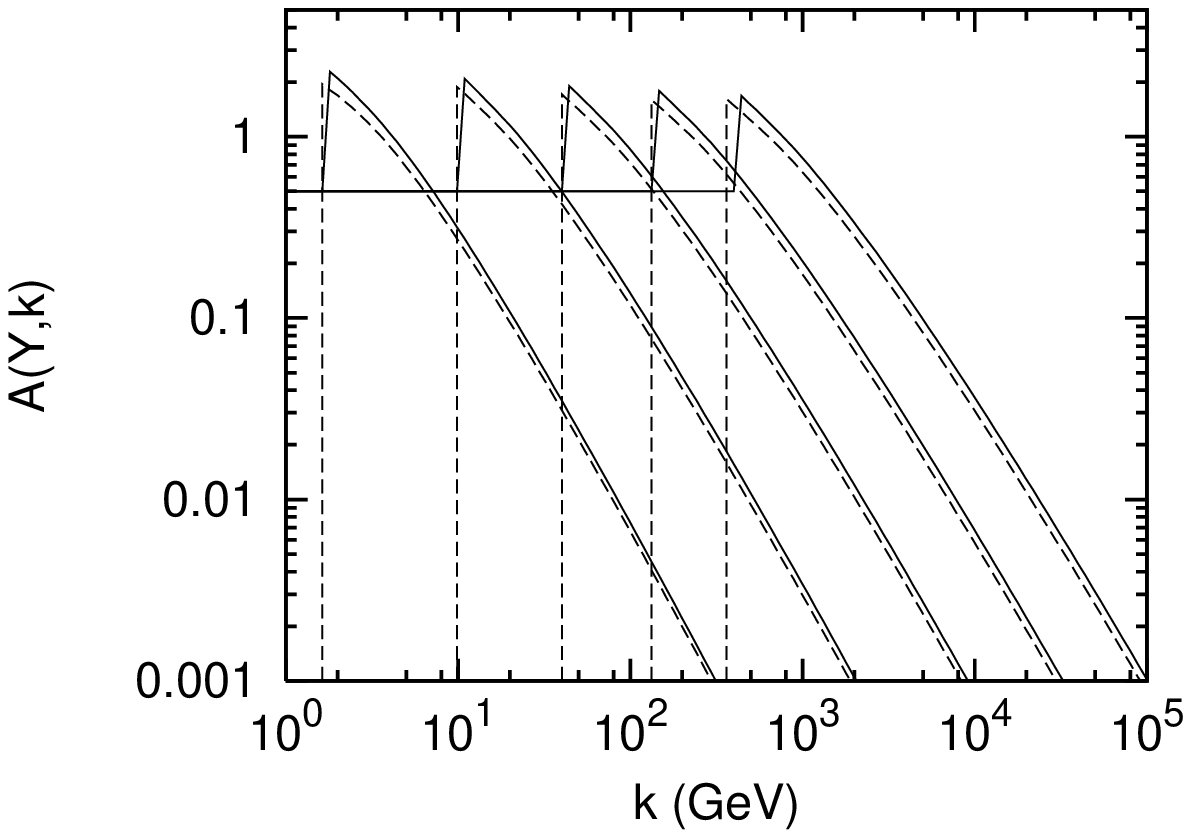}
  \includegraphics[angle=270, scale=0.57]{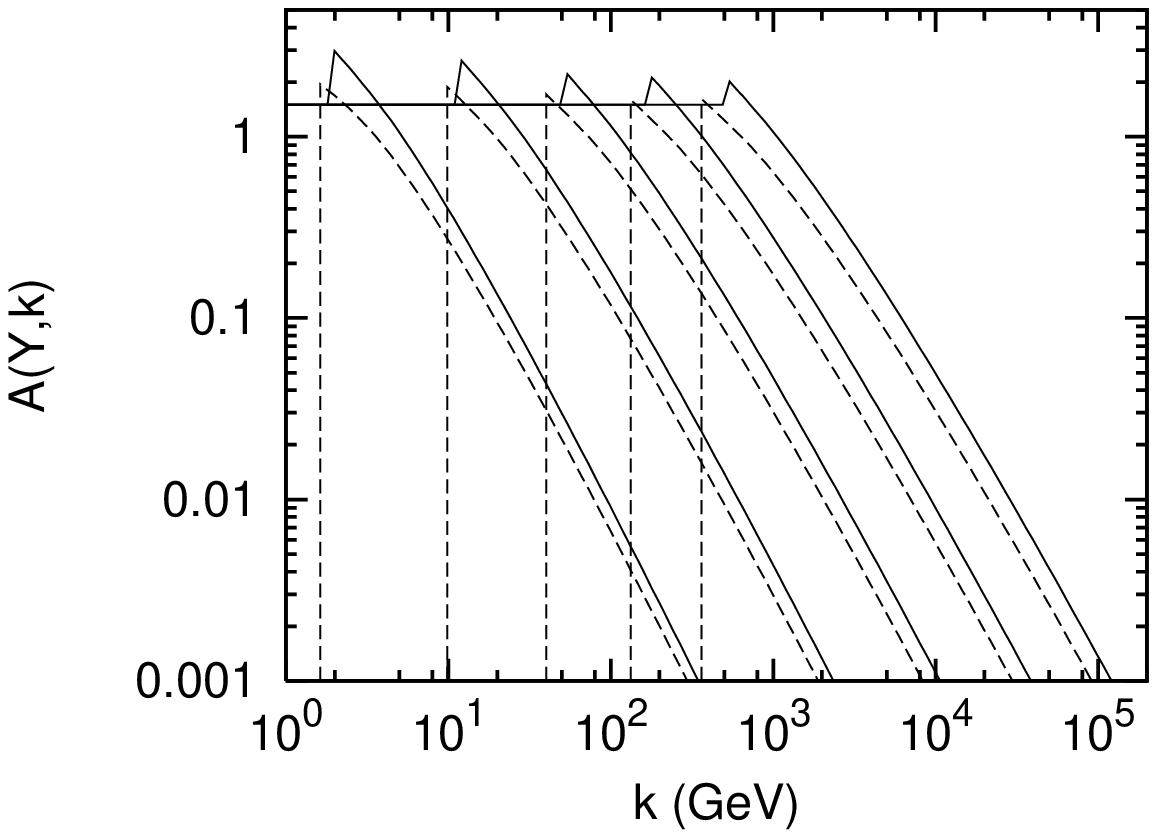}}
  \caption { \sl The solutions to \eqnum{eq:ccfmdiffeq3} with different
    boundary conditions for a running $\abar$, and for $Y=10, 20, 30, 40$ and 50.
    Left: $\mcal{A} = 0.5$ (solid lines)
    compared to the default $\mcal{A} = 0$ (dashed lines).
    Right: $\mcal{A} = 1.5$ (solid lines)
    compared to the default $\mcal{A} = 0$ (dashed lines).
    \label{fig:ccfmeqplot5} }
}
\end{figure}

Let us finally consider different versions of the saturation boundary, to
see whether this might affect our results. So far we have always applied
a totally absorptive boundary to mimic the non--linear physics of
saturation. From a theoretical viewpoint, it should be no fundamental
difference between enforcing $\mcal{A}$ to vanish behind the saturation
front, or fixing it to some non--zero constant value of order one: various
choices for this value must lead to the same front dynamics at
asymptotically high energies. One may however wonder whether differences
can be important in practice, for non asymptotic energies. We have
therefore tried two other boundary conditions as well, namely
$\mcal{A}=0.5$ and respectively $\mcal{A} = 1.5$. The corresponding
results are shown in Fig.~\ref{fig:ccfmeqplot5}, where they are also
compared to the results obtained with the absorptive boundary condition
$\mcal{A}=0$. The differences are find to be rather minor and correspond
more or less to rescaling the gluon distribution. In particular, the
energy dependence of $Q_s$ and the shape of $\mcal{A}$ come out the same
in all cases and for values of $Y$, including the relatively small ones.

\section*{Acknowledgements}

We would like to thank G\"osta Gustafson and Gavin Salam for valuable
discussions.  This work is supported in part by
Agence Nationale de la Recherche via the programme ANR-06-BLAN-0285-01.

\appendix

\section{On the correct form of $\Delta_{ns}$}
\label{sec:correctnonsud}\setcounter{equation}{0}

In this appendix we would like to make clear the correct form of the
non-Sudakov form factor $\Delta_{ns}$, as it is usually written in a way
which is not entirely correct. In the literature one usually finds the
formula
\begin{eqnarray}
\Delta_{ns}(k) &=& \exp \left( -\abar \int_{z_k}^{1}\frac{\rmd z}{z}
\int \frac{\rmd q^2}{q^2} \theta(Q_k^2 - q^2) \theta(q^2 -
z^2p_k^2) \right ) \nonumber \\
&=&  \exp \left( -\abar \ln\left(\frac{z_0}{z_k}\right )
\ln\left(\frac{Q_k^2}{z_0z_kp_k^2}\right) \right),
\label{eq:wrongnonsud}
\end{eqnarray}
where
\begin{equation*}
z_0 = \left \{
\begin{array}{rrl}
   1 & \text{if} & Q_k/p_k > 1 \\
   Q_k/p_k & \text{if} & z_k < Q_k/p_k \leq 1 \\
   z_k & \text{if} &  Q_k/p_k \leq z_k
\end{array} \right .
\end{equation*}
In this formula, the integrand is always negative so that $\Delta_{ns}$
always gives a suppression. Although this seems reasonable, it is,
however, not correct.
\begin{figure}[t] {\centerline{
    \includegraphics[angle=0, scale=0.6]{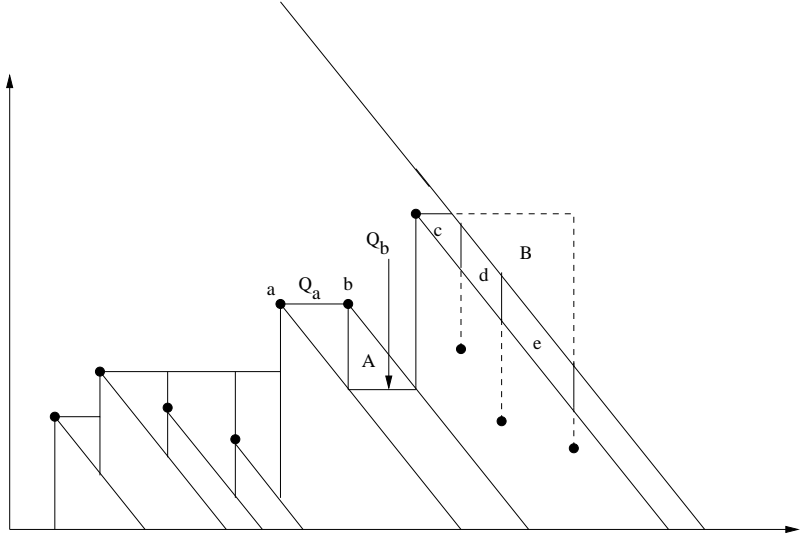}}
\caption { \sl\label{fig:ccfmhardnsoft2} The representation of the situation
  where the real gluon (gluon $b$) takes a large fraction of the
  transverse momentum of the virtual propagator emitting it ( $Q_a$).
  The next propagator $Q_b$ satisfies $Q_b < q_b$. In this case the
  correct non-Sudakov is obtained by integrating over region $A$
  with opposite sign.}
}
\end{figure}

When we identified the areas $A_k$ in Fig.~\ref{fig:ccfmhardnsoft} we
assumed $Q_k > z_kq_k$. Consider now the set of emissions in
Fig.~\ref{fig:ccfmhardnsoft2}. Here we also show the maximum allowed
angle $\bar{\xi}$. Note that the region $B$ would have been included in
BFKL but is excluded in CCFM. The smaller regions $c$, $d$ and $e$ are
included in CCFM. The interesting region here is the one marked by $A$.
It is contained within the region $C_b$ and therefore also in the Sudakov
associated with the gluon $b$. When $b$ is emitted from $Q_a$, it  takes
almost all transverse momentum so that $Q_a \approx q_b \gg Q_b$. The
region $A$ is thus bounded from below by $Q_b$ as shown in the figure.
Here indeed $z_bq_b > Q_b$, and if we set $\Delta_{ne}=1$ here so that it
does not contribute then we have only the contribution from the Sudakov
form factor in this region. However, there should actually be no net
contribution from the form factors in this region. To see this one should
remember the result in Fig.~\ref{fig:bfkl2emiss}; when we multiply
$S_{ne}$ and $S_{eik}$, what is left over is a region bounded from above
by the virtual propagators $Q_k$.

There should therefore be no net contribution from the form factors in
region $A$. Thus in this region the Sudakov and the non-Sudakov must
cancel each other. If this is to happen we have to set
\begin{eqnarray}
\Delta_{ne} = \exp \left( +\bar{\alpha} A  \right )
\end{eqnarray}
here and not $\Delta_{ne} = 1$.

\begin{figure}[t] {\centerline{
    \includegraphics[angle=0, scale=0.6]{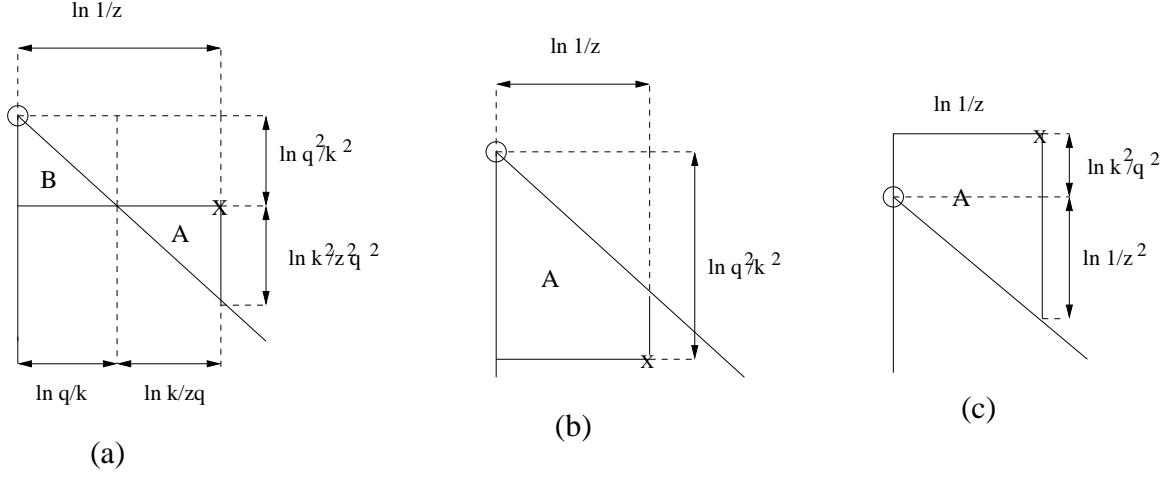}}
\caption { \sl\label{fig:nonsudpic}  Closer inspection of the phase
  region over which the non-Sudakov is integrated. The correct result
  is obtained by \emph{subtracting} the regions $B$ (in figure $(a)$)
  and $A$ (in figure $(b)$). The results are given in the text. }
}
\end{figure}

Actually let us from geometrical considerations derive the formula for
$\Delta_{ns}$ more carefully. In
Fig.~\ref{fig:nonsudpic} we show the phase space over which the
non-Sudakov is integrated. In figure $(a)$ we have (denoting the virtual
gluon momentum by $k$ and the real gluon momentum by $q$)
\begin{equation}
\Delta_{ns} = \exp ( -\abar (B-A) )
\end{equation}
where the areas $A$ and $B$ are marked in the figure. We see that
\begin{eqnarray}
B-A = \ln^2\frac{k}{zq} - \ln^2 \frac{k}{q}
 = \ln\frac{1}{z}\ln\frac{k^2}{zq^2}.
\end{eqnarray}
For figure $b$  we instead have
\begin{equation}
\Delta_{ns} = \exp ( -\abar (-A) ),
\end{equation}
and
\begin{equation}
A=\ln \frac{1}{z}\ln \frac{q^2}{k^2}
- \ln^2 \frac{1}{z} = - \ln\frac{1}{z}\ln
\frac{k^2}{zq^2},
\end{equation}
while for figure $(c)$ we have
\begin{equation}
\Delta_{ns} = \exp ( -\abar A ),
\end{equation}
and
\begin{equation}
A= \ln\frac{1}{z}\ln \frac{k^2}{q^2}
+ \ln^2\frac{1}{z} = \ln\frac{1}{z}\ln \frac{k^2}{zq^2}.
\end{equation}
Thus $\Delta_{ns}$ is indeed given by the formula in \eqref{eq:nonsud}.
As remarked before we see that  $\Delta_{ns}$ always gives a
suppression if $k^2> zq^2$, that is if the kinematical constraint is
assumed to hold.

\section{Non leading effects in the CCFM equation}
\setcounter{equation}{0}

When deriving the formulas for $\Delta_{ns}$ and $\Delta_s$ in
\eqref{eq:nonsud} and \eqref{eq:sud}  from the original form factors in
\eqref{eq:eik} and \eqref{eq:noneik}, recoils were not taken into effect.
In this section we  will propose modifications to
\eqref{eq:nonsud} and \eqref{eq:sud} in order to more properly take into
account recoil effects. Although such effects are formally suppressed,
experience tells us that they can be quite important for phenomenology.
For example in reference \citep{Bottazzi:1998rs}, $\Delta_{ns}$ was
modified beyond the leading order in a simple way, and it turns out that
the modification is significant for phenomenology. Besides, in the Monte
Carlo implementation in CASCADE full energy-momentum conservation is
already taken into account and one has then no reason not to modify the
virtual form factors accordingly.

\begin{figure}[t] {\centerline{
    \includegraphics[angle=0, scale=0.7]{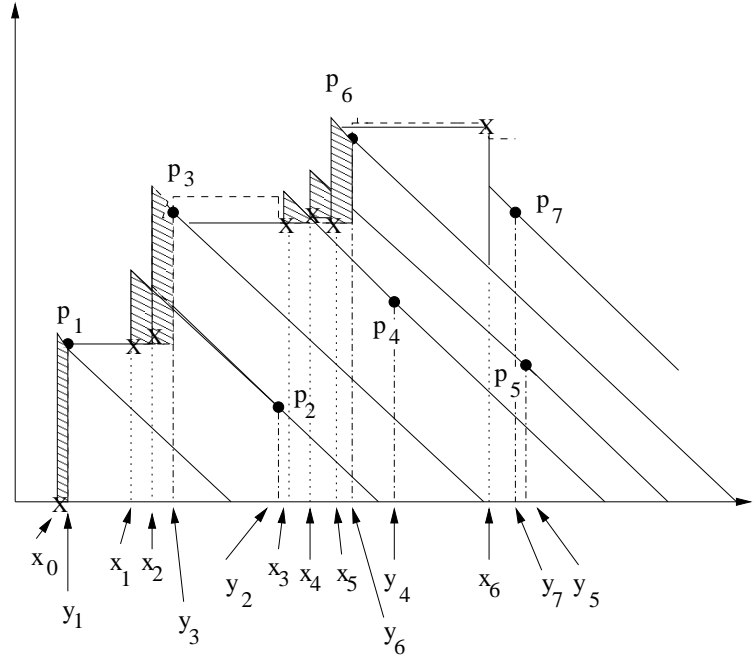}}
\caption { \sl\label{fig:recoilphase} A CCFM chain in the phase space,
  including also recoils in the emissions. The shaded regions
  indicate the difference in phase space between the original
  expression in \eqref{eq:eik} and \eqref{eq:noneik}, and the
  form factors $\Delta_{ns}$ and $\Delta_s$. }
}
\end{figure}

For the sake of demonstration we will distinguish between the momenta of
the $t$--channel gluons in the angular and, respectively, energy ordered
cascades, that we shall denote as $K_i$ and, respectively, $Q_i$. (Recall
that $K_i$ and $Q_i$ differ by soft gluon effects which have been
neglected in the main text.) In Fig.~\ref{fig:recoilphase} we again show
a gluon chain, but this time including the recoils that were so far
neglected. The shaded areas here show the difference in phase space
between the original expressions in \eqref{eq:eik} and \eqref{eq:noneik},
and $\Delta_{ns}$ and $\Delta_s$. The horizontal dashed lines in the
figure represent $Q_i$ while the $K_i$ are marked by crosses as before.
So actually also the region between the dashed horizontal lines and the
solid lines represent the differences although we have not shaded them.
The problem is that it is difficult to take into account this difference
since we cannot construct the exact $Q_i$ until the whole chain has been
generated, as the final energy ordering is a priori not known. We will,
however, see how we can take into account the shaded regions by modifying
$\Delta_{ns}$ and $\Delta_s$.

What we see is that one generally overestimates the phase region included
in $\Delta_{ns}$ and $\Delta_s$. This means that one overestimates the
suppression coming from these factors. Actually in some of the shaded
regions in the figure both $\Delta_{ns}$ and  $\Delta_s$ overestimate the
phase region so we get double extra suppression. Let us first concentrate
on the soft emissions. For the soft emission $p_i$, the $y$ integral is
integrated up to $x_{i-1}$. Now this will always cause an overestimate
since the subsequent emissions are emitted from $K_i$ with energy $x_i <
x_{i-1}$, and when eventually the next hard gluon is emitted it will have
energy less than $x_i$ itself. Thus we see that it would be better to
have the energy integral in $\Delta_s$ for the soft emission integrated
up to $x_i$, instead of $x_{i-1}$. Ideally we would like to have had the
integral up to the energy of the next hard gluon, but of course we do not
know what that energy will be beforehand. Thus we propose that
\begin{eqnarray}
\Delta_s^{soft}(i) \to \exp \left ( -\abar \int _{\xi_{i-1}}^{\xi_i} \frac{\rmd  \xi}{\xi}
\int_{\epsilon}^{x_{i}} \frac{\rmd y}{y}  \right )
\label{eq:newsoftsud}
\end{eqnarray}
for soft emissions. Again writing this in terms of momenta we get
\begin{eqnarray}
\Delta_s^{soft}(k) = \exp \left ( -\abar
\int_{z_{k-1}^2p_{k-1}^2}^{p_{k}^2}\frac{\rmd p^2}{p^2} \int_{\epsilon'}^{z_i} \frac{\rmd y}{y}  \right ).
\label{eq:newsoftsud2}
\end{eqnarray}
which should be compared to \eqref{eq:sud} (again $\epsilon'=q_0/p$).
 We see that the difference in the $y$
integral is
\begin{eqnarray}
\int_{\epsilon'}^1 \frac{\rmd y}{y} - \int_{\epsilon'}^{z_i} \frac{\rmd y}{y}
= \ln\frac{1}{z_i}
\end{eqnarray}
where of course for soft emissions $z_i \approx 1$.
\begin{figure}[t]{\centerline{
    \includegraphics[angle=0, scale=0.7]{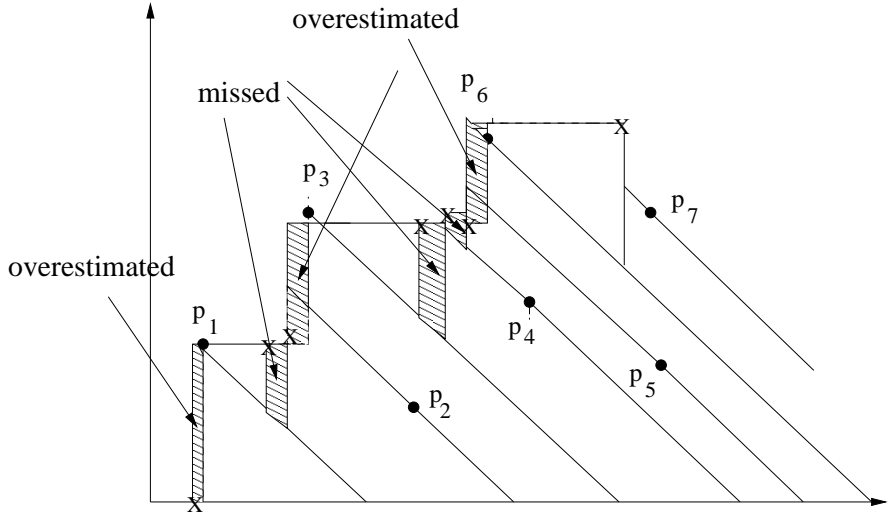}}
\caption { \sl\label{fig:recoilphase2} The shaded regions marked
  ``missed'' are regions which contribute to the original
  form factors but not to the new form factors, after the Sudakov for
  the soft emissions has been mofified.
  Regions marked ``overestimated'' are regions where we have too
  much suppression due to the new form factors. These differences can be
  removed by modifying the non-Sudakov and Sudakov form factors.}
}
\end{figure}

However, now we can also see that there is a region between $x_{i-1}$ and
$x_i$ which is missed since the contribution from $\Delta_s^{soft}$ has
been removed from this region, as illustrated in
Fig.~\ref{fig:recoilphase2}.
 Of course by removing that contribution
we get rid of the potentially large overestimate of the phase region. The
missing parts would have had zero area if there were no recoils, since
then $x_{i-1}$ and $x_i$ would have been equal. Thus we can fill in this
region by associating a non-Sudakov $\Delta_{ns}^{soft}$ with the soft
emissions as well. It is easy to see from Fig.~\ref{fig:recoilphase2}
that the missed part extends down from $K_{i-1}$ to the angle of emission
$i-1$, \emph{i.e.} to $\xi_{i-1}$. Thus for each soft emission we
associate
\begin{eqnarray}
\Delta_{ns}^{soft}(i) = \exp \left ( -\abar \int_{z_i}^1 \frac{\rmd z}{z}
\int_{z^2p_{i-1}^2}^{K_{i-1}^2} \frac{\rmd p^2}{p^2} \right )
\label{eq:softnonsud}
\end{eqnarray}
which we have written in terms of the rescaled momenta as before. Again
if $K_{i-1} <zp_{i-1} $, \eqref{eq:softnonsud} will give an enhancement
in some region but this is correct since in that case it will cancel an
oversuppression coming from $\Delta_s$.

From Fig.~\ref{fig:recoilphase2} we can see that there are still some
regions in which we have an oversuppression coming from both $\Delta_s$
and from $\Delta_{ns}$ associated with the hard emissions. Actually
depending on the kinematics (when $K_i < z_ip_i$), $\Delta_{ns}$ will
cancel the oversupression from $\Delta_s$ in some small region. This has
been taken into account in Fig.~\ref{fig:recoilphase2}. As one can see
one overestimates the phase region when $q_i > K_{i-1}$. Therefore in
this case we can modify the form factors as follows
\begin{eqnarray}
\Delta_s^ {hard}(i) &\to& \exp \left ( -\abar
\int_{\xi_{i-1}}^{\xi_{i}}\frac{\rmd \xi}{\xi} \int_{\epsilon}^{y_i} \frac{\rmd y}{y}  \right )
\,\,\,\,\, \mathrm{if} \,\,\, q_i > K_{i-1}     \nonumber \\
&=&  \exp \left ( -\abar
\int_{z_{i-1}^2p_{i-1}^2}^{p_{i}^2}\frac{\rmd p^2}{p^2} \int_{\epsilon'}^{1-z_i} \frac{\rmd y}{y}  \right )
  \label {eq:newhardsud}\\
\Delta_{ns}^{hard}(i) &\to& \exp \left ( -\abar \int_{x_i}^{y_i} \frac{\rmd y}{y}
\int_{\xi_i}^{\xi(K_{i}^2)} \frac{\rmd \xi}{\xi} \right )\,\,\,\,\, \mathrm{if} \,\,\, q_i > K_{i-1}
\nonumber \\
&=& \exp \left ( -\abar \int_{\frac{z_i}{1-z_i}}^{1} \frac{\rmd z}{z}
\int_{z^2p_i^2}^{K_{i}^2} \frac{\rmd q^2}{q^2} \right )
\label{eq:newhardnonsud}
\end{eqnarray}
so that in each factor we integrate in energy up to $y_i$ instead of
integrating up to $x_{i-1}$. In this way we minimize the overestimation
of the true phase region.

To summarize, the standard splitting probability for the emission of the
$i$th gluon which is given by
\begin{eqnarray}
dP(i) = \abar \,dz_i\frac{\rmd ^2p_i}{\pi p_i^2}\left(
\frac{\Delta_{ns}(i)}{z_i} + \frac{1}{1-z_i}\right ) \Delta_s(i)
\theta(p_i- z_{i-1}p_{i-1})
\end{eqnarray}
with $\Delta_s(i)$ given in \eqref{eq:sud} and $\Delta_{ns}(i)$ in
\eqref{eq:nonsud}, is now replaced by
\begin{eqnarray}
dP(i) = \abar \, dz_i\frac{\rmd ^2p_i}{\pi p_i^2}\left(
\theta(K_{i-1} - (1-z_i)p_i)\left(\frac{\Delta_{ns}(i)\Delta_s(i)}{z_i} +
\frac{\Delta_{ns}^{soft}(i)\Delta_{s}^{soft}(i)}{1-z_i}\right ) \right. + \nonumber \\
\left. \theta((1-z_i)p_i - K_{i-1}) \left(\frac{\Delta_{ns}^{hard}(i)\Delta_s^{hard}(i)}
{z_i} + \frac{\Delta_{ns}^{soft}(i)\Delta_{s}^{soft}(i)}{1-z_i}\right )
\right )\theta(p_i- z_{i-1}p_{i-1})
\label{eq:newsplitprob}
\end{eqnarray}
where $\Delta_{s}^{soft}(i)$, $\Delta_{ns}^{soft}(i)$,
$\Delta_s^{hard}(i)$ and $\Delta_{ns}^{hard}(i)$ are given by
\eqref{eq:newsoftsud2}, \eqref{eq:softnonsud},
  \eqref{eq:newhardsud} and \eqref{eq:newhardnonsud} respectively.

It is really non-trivial to implement these changes in the type of
numerical procedure that we use in this paper. However, they are suitable
for implementing in a Monte Carlo procedure and as such it would be
straightforward to implement them in CASCADE, and it would be interesting
to see how large the effects are. One can also put in theta functions to
make sure that for example $z < 0.5$ for hard emissions, and $z > 0.5$
for soft emissions.

\section{Rewriting $\Delta_{ns}$ in CCFM and the comparison to BFKL}
\setcounter{equation}{0}

\begin{figure}[t]{\centerline{
    \includegraphics[angle=0, scale=0.6]{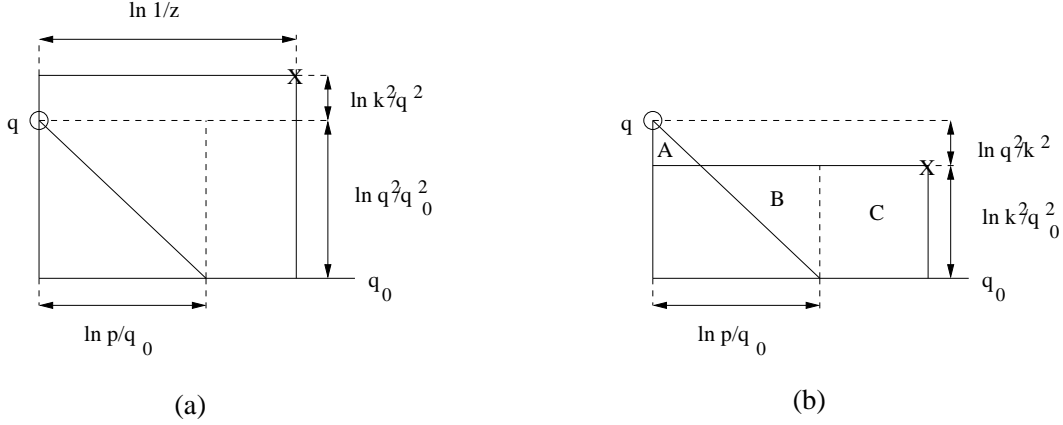}}
\caption { \sl\label{fig:nonsudpic2} The phase space for $\Delta_{ns}$ in
CCFM
  in case we take very large steps in $z$, such that $z < q_0/p$. }
}
\end{figure}

In the literature one usually finds the statement that
\eqnum{eq:ccfminteq2} reduces to the BFKL equation in the high energy
limit. In this appendix we would like to elaborate more on this. The key
point to this statement is the derivation of $\Delta_{ns}$ appropriate in
the formal high energy limit, where energy is infinite, and importantly
$\abar \to 0$ for \emph{fixed} $q_0$.

Previously we derived $\Delta_{ns}$ following figure \ref{fig:nonsudpic}.
In this case, however,  we must consider what happens in case $z$ is so
small that the diagonal line through $q$, on which $\xi$ is constant and
equal to the angle of the real gluon $q$, cuts the line representing the
soft cutoff $q_0$ before cutting the vertical line through $k$. This
situation is illustrated in Fig.~\ref{fig:nonsudpic2}. The horizontal
line at the bottom just indicates the momentum cut $q_0$. What happens
here is that $z < q_0/p$ where $p$ is the rescaled momentum of the real
emitted gluon $q$. Actually since we now consider the small $z$ limit,
there is no need to make a distinction between the rescaled and the
regular momenta, thus one might as well set $p=q$. From figure $(a)$, one
can see that $\Delta_{ns}$ is integrated over a region of total size
\begin{eqnarray}
A &=& \ln\frac{q^2}{q_0^2}\cdot \left( \ln\frac{1}{z}
 - \frac{1}{2}\ln\frac{p}{q_0} \right ) +
\ln\frac{k^2}{q^2} \ln \frac{1}{z}
\nonumber \\
&=& \ln\frac{1}{z}\ln\frac{k^2}{q_0^2}
- \ln^2 \frac{q}{q_0}
\label{eq:modnonsud}
\end{eqnarray}
where in the last equality we have just set $p=q$. $\Delta_{ns}$ is then
given by $\exp(-\abar A)$. In the case shown in figure $(b)$ we instead
have
\begin{eqnarray}
\Delta_{ns} = \exp(-\abar (-A+B+C)),
\end{eqnarray}
where
\begin{eqnarray}
-A+B+C &=& \ln\frac{1}{z}\ln \frac{k^2}{q_0^2}
+ \ln  \frac{p}{q_0} \ln  \frac{k^2}{pq_0}
 - \ln  \frac{p}{q_0} \ln \frac{k^2}{q_0^2}
\nonumber \\
&=& \ln\frac{1}{z}\ln \frac{k^2}{q_0^2}
- \ln^2 \frac{q}{q_0}
\end{eqnarray}
where we have again set $p=q$ in the last equality. We now remember that
the virtual form factor is in BFKL given by \eqref{eq:bfklnoneik}, which
equals
\begin{eqnarray}
\exp \left( - \abar \ln\frac{1}{z}\ln\frac{k^2}{q_0^2}
\right ) .
\end{eqnarray}
Assuming now that the typical $z$ are so small that the result in
\eqref{eq:modnonsud} is valid, we can write the CCFM non-Sudakov form
factor as
\begin{eqnarray}
\Delta_{ns}(z, k, q) = \Delta(z, k) \cdot \Delta^R(q)
\end{eqnarray}
where $\Delta(z, k)$ is the BFKL form factor, and
\begin{eqnarray}
\Delta^R(q) = \exp \left( \abar \ln^2 \frac{q}{q_0}
\right )
\label{eq:deltar}
\end{eqnarray}
with the "$R$" standing for ``Residual''.  In this case
\eqnum{eq:ccfminteq2} can be written as
\begin{eqnarray}
\mathcal{A}(Y, k, \bar{q}) = \abar \int_0^Y dy
\int \frac{\rmd ^2q}{\pi q^2}\, \theta (Y - y + \ln(\bar{q}/q))\Delta(Y-y, k)\Delta^R(q)
\, \mathcal{A}(y, k', q),
\label{eq:modccfminteq2}
 \end{eqnarray}
where $\Delta$ is the BFKL non-Sudakov form factor, while $\Delta^R(q)$
is given in \eqref{eq:deltar} and is independent of $Y$. Now of course if
one considers the leading contributions in $Y-y$, then the angular
ordering theta function can be neglected and, since we moreover have
$\abar \to 0$ one can let $\Delta^R \to 1$. In that case obviously the
BFKL equation is recovered.

It is obvious from the derivation here that the form for $\Delta_{ns}$ is
appropriate only for very high energies, and for $\abar \to 0$ with
fixed cutoff $q_0$.  Thus if one makes a certain calculation, for example
jet production rates using $q_0$ as a resolution scale, one needs to be
careful considering the limit when $q_0\to 0$. Obviously in that case we
gradually go from the situation in Fig.~\ref{fig:nonsudpic2} to the
situation in Fig.~\ref{fig:nonsudpic}. This means that we go from the
result \eqref{eq:modnonsud} to the result in \eqref{eq:nonsud}, in which
case the similarity to BFKL is somewhat lost. For similar discussions see
also \citep{Salam:1999ft, Forshaw:1998uq, Marchesini:1994wr}. Of course
in a sense this is a bit artificial since from the beginning we had that
$S_{ne}^2\cdot S_{eik}^2=\Delta^{BFKL}$, but remember also the discussion
at the end of section \ref{sec:angord}.

\section{Increasing the accuracy of the real-virtual cancellations}
\setcounter{equation}{0} \label{sec:betteraccuracy}

In this final appendix, we would like discuss a bit more carefully the
definition of the $k_\perp$-conserving emissions, and the real-virtual
cancellations discussed in section \ref{sec:inclccfm}. We were not very
careful when defining  the $k_\perp$-conserving emissions, and especially
the situation  when the momenta of these emissions, $q$, become
comparable to $k$ was not discussed. To be more careful, one would have
liked to define $k_\perp$-conserving emissions as emissions with say $q <
a k$ for some $a$ which is strictly smaller than 1, as this would give a
better accuracy. As the procedure in section \ref{sec:inclccfm} gives
rise to an intercept higher than BFKL, one might try to thus increase the
accuracy in the definition of the $k_\perp$-conserving emissions to
derive a new equation. Indeed we saw in section \ref{sec:anmdim} that a
better agreement with BFKL was obtained in the asymptotic limit if a
constant negative term proportional to the gluon distribution is added to
the equation. As such a contribution is related to the virtual
corrections, we might indeed guess that it has something to do with the
real-virtual cancellations.

The issue of improving the accuracy in the definition of the
$k_\perp$-conserving emissions was discussed first in
\citep{Salam:1999ft}, again considering only the possibility that $k >
k'$. In our case we also look at the opposite case, again with the
kinematical constraint included. The procedure is the same as before, and
one can easily derive new evolution equations. As we later discuss,
however, this procedure does not simplify the problem of solving the more
exclusive equation in \eqref{eq:ccfminteq2}, although it sheds some light
on the physical origin of the discrepancy between BFKL and the asymptotic
behaviour of equations  \eqref{eq:ccfmdiffeq3} and
\eqref{eq:ccfmdiffeq4}.

\begin{figure}[t]{\centerline{
    \includegraphics[angle=0, scale=0.6]{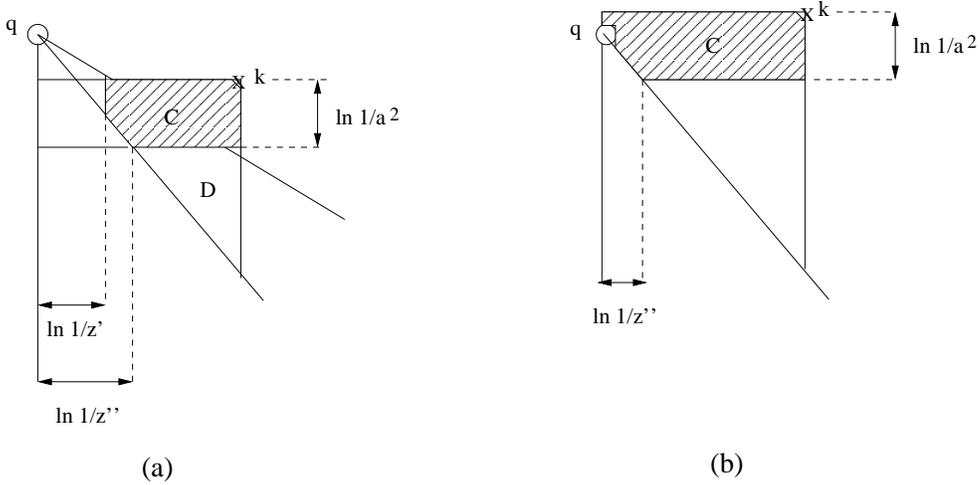}}
\caption { \sl\label{fig:nonsudpickc2} (a): Region $D$ is the region
  where we have real $k_\perp$-conserving emissions when the accuracy of the
  definition of $k_\perp$-conserving emissions are improved. The shaded region
  is than the region left over by the real-virtual cancellation. Here $q > k$.
  It is seen that $z'= k^2/q^2$ and $z''=ak/q$. We have assumed $z' > z''$, and
  the case $z' < z''$ is discussed in the text.
  (b): Same as in figure (a) but now for $q < k$. }
}
\end{figure}

Since now the $k_\perp$-conserving emissions have momenta $a$ times
smaller than the corresponding virtual propagator, they will not
completely cancel $\Delta_{ns}$, but there will be a region left over
from the cancellation. Assume first that $k < q$. Remember from
Fig.~\ref{fig:nonsudpickc} that  $\Delta_{ns} = \exp(-\abar A)$ where $A$
was defined in the figure. The region $A$ from Fig.~\ref{fig:nonsudpickc}
is then the sum of the regions $C$ and $D$ in
Fig.~\ref{fig:nonsudpickc2}(a), $A=C+D$. Region $D$ is the region where
we have the real $k_\perp$-conserving emissions, and summing over these
emissions we get a factor $\exp(\abar\, D)$ and what is left from the
multiplication with $\Delta_{ns}$ is therefore $\exp(-\abar \, C)$. From
Fig.~\ref{fig:nonsudpickc2}(a) we see that
\beq
C &=& \ln\frac{1}{a^2}\ln\frac{z'}{z} - \frac{1}{2} \ln
\frac{z'^2q^2}{a^2k^2} \ln\frac{z'}{z''}
\nonumber \\
&=& \ln \frac{1}{a^2} \ln\frac{k^2}{zq^2} - \ln^2 \frac{k}{aq} =
\ln\frac{1}{z} \ln \frac{1}{a^2} - \ln^2\frac{q}{ak}.
\label{eq:C}
\eeq
If instead $k > q > ak$, we see from Fig.~\ref{fig:nonsudpickc2}(b) that
region $C$ is again given by the formula above. Thus this is the region
left over by the cancellations. Actually, in
Fig.~\ref{fig:nonsudpickc2}(a) we assumed that $z' > z''$ which means
that $a < k/q$. If instead $a > k/q$ we get a slightly different
situation, since the region $C$ would then be replaced by a rectangular
region. In that case it is easy to see that we get
\beq
C' = \ln \frac{1}{a^2}\ln\frac{k^2}{zq^2} =
 \ln\frac{1}{z} \ln \frac{1}{a^2} - \ln^2\frac{q}{ak} +  \ln^2\frac{k}{aq}
\label{eq:Cprime}
\eeq
which contains an extra term compared to the formula for $C$. Since in
this case $k < q$, we can neglect the last term, however. Actually we
could well approximate the regions $C$ and $C'$ by the first term only
(which contains ln$(1/z)$). This is indeed appropriate as we will below
only study the asymptotic limit. We see that this term will generate an
additional contribution in the right--hand side of the differential
equation which reads
\beq
-\abar \ln\frac{1}{a^2} \, \mathcal{A}(Y, k).
\eeq
This is infact the term we were looking for in order to reduce the Mellin
space eigenfunction. However, note also that we have real emissions left
over in the same region and these will modify the real kernel as well. We
will write down the asymptotic equation below and study its Mellin
transform.

\begin{figure}[t]{ \centerline{
    \includegraphics[angle=270, scale=0.55]{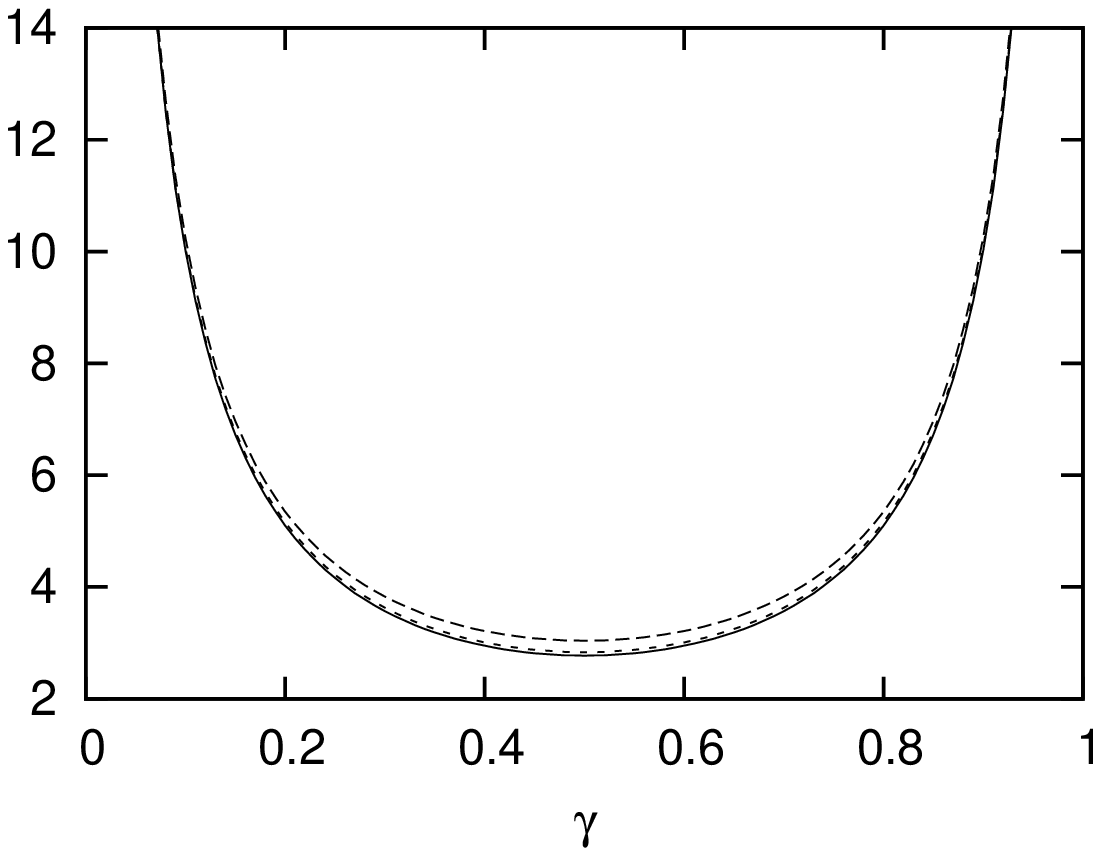}
    \includegraphics[angle=270, scale=0.55]{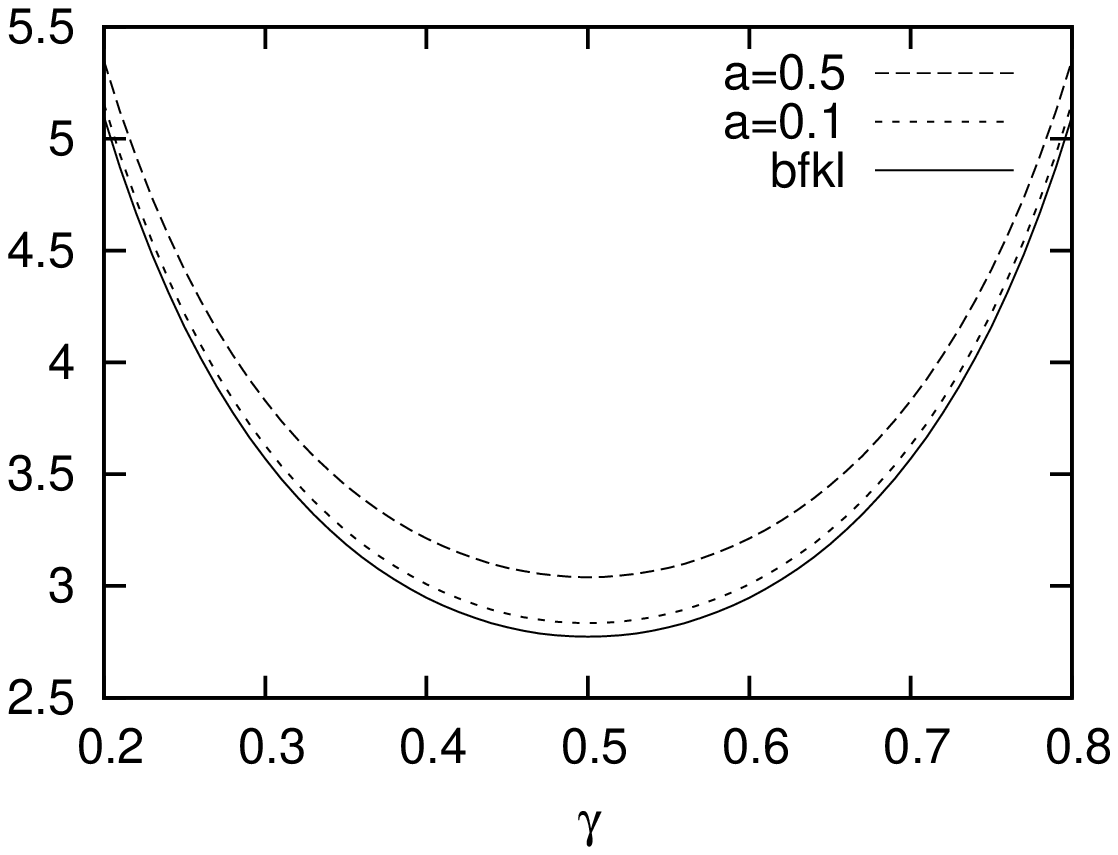}}
    \caption { \sl\label{fig:chiplota} The Mellin space
      eigenfunction of \eqnum{eq:neweq} for $a=0.5$ (long dashed lines)
      and $a=0.1$ (short dashed lines) for $\abar \to 0$ together with the BFKL eigenfunction (solid lines).}
}
\end{figure}
Before doing that, let us comment on the procedure of generating new
equations. As an explicit constraint we can obviously use $q >
\mathrm{min}(ak, ak')$. The problem now, however, is that we can no
longer justify the step from \eqref{eq:firststep} to
\eqref{eq:secondstep} where we dropped all the dependence on the third
parameter in the argument of $\mcal{A}$. The dependence on $\bar{q}$
drops again on account of the kinematical constraint, but we can no
longer guarantee that $q \geq k'$. If we nevertheless assume this then we
lose the accuracy gained by introducing $a$. Infact if it was possible to
drop all dependence on the third parameter by just using the kinematical
constraint, then we would not need to bother cancelling $\Delta_{ns}$
since it would have been equally easy to solve \eqref{eq:ccfminteq2}
directly. As far as the asymptotic behaviour is concerned, however, we
are justified to go from \eqref{eq:firststep} to \eqref{eq:secondstep}.
(Unless we are looking at the $\abar \to 0$ limit, we would not
be able to drop the second and third terms in \eqref{eq:C} and \eqref{eq:Cprime}
anyway.)
Thus to study the intercept we make the same steps as before and arrive
at the asymptotic equation
 \beq
  \partial_Y\mathcal{A}(Y, k) = \abar \int \frac{\rmd k'^2}{\vert k^2-k'^2\vert} \,
h_a(\kappa) \mathcal{A}(Y,k') -\abar \ln\frac{1}{a^2} \, \mathcal{A}(Y,
k), \label{eq:neweq}
\eeq
where now
\beq
h_a(\kappa) = 1
-\frac{2}{\pi}\arctan\left(\frac{1+\sqrt{\kappa}}{1-\sqrt{\kappa}}\sqrt{
\frac{2\sqrt{\kappa}-1-(1-a^2)\kappa}{2\sqrt{\kappa}+1+(1-a^2)\kappa}}
\right ) \theta(2\sqrt{\kappa}-(1-a^2)\kappa-1), \label{eq:ha}
\eeq
and $\kappa$ is defined as before. 

It is straightforward to study the Mellin space eigenfunction
numerically. In Fig.~\ref{fig:chiplota}
we plot the eigenfunction for $a=0.5$ and $a=0.1$ together with the BFKL
result. As $a$ decreases we indeed see that the curves move toward the
BFKL eigenfunction. We basically get a constant shift around $\gamma =
0.5$, and the behaviour at the endpoints $\gamma = 0, 1$ is not changed.
For the intercept, one goes from 3.23 at $a=1$ to 3.02 at $a=0.5$, and
2.83 at $a=0.1$. The saddle point occurs at $\gamma = 0.5$ in all cases.
The saturation saddle point moves from $\gamma_s \approx 0.35$ at $a=1$
towards $\gamma_s \approx 0.37$ at $a=0.1$.

\bibliographystyle{utcaps}
\bibliography{refs}

\end{document}